\documentclass[12pt]{article}
\pdfoutput=1
\usepackage{amssymb}
\usepackage{amsmath}
\usepackage{amsthm}
\usepackage{color}
\usepackage{diagbox}
\usepackage[section]{placeins}
\usepackage[figuresright]{rotating}
\usepackage{rotating}
\usepackage{multirow}
\usepackage{caption}
\usepackage{subcaption}
\usepackage{tikz}
\usepackage{pgfplots}
\usepackage{graphics}
\usepackage{cite}
\usepackage{float}
\usepackage{authblk}
\usepackage{mathtools}
\usepackage{mathrsfs}
\usepackage[raggedright]{titlesec}
\usepackage{bm}
\usepackage{easyReview}
\usepackage[colorlinks = true,
linkcolor = blue,
urlcolor  = blue,
citecolor = blue,
anchorcolor = blue]{hyperref}
\usepackage{siunitx}
\begin{document}



\def\e{{\rm e}}
\def\i{{\rm i}}
\def\no{\nonumber}
\global\long\def\bra#1{\langle #1 |}
\global\long\def\ket#1{| #1 \rangle }
\global\long\def\kket#1{|| #1 \rangle\rangle}
\global\long\def\braket#1#2{\left\langle #1|#2\right\rangle }

\newtheorem{Theorem}{Theorem}
\newtheorem{Definition}{Definition}
\newtheorem{Proposition}{Proposition}
\newtheorem{Lemma}{Lemma}
\newtheorem{Corollary}{Corollary}
\newtheorem{remark}{Remark}
\newtheorem{hypothesis}{Hypothesis}
\newtheorem*{hypothesis*}{Hypothesis}
\newtheorem*{remark*}{Remark}
\renewcommand{\theequation}{\thesection.\arabic{equation}}
\renewcommand{\thetable}{\thesection.\arabic{table}}
\renewcommand{\thehypothesis}{\thesection.\arabic{hypothesis}}
\renewcommand{\thefigure}{\thesection.\arabic{figure}}
\begin{titlepage}
\title{Bethe-root configurations and spectral degeneracy in the open XXZ chain with degenerate boundaries}

\author{Shizhang Wu$^{a,c}$, Shu Chen$^{a,c}$, Junpeng Cao$^{a,b,c}$, Xin Zhang$^{a}$\thanks{xinzhang@iphy.ac.cn}}

\affil{$^a$ Beijing National Laboratory for Condensed Matter Physics,
Institute of Physics, Chinese Academy of Sciences, Beijing 100190, China  \\$^b$ Peng Huanwu Center for Fundamental Theory, 710127, Xi’an, China\\$^c$ School of Physical Sciences, University of Chinese Academy of Sciences, 100049, Beijing, China}

\end{titlepage}

\maketitle
\begin{abstract}
We investigate the Bethe-root configurations in the spin-1/2 XXZ chain with degenerate open boundaries. The physical solutions of the Bethe Ansatz equations are classified into three types at the fully degenerate points and two types at the partially degenerate points, for which we propose counting formulas for the number of physical solution sets of each type. Furthermore, we observe spectral degeneracies at the fully degenerate points and show that they can be naturally explained by the presence of specific phantom strings in the Bethe roots. {At degenerate points, the Bethe states corresponding to different Bethe-root configurations also behave differently, which provides another justification for the proposed classification of Bethe roots. We further discuss the Bethe-root classification and the resulting spectral degeneracies in the diagonal limit and at the quasi-degenerate points}. 
\end{abstract}

\section{Introduction}
The spin-$\frac{1}{2}$ Heisenberg model is a paradigmatic model in quantum integrability. Its exact solvability, rich algebraic structure, and deep connections with quantum many-body physics, statistical mechanics, and representation theory have kept it at the center of research for decades \cite{Bethe:1931, BaxterBook, Korepin, faddeev:1996}. Introducing boundaries further enriches the picture. The one-dimensional open Heisenberg model, governed by the reflection equation, provides a controlled arena for investigating quantum integrability and Bethe Ansatz techniques \cite{Cherednik:1984,Sklyanin:1988, Alcaraz:1987,Vega:1993,deVega:1994}.

For the Heisenberg model with generic boundary magnetic fields, although integrability has been established \cite{Cherednik:1984, Sklyanin:1988,deVega:1994,Inami:1994,Ghoshal:1994}, the $U(1)$ symmetry is broken. Consequently, one cannot easily identify a reference (vacuum) state and the associated generators required to construct the exact eigenstates and their corresponding eigenvalues. The eigenvalue and eigenstate problem of the open Heisenberg model was systematically addressed by an analytic approach---the off-diagonal Bethe Ansatz (ODBA) method \cite{OffDiagonal,Cao2013,Cao2013:2,Zhang:2014,Zhang:2015}. In contrast to the conventional Bethe Ansatz equations (BAEs), which take a symmetric form, the equations resulting from the ODBA method are considerably more involved and cannot be solved in a simple way. {Moreover, many other approaches have been developed over the past years to solve the Heisenberg chain with open boundaries and to obtain its spectrum, eigenstates, correlation functions and thermodynamic properties, including the separation of variables (SoV) method \cite{Niccoli:2012,Faldella:2014,Kitanine:2014,Niccoli2022}, the modified algebraic Bethe ansatz
\cite{Belliard:2013:XXX,Belliard:2014:1,Belliard:2014:2,Belliard:2021}, the representation-theoretic approach \cite{Chernyak:2022,Chernyak:2023}, the non-linear integral equations \cite{Frahm:2026}, the $t$-$W$ relation \cite{Qiao:2021} and the $Q$-system \cite{Jiang:2026}.}

Nevertheless, in certain special cases, the conventional BAEs can be recovered and the elegant structure of these equations enables the study of the physical properties of the model for both finite-size systems and in the thermodynamic limit. These special cases fall into two classes: (1) the XXX and XXZ models with diagonal boundary fields, for which the system possesses $U(1)$ symmetry \cite{Alcaraz:1987,Sklyanin:1988,Kitanine:2007,Kitanine:2008}; and (2) the XXX, XXZ, and XYZ models with constrained non-diagonal boundary fields \cite{Fan1996,Cao2003,Nepomechie:2003_1,Nepomechie:2003_2,Jan2005,Yang2006,Crampe:2010,Belliard:2013,Zhang2021Phantom,Zhang2021CBA,Chernyak:2023,Zhang:2022}. The tractability of the BAEs largely accounts for the sustained interest in these classes of open Heisenberg models.

Motivated by the above, in this paper we focus on the open XXZ model with boundary fields admitting conventional BAEs, covering both the constrained non-diagonal and diagonal cases. In Ref. \cite{Jiang:2024}, the authors studied the XXZ model with diagonal boundary fields, where they classified the Bethe-root configurations and observed spectral degeneracies. However, whether similar structures exist for constrained non-diagonal boundaries and how the root configurations are related to the resulting spectral degeneracies remain unclear.

In this paper, we present our main results on the XXZ model with constrained non-diagonal boundary fields. The exact solutions of the model are parameterized by two sets of Bethe Ansatz equations, namely \eqref{BAE_G_M^+_x} and \eqref{BAE_G_M^-_x}, which correspond to the invariant subspaces $G_M^+$ and $G_M^-$, respectively \cite{Zhang2021Phantom,Zhang2021CBA}. We demonstrate that at degenerate points (fully or partially degenerate), the Bethe roots in \eqref{BAE_G_M^+_x} may be pinned by the boundary, leading to a natural classification based on the number of pinned roots. At the fully degenerate points, the Bethe roots are classified into three types, whereas only two types remain at the partially degenerate points. The degenerate boundaries either reduce the number of unpinned roots or simplify the form of the BAEs. In both cases, we propose explicit counting formulas for the number of physical solutions of each type.

Another interesting phenomenon at the fully degenerate points is the spectral degeneracy. We observe that the BAEs for two spaces $G_M^\pm$ overlap. For one set of BAEs, there exist solutions in which all Bethe roots are finite. For the other set, the solutions consist of a copy of the Bethe roots from the former set together with a so-called phantom string \cite{Zhang2021:SHS,Zhang2021Phantom}, which does not contribute to the energy. This results in the degeneracy of certain energy levels. Apart from its role in explaining the degeneracy, the phantom string is also essential for counting the number of solutions. At the partially degenerate points, by contrast, such phantom strings are absent and the spectral degeneracy disappears.

{At degenerate points, the eigenstate of the Hamiltonian (the Bethe state) also exhibits interesting features. In the generic case, the Bethe state in the subspace $G_M^+$ can be expanded in terms of a set of factorized chiral vectors with kinks, where the maximum number of kinks equals $M$ \cite{Zhang2021Phantom,Zhang2021CBA}. For one class of degenerate point, when one or two Bethe roots are pinned by the boundary, the maximum number of kinks in the corresponding Bethe state may decrease, signaling the existence of internal subspaces in $G_M^+$. For the other class, the maximum number of kinks remains the same, whereas the wave function becomes simplified in the presence of a boundary-pinned root.}

We also discuss the Bethe-root configurations and spectral degeneracies in the diagonal limit. Unlike the non-diagonal case, the loss of boundary degrees of freedom reduces the number of possible types in the classification of Bethe roots. Furthermore, the degeneracy can also be explained by the phantom string.

{Moreover, we find that phantom strings and boundary-pinned roots also exist in other cases, which we call quasi-degenerate points and which are generalizations of the degenerate points.
We explore the boundary-induced string solution of the BAEs at these points. At the quasi-degenerate points, spectral degeneracy may also occur and can be understood in terms of the phantom string solution.}

The paper is organized as follows. Section \ref{sec:model} introduces the open XXZ chain with constrained non-diagonal boundary fields and reviews the corresponding Bethe ansatz equations. In Section \ref{sec:fully:degeneration}, the fully degenerate points are studied: the solutions of the BAEs are classified into three types, counting formulas are derived, and spectral degeneracies are explained via phantom strings. Section \ref{sec:partially:degeneration} addresses the partially degenerate points, where the analogous classification and counting are obtained. The Bethe state at the degenerate points is studied in Section
\ref{sec:BetheState}. Section \ref{sec:diagonal:boundary} discusses the diagonal boundary case, yielding the Bethe-root configuration and spectral degeneracy. The Bethe-root configuration at the quasi-degenerate points is further studied in Section \ref{sec:quasi:degenerate}. Conclusions and outlook are presented in Section \ref{sec:summary:outlook}. Supplemental formulas and numerical results are collected in the appendix.

\section{Open XXZ chain with constrained non-diagonal boundaries}\label{sec:model}
\subsection{Hamiltonian}
The Hamiltonian of the XXZ spin chain with open boundaries is 
\begin{align} 
H=\sum_{j=1}^{N-1}\left[ \sigma_j^x\sigma_{j+1}^x+\sigma_j^y\sigma_{j+1}^y+\cosh \eta\,\sigma_j^z\sigma_{j+1}^z-\cosh\eta\, \mathbb{I}\right]+\vec{h}_1\cdot \vec{\sigma}_1 + \vec{h}_N\cdot \vec{\sigma}_N,\label{Ham_1}
\end{align}
where $\{\sigma^\alpha\}$ are Pauli matrices, $N$ is the number of sites, and the boundary magnetic fields are given by
\begin{align} 
&\vec{h}_1=\frac{\sinh\eta}{\sinh\alpha_1^- \sinh\alpha_2^-}(-\i \cosh\theta_-,\,\sinh\theta_-,\,\cosh\alpha_1^- \cosh\alpha_2^-),\label{h_1}\\[4pt]
&\vec{h}_N=\frac{\sinh\eta}{\sinh\alpha_1^+ \sinh\alpha_2^+}(-\i \cosh\theta_+,\,\sinh\theta_+,\, -\cosh\alpha_1^+ \cosh\alpha_2^+).\label{h_N}
\end{align}

The XXZ model is one of the best-known integrable systems. In the generic case, the non-diagonal boundary fields break the $U(1)$ symmetry of the model, thereby preventing the construction of conventional homogeneous $T$-$Q$ relations and Bethe Ansatz equations (BAEs). 

The exact spectrum of the open XXZ chain was obtained by the off-diagonal Bethe Ansatz method, where an inhomogeneous term is added to the $T$-$Q$ relation \cite{OffDiagonal,Cao2013}, see Appendix \ref{App:A} for details. The presence of this inhomogeneous term prevents the corresponding BAEs from taking the symmetric form as in the periodic and diagonal cases. Consequently, solving the BAEs becomes more complicated, and the analysis of the exact solutions of the model remains a challenging task.

It has been demonstrated, however, that the $T$-$Q$ relation and BAEs can take conventional form (see Section~\ref{sec:BAE}) under the following constraint
\begin{align}
(N-1-2M)\eta = \alpha_1^- +\alpha_2^- +\alpha_1^+ +\alpha_2^+ +\mathrm{i}\pi \pm (\theta_- -\theta_+) \quad\mod{2\mathrm{i}\pi}, \label{special_boundaries}
\end{align}
where $M$ is an integer,
even though the constraint \eqref{special_boundaries} does not restore the $U(1)$ symmetry \cite{Cao2003, Nepomechie:2003_1, Nepomechie:2003_2, OffDiagonal}. This fact makes these boundary-constrained models a bridge between the well-studied diagonal case and models with generic boundary fields. As such, we study the open XXZ chain under the constraint \eqref{special_boundaries} in this paper.

Since the Hamiltonian is invariant under the substitution
$$\alpha_k^\pm\to -\alpha_k^\pm, \quad \theta_\pm\to \theta_\pm,$$
Eq.~\eqref{special_boundaries} still holds upon replacing $M$ with $M'=N-1-M$. In the following sections, we will demonstrate that the integers $M$ and $M'$ also represent the total number of Bethe roots in the BAEs, provided that they are non-negative.

\subsection{Bethe Ansatz equations}\label{sec:BAE}
Define the following functions
\begin{align}
&S(u,v)=\frac{\sinh(u-v-\eta)}{\sinh(u-v+\eta)}\frac{\sinh(u+v-\eta)}{\sinh(u+v+\eta)}, \label{def S}\\
&F(u)=\prod_{\sigma=\pm}\prod_{l=1}^2{\sinh(u-\alpha_l^\sigma-\tfrac{\eta}{2})}.\label{def:F}
\end{align}

Under the condition (\ref{special_boundaries}), the exact solutions of the model described in Eq. \eqref{Ham_1} can be described by the following BAEs \cite{Cao2003,Nepomechie:2003_1,Zhang2021CBA}
\begin{align}
&\left [\frac{\sinh(x_j+\frac{\eta}{2})}{\sinh(x_j-\frac{\eta}{2})} \right ]^{2N} \frac{F(x_j)}{F(-x_j)}\prod_{k\ne j}^{M} S(x_j,x_k)=1,\quad j=1,\ldots,M,\label{BAE_G_M^+_x}
\end{align}
where $M$ is given by Eq. \eqref{special_boundaries}.
A physical solution of the BAEs \eqref{BAE_G_M^+_x} should satisfy the following selection rule   
\begin{align}
\prod_{j}\sinh(2x_j)\prod_{k\neq l}\sinh(x_l-x_k)\sinh(x_l+x_k)\neq 0.\label{selection:rule}
\end{align}
The energy of the model in terms of the Bethe roots $\bm{x}=\{x_1,\ldots,x_M\}$ is given by 
\begin{align}
E^+(\bm{x})=\sum_{j=1}^{M} \varepsilon(x_j)+E_0,\label{E_G+_x}
\end{align}
where 
\begin{align}
\varepsilon(u)&=\frac{4\sinh^2\eta}{\cosh(2u)-\cosh\eta},\\
E_0&=-\sinh\eta\sum_{\sigma=\pm}\sum_{k=1}^2\coth\alpha_k^\sigma.
\end{align}
The BAEs in \eqref{BAE_G_M^+_x} are only valid when $M\geq 0$. In the case $M=0$, $\bm{x}$ is an empty set and one obtains the energy level $E=E_0$.

There exists another set of BAEs for the Bethe roots $\bm{z}=\{z_1, \dots, z_{M'}\}$
\begin{align}
&\left [\frac{\sinh(z_j+\frac{\eta}{2})}{\sinh(z_j-\frac{\eta}{2})} \right ]^{2N} \frac{\bar F(z_j)}{\bar F(-z_j)}\prod_{k\ne j}^{M'}S(z_j,z_k)=1,\quad j=1,\ldots,M'.\label{BAE_G_M^-_x}
\end{align}
where $M'=N-1-M$, $S(u,v)$ is defined in (\ref{def S}), and 
\begin{align}
\bar F(u)=\prod_{\sigma=\pm }\prod_{l=1}^2{\sinh(u+\alpha_l^\sigma-\tfrac{\eta}{2})}.\label{def F_bar}
\end{align}
The corresponding energy reads
\begin{align} 
E^-(\bm z)=\sum_{j=1}^{M'} \varepsilon(z_j)-E_0.\label{E_G-_x}
\end{align}

Analogously, the BAEs in \eqref{BAE_G_M^-_x} are only valid when $M'\geq 0$. When $M'=0$, one obtains the energy level $E=-E_0$.

We consider the case where $M$ takes values from $0$ to $N-1$. In this case, the BAEs in \eqref{BAE_G_M^+_x} and \eqref{BAE_G_M^-_x} are both valid and correspond to the invariant subspaces $G_M^+$ and $G_M^-$, respectively, whose dimensions are $d_+=\sum_{k=0}^M \binom{N}{k}$ and $d_-=\sum_{k=0}^{M'} \binom{N}{k} = 2^N - d_+$ \cite{Zhang2021Phantom, Zhang2021CBA}. Numerical results indicate that the total number of physical solutions of the BAEs equals the dimension of each subspace, and that these two sets of BAEs constitute the complete spectrum of the transfer matrix and hence of the Hamiltonian. However, an analytic proof of this observation remains an open question.

The parameters $\theta_\pm$ can always be adjusted to satisfy Eq. \eqref{special_boundaries} without restricting the choice of $\alpha_k^\pm$ and $\eta$. For a generic choice of $\alpha_k^\pm$ and $\eta$, all energy levels are non-degenerate and $M$ is uniquely determined by Eq.~\eqref{special_boundaries}. When $\eta$ is a root of unity ($\eta = \i\pi j/k$, $j,k\in\mathbb{Z}$), however, $M$ may take different values. A simple example is the XX model ($\eta = \i\pi/2$), where $M$ can be either even or odd \cite{Liu:2026}. Such root-of-unity cases are excluded in the present work.

\begin{remark*}
One may select either the ``$+$" or ``$-$" sign in Eq.~\eqref{special_boundaries}. It is easily verified that the Bethe Ansatz equations and the energy expression are identical in both cases, implying that the spectrum of the system is invariant under the substitution $\theta_\pm\to -\theta_\pm$. Moreover, the same invariance holds for the generic boundary case, since the $\theta_\pm$-dependent term in the BAEs is $\cosh(\theta_--\theta_+)$ (see Appendix~\ref{App:A}).
\end{remark*}

\section{Bethe-root configuration at fully degenerate points}\label{sec:fully:degeneration}
\setcounter{equation}{0}

\subsection{Fully degenerate boundary points}

We focus on the BAEs \eqref{BAE_G_M^+_x} for the subspace $G_M^+$. These BAEs constitute a set of coupled nonlinear equations, and all Bethe roots depend on the boundary parameters in a complicated manner. The boundary dependent term in Eq. \eqref{BAE_G_M^+_x} is given by the ratio \(F(u)/F(-u)\), which is a trigonometric function constructed from the boundary parameters. When \(F(u)\) and \(F(-u)\) share common zeros, the numerator and denominator vanish simultaneously at those points, and we refer to this situation as the \textit{degenerate points}. As we shall see, this degeneracy plays a crucial role in simplifying the structure of the BAEs and leads to a straightforward classification of their solutions (see Section \ref{sec:classification}).

Let us first consider the \textit{fully degenerate points} where
\begin{align}
F(u)=F(-u).\label{special_case_x_j}
\end{align}
This yields the following constraints on the boundary parameters
\begin{align}
\begin{aligned}
({\rm i}):&\quad \alpha_1^- +\alpha_2^- +\eta=\i \pi m_1,\quad \alpha_1^+ +\alpha_2^+ +\eta=\i \pi m_2,\\
({\rm ii}):&\quad \alpha_1^- +\alpha_1^+ +\eta=\i\pi n_1,\quad \alpha_2^- +\alpha_2^+ +\eta=\i \pi n_2,
\end{aligned}\label{Constraints_bp}
\end{align}
where $m_1,m_2,n_1,n_2\in \mathbb{Z}$.

Under Eqs. \eqref{special_boundaries} and \eqref{Constraints_bp}, the boundary fields take the form
\begin{align}
({\rm i}):\quad &\vec{h}_1=\frac{\sinh\eta}{\sinh\alpha_1^- \sinh(\alpha_1^- +\eta)}(\i \cosh\theta_-,\, -\sinh\theta_-,\, -\cosh\alpha_1^- \cosh(\alpha_1^- +\eta)),\no\\[4pt]
&\vec{h}_N=\frac{\sinh\eta}{\sinh\alpha_1^+ \sinh(\alpha_1^+ +\eta)}(\i \cosh\theta_+,\, -\sinh\theta_+,\, \cosh\alpha_1^+ \cosh(\alpha_1^+ +\eta)),\no\\
&(N+1-2M)\eta = \i \pi\pm (\theta_--\theta_+)\quad \mod{2\i \pi},\label{constraint:case1}\\
({\rm ii}):\quad &\vec{h}_1=\frac{\sinh\eta}{\sinh\alpha_1^- \sinh\alpha_2^-}(-\i \cosh\theta_-,\,\sinh\theta_-,\,\cosh\alpha_1^- \cosh\alpha_2^-),\no\\[4pt]
&\vec{h}_N=\frac{\sinh\eta}{\sinh(\alpha_1^- +\eta) \sinh(\alpha_2^- +\eta)}(-\i \cosh\theta_+,\,\sinh\theta_+,\, -\cosh(\alpha_1^- +\eta) \cosh(\alpha_2^- +\eta)),\no\\
&(N+1-2M)\eta = \i \pi\pm (\theta_--\theta_+)\quad \mod{2\i \pi}.\label{constraint:case2}
\end{align}
Here, we make some minor modifications to $\theta_\pm$ compared to the original ones.
In both cases \eqref{constraint:case1} and \eqref{constraint:case2}, the boundary fields depend on three free boundary parameters and an adjustable integer $M$. The Hermiticity of the model is not required in our study. Nevertheless, we present the Hermiticity conditions in Appendix \ref{App:B} for completeness.

The spectra of the Hamiltonians described in either Eq.~\eqref{constraint:case1} or Eq.~\eqref{constraint:case2} depend only on $|\theta_--\theta_+|$ and are thus invariant under the substitution $M\to N+1-M$.

\subsection{Classification of solutions to BAEs (\ref{BAE_G_M^+_x})}\label{sec:classification}

Once Eq. (\ref{Constraints_bp}) is satisfied, the zeros of $F(u)$ form two sets of opposite numbers, namely 
\begin{align}
\{\alpha_1^-+\tfrac{\eta}{2},\alpha_1^++\tfrac{\eta}{2},\alpha_2^-+\tfrac{\eta}{2},\alpha_2^++\tfrac{\eta}{2}\}=\{w_+,-w_+,w_-,-w_-\}.
\end{align}
One can rewrite BAEs (\ref{BAE_G_M^+_x}) as
\begin{align}
\left [\frac{\sinh(x_j+\frac{\eta}{2})}{\sinh(x_j-\frac{\eta}{2})} \right ]^{2N} \prod_{k\ne j}^{M} S(x_j,x_k) F(x_j)=F(-x_j),\quad j=1,\ldots,M.\label{BAE_G_M^+_x_equiv}
\end{align}
It is evident that if $x_j$ coincides with a common zero of both $F(u)$ and $F(-u)$,  Eq. \eqref{BAE_G_M^+_x_equiv} is automatically satisfied.

Since the boundary factor possesses only two independent common zeros, $\{w_+,w_-\}$, and multiple occupation of these special points is forbidden by the selection rule in Eq. \eqref{selection:rule}, at most two Bethe roots can be pinned at these boundary zeros. As a result, the solutions  of the BAEs \eqref{BAE_G_M^+_x} in the subspace $G_M^+$ admit an exhaustive classification into three distinct types, determined by whether none, one, or two Bethe roots are pinned by the boundary.

\paragraph{Type I} None of the Bethe roots are located at the zeros of $F(u)$, and they satisfy the following BAEs
\begin{align}
&\left [ \frac{\sinh(x_j+\frac{\eta}{2})}{\sinh(x_j-\frac{\eta}{2})} \right ]^{2N}\prod_{k\ne j}^{M} S(x_j,x_k)=1,\quad j=1,\ldots,M.\label{iBAE_I_x}
\end{align}
The corresponding energy is given by
\begin{align}
E^+_{\rm I}(\bm{x})&=\sum_{j=1}^{M} \varepsilon(x_j)+E_0.\label{iBAE_Energy}
\end{align}

The BAEs \eqref{iBAE_I_x} no longer contain any boundary parameters; therefore the Bethe roots $\{x_1,\ldots,x_M\}$ depend only on $N$, $M$, and $\eta$, and the boundary dependence of the energy enters solely through $E_0=-\sinh\eta(\coth\alpha_1^- +\coth\alpha_2^- +\coth\alpha_1^+ +\coth\alpha_2^+)$. In other words, the Hamiltonians $H-E_0(\{\alpha_k^\pm\})\,\mathbb{I}$ with different choices of $\{\alpha_k^\pm\}$ share common energy levels, and the total number of such energy levels equals the number of solutions of \eqref{iBAE_I_x}. This demonstrates the spectrum-preserving property of the system.

\paragraph{Type II} One of the Bethe roots $x_{M}$ coincides with one zero of $F(u)$, denoted as $w_\sigma$, while the remaining $M-1$ Bethe roots $\{x_1,\ldots,x_{M-1}\}$ satisfy the following BAEs
\begin{align}
\left[\frac{\sinh(x_j+\frac{\eta}{2})}{\sinh(x_j-\frac{\eta}{2})}\right]^{2N} S(x_j,w_\sigma)\prod_{k\ne j}^{M-1} S(x_j,x_k)&=1,\quad j=1,\ldots,M-1.\label{iBAE_II_x}
\end{align}
The energy is given by
\begin{align}
E^+_{\rm II}(\bm x)=\sum_{j=1}^{M-1} \varepsilon(x_j)+\varepsilon(w_\sigma)+E_0.
\end{align}
The Bethe roots $\{x_1,\ldots,x_{M-1}\}$ in \eqref{iBAE_II_x} are independent of $w_{-\sigma}$, while the dependence of the energy on the boundary parameter $w_{-\sigma}$ enters only through $E_0$.

\paragraph{Type III} Two of the Bethe roots are pinned at the two independent zeros of $F(u)$, i.e., $\{x_{M-1},x_M\}=\{w_+,w_-\}$, while the remaining $M-2$ Bethe roots $\{x_1,\ldots,x_{M-2}\}$ satisfy the following BAEs
\begin{align}
\left [ \frac{\sinh(x_j+\frac{\eta}{2})}{\sinh(x_j-\frac{\eta}{2})} \right ]^{2N} S(x_j,w_+)S(x_j,w_-)\prod_{k\ne j}^{M-2} S(x_j,x_k)&=1,\quad j=1,\ldots,M-2.\label{iBAE_III_x}
\end{align}
Due to the identity
\begin{align}
S(x,w_+)S(x,w_-)=\frac{\bar F(x)}{\bar F(-x)},
\end{align}
Eq. \eqref{iBAE_III_x} can be rewritten as 
\begin{align}
\left [ \frac{\sinh(x_j+\frac{\eta}{2})}{\sinh(x_j-\frac{\eta}{2})} \right ]^{2N} \frac{\bar F(x_j)}{\bar F(-x_j)}\prod_{k\ne j}^{M-2} S(x_j,x_k)&=1,\quad j=1,\ldots,M-2.\label{iBAE_III_x2}
\end{align}
The energy in terms of the Bethe roots $\bm{x}$ reads
\begin{align}
E^+_{\rm III}(\bm x)&=\sum_{j=1}^{M-2} \varepsilon(x_j)+\varepsilon(w_+)+\varepsilon(w_-)+E_0.
\end{align}
From Eq. \eqref{Constraints_bp}, one finds $\varepsilon(w_+)+\varepsilon(w_-) = -2E_0$, which leads to another equivalent expression of $E^+_{\rm III}(\bm x)$ 
\begin{align}
E^+_{\rm III}(\bm x)&=\sum_{j=1}^{M-2} \varepsilon(x_j)-E_0.\label{en_III2}
\end{align}

The classification of the three types of BAEs applies only for $M\geq 2$. For $M=0$, the Bethe root set is empty, giving the single energy level $E=E_0$. For $M=1$, the type I and type II BAEs in \eqref{iBAE_I_x} and \eqref{iBAE_II_x} remain: the latter admits the unique solution $\bm{x}=\{w_\sigma\}$, while the former reduces to a single-variable equation and yields $N-1$ physical solution sets.

{For finite-size systems, BAEs are easier to solve completely at the degenerate points than in
the generic case. This simplification occurs through two mechanisms. For the type II and type III BAEs \eqref{iBAE_II_x} and \eqref{iBAE_III_x}, pinned roots reduce the number of unfixed Bethe roots, lowering the degrees of freedom and rendering the equations
considerably more tractable. For the type I BAEs \eqref{iBAE_I_x}, by contrast, the number of unfixed roots remains the same, but the equations become independent of the boundary parameters, which also facilitates their solution. In the framework of the algebraic Bethe ansatz, the type II and type III
Bethe roots reduce the degree of the $Q$-functions in the deformed $T$-$Q$ relation, while the type I Bethe roots give rise to a common boundary-related polynomial in the eigenvalue $\Lambda(u)$ of the
transfer matrix, thereby simplifying the remaining part of $\Lambda(u)$.}

It should be noted that the zeros of $F(u)$ are not the zeros of $\bar{F}(u)$ defined in \eqref{def F_bar}. Thus, the BAEs for $G_M^-$ \eqref{BAE_G_M^-_x} do not possess a classification analogous to that of the BAEs \eqref{BAE_G_M^+_x}.

\subsection{Existence of phantom strings and spectral degeneracy}\label{sec:degeneracy}

At the fully degenerate points given in Eq.~\eqref{Constraints_bp}, the type III BAEs for the subspace $G_M^+$, given by Eq.~\eqref{iBAE_III_x2}, overlap with the BAEs for the subspace $G_M^-$, given by Eq.~\eqref{BAE_G_M^-_x}. This overlap is realized through the phantom string proposed in Ref.~\cite{Zhang2021Phantom}, which carries zero energy and consequently leads to degeneracy of the corresponding energy levels.

\paragraph{Existence of phantom strings}
The BAEs in \eqref{iBAE_III_x2} and \eqref{BAE_G_M^-_x} share the same functional form, with the only difference being the number of Bethe roots.
Let us first consider the case $M-2>M'$ (i.e., $M> \frac{N+1}{2}$). For certain solutions of the BAEs \eqref{iBAE_III_x2}, the Bethe roots $\{x_1,\ldots,x_{M-2}\}$ can be partitioned into two subsets
\begin{align}
&\mbox{finite roots}:\,\,\{x_1,\ldots,x_{M'}\}=\{z_1,\ldots,z_{M'}\},\\
&\mbox{phantom string}:\,\,\{x_{M'+1},\ldots,x_{M-2}\}=\left\{\kappa+\tfrac{\i \pi }{p},\ldots,\kappa+\tfrac{\i \pi p}{p}\right\},\no\\[2pt]
&\hspace{3.3cm} p=2M-N-1,\quad\kappa\to +\infty.\label{phantom:string}
\end{align}
\begin{proof}
For the phantom root $x_j$, the LHS (left-hand side) of \eqref{iBAE_III_x2} becomes
\begin{align}
&\e^{2(\alpha_1^++\alpha_1^-+\alpha_2^++\alpha_2^-+2\eta)}\prod_{k\ne j}^{p}\frac{\sinh(\i\pi\,\frac{j-k}{p}-\eta)} {\sinh(\i\pi\,\frac{j-k}{p}+\eta)}
=\e^{2(\alpha_1^++\alpha_1^-+\alpha_2^++\alpha_2^-+2\eta)}\overset{\eqref{Constraints_bp}}{=}1,\label{proof:phantom}
\end{align}
and the BAEs \eqref{iBAE_III_x2} are satisfied automatically. For the remaining finite root $x_j$, the LHS of \eqref{iBAE_III_x2} is\begin{align}\left[\frac{\sinh(x_j+\frac{\eta}{2})}{\sinh(x_j-\frac{\eta}{2})} \right ]^{2N} \frac{\bar F(x_j)}{\bar F(-x_j)}\prod_{k\ne j}^{M'} S(x_j,x_k),\label{iBAE_III_x3}
\end{align}
which is identical to the LHS of \eqref{BAE_G_M^-_x}. As a consequence, $\{x_1,\ldots,x_{M'}\}=\{z_1,\ldots,z_{M'}\}$.
\end{proof}
It is worth noting that the Bethe-root configuration with a phantom string is not the only possibility. When $M-2>M'$, the BAEs \eqref{iBAE_III_x2} also admit solutions in which all Bethe roots are finite.

For the case $M<\frac{N+1}{2}$, a similar situation arises. Specifically, in certain solutions, $N+1-2M$ of the elements in $\{z_1, \ldots, z_{M'}\}$ form a phantom string, while the remaining finite part coincides with $\{x_1,\ldots,x_{M-2}\}$.

{Although phantom strings are difficult to capture, they are an inherent feature of the degenerate points rather than a numerical artifact. To
illustrate their existence more clearly, we choose points sufficiently close to the degenerate point. Some Bethe roots are then observed to acquire large real parts and to be nearly equally spaced along the
imaginary axis in the complex plane. As the selected point approaches the degenerate point, the real parts of these quasi string configurations grow
without bound; see Figure \ref{Fig:phantom:string} for an example. Further numerical evidence for phantom strings is provided in Appendix \ref{App:data:1}. In summary, the phantom string solutions arise as a limiting case of the corresponding solutions in the non-degenerate points when the parameters approach the degenerate point.}

\begin{figure}[htbp]
\begin{subfigure}{0.45\textwidth}
\begin{tikzpicture}
\begin{axis}[
  width=\textwidth,
  height=0.75\textwidth,
  axis lines=middle,
  xlabel={${\rm Re}$},
  ylabel={${\rm Im}$},
  xlabel style={at={(axis cs:4.8,0)}, anchor=west},
  ylabel style={at={(axis cs:0,1.2)}, anchor=south},
  xmin=0, xmax=4.8,
  ymin=-1.2, ymax=1.2,
  ytick={pi/3, -pi/3,0},
  yticklabels={$\frac{\pi}{3}$, $-\frac{\pi}{3}$, 0},
  grid=none,
  only marks,
  axis line style={-}
]
\addplot[blue, mark=asterisk, mark size=3pt]  coordinates {
(1.01488,0.68990)
(1.01488,2.45170-pi)
(1.12325,3.14159-pi)
};

\addplot[green, mark=asterisk, mark size=3pt]  coordinates {
(1.84318,0.98913)
(1.84318,2.15247-pi)
(1.91490,3.14159-pi)
};

\addplot[red, mark=asterisk, mark size=3pt]  coordinates {
(2.62930,1.03608)
(2.62930,2.10551-pi)
(2.64721,3.14159-pi)
};

\addplot[black, mark=asterisk, mark size=3pt]  coordinates {
(3.40148,1.04487)
(3.40148,2.09672-pi)
(3.40545,3.14159-pi)
};

\addplot[purple, mark=asterisk, mark size=3pt]  coordinates {
(4.17004,1.04670)
(4.17004,2.09489-pi)
(4.17091,3.14159-pi)
};

\draw[dashed, thick, gray,line width=0.5pt] (axis cs:0,pi/3) -- (axis cs:5,pi/3);

\draw[dashed, thick, gray,line width=0.5pt] (axis cs:0,-pi/3) -- (axis cs:5,-pi/3);

\end{axis}
\end{tikzpicture}
\caption{$c=1.7$}
\end{subfigure}
~~~~
\begin{subfigure}{0.45\textwidth}
\begin{tikzpicture}
\begin{axis}[
  width=\textwidth,
  height=0.7\textwidth,
  axis lines=middle,
  xlabel={${\rm Re}$},
  ylabel={${\rm Im}$},
  xlabel style={at={(axis cs:4.8,0)}, anchor=west},
  ylabel style={at={(axis cs:0,3)}, anchor=south},
  xmin=0, xmax=4.8,
  ymin=0, ymax=3,
  ytick={pi/6, pi/2, 5*pi/6},
  yticklabels={$\frac{\pi}{6}$, $\frac{\pi}{2}$, $\frac{5\pi}{6}$},
  grid=none,
  only marks,
  axis line style={-}
]
\addplot[blue, mark=asterisk, mark size=3pt]  coordinates {
(1.48645,0.47399)
(0.82652,1.57080)
(1.48645,2.66760)
};

\addplot[green, mark=asterisk, mark size=3pt]  coordinates {
(1.95635,0.48590)
(1.86813,1.57080)
(1.95635,2.65569)
};

\addplot[red, mark=asterisk, mark size=3pt]  coordinates {
(2.69953,0.51436)
(2.68247,1.57080)
(2.69953,2.62723)
};

\addplot[black, mark=asterisk, mark size=3pt]  coordinates {
(3.46256,0.52156)
(3.45898,1.57080)
(3.46256,2.62004)
};

\addplot[purple, mark=asterisk, mark size=3pt]  coordinates {
(4.22915,0.52316)
(4.22839,1.57080)
(4.22915,2.61844)
};

\draw[dashed, thick, gray,line width=0.5pt] (axis cs:0,pi/2) -- (axis cs:5,pi/2);
\draw[dashed, thick, gray,line width=0.5pt] (axis cs:0,pi/6) -- (axis cs:5,pi/6);
\draw[dashed, thick, gray,line width=0.5pt] (axis cs:0,5*pi/6) -- (axis cs:5,5*pi/6);

\end{axis}
\end{tikzpicture}
\caption{$c=-2.9$}
\end{subfigure}
\caption{Numerical evidence for the  quasi-phantom string solution of BAEs \eqref{BAE_G_M^-_x} with $N=6$, $M=2$, $\eta=0.5\rm{i}$, $\alpha_1^-=0.4\rm{i}$, $\alpha_1^+=0.7\rm{i}$, $\alpha_2^-=-\eta-\alpha_1^- -\i c\epsilon$, $\alpha_2^+=-\eta-\alpha_1^+ -\i\epsilon$, $\theta_+=0.9\rm{i}$ and $\theta_-=(N-1-2M)\eta-(\alpha_1^- +\alpha_2^- +\alpha_1^+ +\alpha_2^+) -\i \pi +\theta_+$. Here, blue, green, red, black and purple dots represent $\epsilon=10^{-1}$, $\epsilon=10^{-3}$, $\epsilon=10^{-5}$, $\epsilon=10^{-7}$ and $\epsilon=10^{-9}$, respectively. In panel (a), the energies corresponding to the Bethe roots (with
descending $\epsilon$) are $1.09145932$, $1.13588479$, $1.13629838$, $1.13630250$ and $1.13630255$. In panel (b), the energies corresponding to the Bethe roots (with descending $\epsilon$) are $1.03724209$, $1.13619970$, $1.13630157$, $1.13630254$ and $1.13630255$.}\label{Fig:phantom:string}
\end{figure}

The solutions containing a phantom string play a crucial role in the counting of the total number of solutions. In the following sections, we will demonstrate the existence of phantom strings in other cases, including the $M>\frac{N+1}{2}$ regime for type I and type II BAEs in \eqref{iBAE_I_x} and \eqref{iBAE_II_x}. The proof follows the same reasoning as in Eqs. \eqref{proof:phantom} and \eqref{iBAE_III_x3}, so we omit the repetitive details for brevity.

\paragraph{Degeneracy} For the BAEs in  \eqref{BAE_G_M^-_x} and \eqref{iBAE_III_x2}, the corresponding energy expressions in terms of the Bethe roots are also identical (see Eqs.~\eqref{E_G-_x} and \eqref{en_III2}). Since the phantom roots carry zero energy $$\varepsilon(\infty)=0,$$ the corresponding energy levels become degenerate. The number of such doubly degenerate levels equals the minimum of the total numbers of solutions of the type III BAEs in the $G_M^+$ subspace and the BAEs in the $G_M^-$ subspace, i.e., $\sum\limits_{k=0}^{{\min}(M-2,\,M')}\binom{N}{k}$. See Section \ref{sec:solution:number} for more details.

\subsection{Counting formulas for classified Bethe root solutions}\label{sec:solution:number}
Determining the number of physical solutions of each set of BAEs is a highly nontrivial problem.
Let us denote by ${\mathcal{N}}_{\rm I}(N,M)$, ${\mathcal{N}}_{\rm II}(N,M)$, and ${\mathcal{N}}_{\rm III}(N,M)$ the numbers of solution sets for the type I, II, and III BAEs, respectively, and by ${\mathcal{N}}_{-}(N,M)$ the total number of solutions of the BAEs \eqref{BAE_G_M^-_x}.  

{Based on sufficient numerical evidence (see Table \ref{Numerical_4_M_Counting} and \ref{App:data:1}), we propose the following conjectures}
\begin{align}
{\mathcal{N}}_{\rm I}(N,M)&=\binom{N}{P}-\binom{N}{P-1},\label{Number1}\\
{\mathcal{N}}_{\rm II}(N,M)&=2\binom{N}{P-1},\label{Number2}\\
{\mathcal{N}}_{\rm III}(N,M)&=\sum_{k=0}^{M-2} \binom{N}{k},\label{Number3}\\
{\mathcal{N}}_{-}(N,M)&=\sum_{k=0}^{N-1-M} \binom{N}{k},\label{Number4}
\end{align}
where $P=\min(M,N+1-M)$. Here, the factor $2$ in \eqref{Number2} originates from the fact that $x_M$ may be chosen as either $w_+$ or $w_-$. Combining the results for the type I, type II, and type III BAEs, the total number of solutions to \eqref{BAE_G_M^+_x} is $\sum_{k=0}^{M}\binom{N}{k}$, which matches the dimension of the subspace $G_M^+$. Similarly, the total number of solutions to the BAEs \eqref{BAE_G_M^-_x} equals the dimension of $G_M^-$.

In order to clarify the counting formulas \eqref{Number1}-\eqref{Number4} and the configuration of Bethe roots, we denote by $\{\ldots\}_A$ the set of Bethe roots, where $A \in \{\mathrm{I}, \mathrm{II}, \mathrm{III}, -\}$ corresponds to the type I, type II, type III BAEs for $G_M^+$ and the BAEs for $G_M^-$, respectively. Moreover, the symbol $\{\text{PS}\}$ indicates a phantom string within the Bethe roots.

\paragraph{$M<\frac{N+1}{2}$ case}
In this case, all Bethe roots appearing in the BAEs for $G_M^+$, including types I, II, and III in \eqref{iBAE_I_x}, \eqref{iBAE_II_x}, and \eqref{iBAE_III_x2}, are finite.
The BAEs for $G_M^-$ \eqref{BAE_G_M^-_x} admit two kinds of solutions
\begin{align}
&({\rm i}):\,\,\{z_1,\ldots,z_{N-1-M}\}_-=\{x_1,\ldots,x_{M-2}\}_{\rm III}\,\cup\,\{{\rm PS}\},\,\mathcal{N}_{\rm III}(N,M) \,\,\mbox{solutions},\label{case1:number:1}\\
&({\rm ii}):\,\,\mbox{all roots in}\,\, \{z_1,\ldots,z_{N-1-M}\}_-\,\, \mbox{are finite},\,\, {\mathcal{N}}_{-}(N,M)-{\mathcal{N}}_{\rm III}(N,M) \,\,\mbox{solutions}.\label{case1:number:2}
\end{align}
The solution in \eqref{case1:number:1} yields the same energy as the set $\{x_1,\ldots,x_{M-2}\}_{\mathrm{III}}$ corresponding to the type III BAEs. Hence, the number of doubly degenerate energy levels is $\mathcal{N}_{\mathrm{III}}(N,M)$.

\paragraph{$M>\frac{N+1}{2}$ case}

For the BAEs for $G_M^-$ \eqref{BAE_G_M^-_x}, all the roots are finite. The type III BAEs  \eqref{iBAE_III_x2} now possess two kinds of solutions
\begin{align}
&({\rm i}):\,\,\{x_1,\ldots,x_{M-2}\}_{\rm III}=\{z_1,\ldots,z_{N-1-M}\}_{-}\,\cup\,\{{\rm PS}\},\,\mathcal{N}_{-}(N,M) \,\,\mbox{solutions},\label{case2:number:1}\\
&({\rm ii}):\,\,\mbox{all roots in}\,\, \{x_1,\ldots,x_{M-2}\}_{\rm III}\,\, \mbox{are finite},\,\, {\mathcal{N}}_{\rm III}(N,M)-{\mathcal{N}}_{-}(N,M) \,\,\mbox{solutions}.\label{case2:number:2}
\end{align}
The solution in \eqref{case2:number:1} and the set $\{z_1,\ldots,z_{N-1-M}\}_-$ give the same energy, which implies that the $\mathcal{N}_{-}(N,M)$ energy levels are doubly degenerate.

For the type II BAEs in \eqref{iBAE_II_x}, the only possible solution is
\begin{align}
\{x_1,\ldots,x_{M-1}\}_{\mathrm{II}}=
\{x'_1,\ldots,x'_{N-M}\}_{\mathrm{II}}\cup \{\mathrm{PS}\},\label{type2:solution}
\end{align}
where $\{x'_j\}$ denote the Bethe roots for distinction. The integer $\mathcal{N}_{\rm II}(N,M)$ is determined by the total number of finite roots $N-M$ with 
$$\mathcal{N}_{\rm II}(N,M)=\mathcal{N}_{\rm II}(N,N-M+1).
$$

The type I BAEs in \eqref{iBAE_I_x} require a more detailed separate discussion. We observe that these equations are ``underdetermined'' in the sense that the Bethe roots corresponding to a given energy level are not unique. Additional evidence for this non-uniqueness can be found in the results for the points sufficiently close to the degenerate points. For example, letting
$
\alpha_2^- = -\alpha_1^- - \eta + \epsilon_1, \,
\alpha_2^+ = -\alpha_1^+ - \eta + \epsilon_2,
$
where $\epsilon_1$ and $\epsilon_2$ take small values, we find that different choices of the ratio $\epsilon_1/\epsilon_2$ make the Bethe roots approach different points in the complex plane, indicating that multiple distinct root configurations can yield the same energy. A useful characterization of the physical solutions of \eqref{iBAE_I_x} can be given by
\begin{align}
\{x_1,\ldots,x_{M}\}_{\mathrm{I}}=
\{x'_1,\ldots,x'_{N-M+1}\}_{\mathrm{I}}\cup \{\mathrm{PS}\}.\label{type3:solution}
\end{align}
Under this characterization, the type I BAEs admit
$$
\mathcal{N}_{\rm I}(N,M)=\mathcal{N}_{\rm I}(N,N-M+1)
$$
sets of physical solutions. 

\paragraph{$M=\frac{N+1}{2}$ case}

When $N$ is odd and $M$ equals $\frac{N+1}{2}$, the solutions of the BAEs exhibit a rather intriguing structure: (i) phantom strings are completely absent from all types of the BAEs; (ii) The type III BAEs in \eqref{iBAE_III_x} coincide with the BAEs for
$G_M^-$ in \eqref{BAE_G_M^-_x}, and their full solution sets are in
one-to-one correspondence, namely
$\{x_1,\ldots,x_{M-2}\}_{\mathrm{III}}
=\{z_1,\ldots,z_{N-1-M}\}_{-}$; (iii) the type I BAEs turn out to admit no physical solutions whatsoever.

\begin{table}[H]
\centering
\caption{Numerical validation of the counting formulas  \eqref{Number1}-\eqref{Number4} for $N=4$ and $N=5$. Here, ``Y'' and ``N'' indicate solutions with and without a phantom string, respectively, and the symbol $/$ denotes an empty set of Bethe roots.}
\label{Numerical_4_M_Counting}
\begin{tabular}{|c|c|c|c|c|c|c|c|c|}
\hline
\diagbox[height=2.45em]{$(N,M)$}{}                        & \multicolumn{2}{c|}{${\mathcal{N}}_{\rm I}(N,M)$}   
& \multicolumn{2}{c|}{${\mathcal{N}}_{\rm II}(N,M)$}      & \multicolumn{2}{c|}{${\mathcal{N}}_{\rm III}(N,M)$}   
& \multicolumn{2}{c|}{${\mathcal{N}}_{-}(N,M)$}\\
\hline
\multirow{2}{*}{$(4,0)$} & N   & $/$   & N  & $/$   & N   & $/$   & N   & $15$\\
\cline{2-9}
                         & Y      & $/$   & Y     & $/$   & Y      & $/$   & Y      & $0$\\
\hline
\multirow{2}{*}{$(4,1)$} & N   & $3$   & N  & $2$   & N   & $0$   & N   & $11$\\
\cline{2-9}
                         & Y      & $0$   & Y     & $0$   & Y      & $0$   & Y      & $0$\\
\hline
\multirow{2}{*}{$(4,2)$} & N   & $2$   & N  & $8$   & N   & $1$   & N   & $4$\\
\cline{2-9}
                         & Y      & $0$   & Y     & $0$   & Y      & $0$   & Y      & $1$\\
\hline
\multirow{2}{*}{$(4,3)$} & N   & $0$   & N  & $0$   & N   & $4$   & N   & $/$\\
\cline{2-9}
                         & Y      & $2$   & Y     & $8$   & Y      & $1$   & Y      & $/$\\
\hline
\multirow{2}{*}{$(5,0)$} & N   & $/$   & N  & $/$   & N   & $/$   & N   & $31$\\
\cline{2-9}
                         & Y      & $/$   & Y     & $/$   & Y      & $/$   & Y      & $0$\\
\hline
\multirow{2}{*}{$(5,1)$} & N   & $4$   & N  & $2$   & N   & $0$   & N   & $26$\\
\cline{2-9}
                         & Y      & $0$   & Y     & $0$   & Y      & $0$   & Y      & $0$\\
\hline
\multirow{2}{*}{$(5,2)$} & N   & $5$   & N  & $10$  & N   & $1$   & N   & $15$\\
\cline{2-9}
                         & Y      & $0$   & Y     & $0$   & Y      & $0$   & Y      & $1$\\
\hline
\multirow{2}{*}{$(5,3)$} & N   & $0$   & N  & $20$  & N   & $6$   & N   & $6$\\
\cline{2-9}
                         & Y      & $0$   & Y     & $0$   & Y      & $0$   & Y      & $0$\\
\hline
\multirow{2}{*}{$(5,4)$} & N   & $0$   & N  & $0$   & N   & $15$  & N   & $/$\\
\cline{2-9}
                         & Y      & $5$   & Y     & $10$  & Y      & $1$   & Y      & $/$\\
\hline
\end{tabular}
\end{table}

From the above analysis, we observe a one-to-one correspondence between the solutions of the BAEs for the $M=m$ and $M=N+1-m$ cases. Specifically, under this correspondence, the type I and type II BAEs in the second case ($M=N+1-m$) are mapped to their counterparts in the first case ($M=m$), while the type III BAEs in either case are mapped to the BAEs for $G_M^-$ in the other case.

A comprehensive schematic diagram summarizing the classification of the BAEs, the corresponding Bethe-root configurations, and the correspondence between the solutions of the BAEs for the $M=m$ and $M=N+1-m$ cases is presented in Figure \ref{fig2}.

{The phantom string is difficult to resolve numerically. Its presence does not affect the energy, and the remaining finite roots satisfy equations of the same form as the original BAEs, with the only difference being the total number of Bethe roots. From the above-mentioned results, the total number of solutions in each case depends only on the number of finite roots (excluding the boundary-pinned roots), denoted as $M_{\rm f}$ (for $G_M^+$) and $M'_{\rm f}$ (for $G_M^-$). The counting formulas can be re-summarized in terms of the number of finite roots as follows
\begin{align}
M\leq\tfrac{N+1}{2}:\quad
&\begin{array}{l@{\quad}l@{\quad}l}
\text{type I BAEs}:   & M_{\rm f}=M,     & \binom{N}{M_{\rm f}}-\binom{N}{M_{\rm f}-1}\ \text{solutions},\\[4pt]
\text{type II BAEs}:  & M_{\rm f}=M-1,   & 2\binom{N}{M_{\rm f}}\ \text{solutions},\\[4pt]
\text{type III BAEs}: & M_{\rm f}=M-2,   & \sum_{k=0}^{M_{\rm f}} \binom{N}{k}\ \text{solutions},\\[4pt]
G_M^{-}\ \text{BAEs}: & M'_{\rm f}=M-2, & \sum_{k=0}^{M'_{\rm f}} \binom{N}{k}\ \text{solutions},\\[4pt]
G_M^{-}\ \text{BAEs}: & M'_{\rm f}=N-1-M, & \sum_{k=N-M'_{\rm f}-2}^{M'_{\rm f}} \binom{N}{k}\ \text{solutions},\\[4pt]
\end{array}\\[4pt]
M>\tfrac{N+1}{2}:\quad
&\begin{array}{l@{\quad}l@{\quad}l}
\text{type I BAEs}:   & M_{\rm f}=N+1-M,     & \binom{N}{M_{\rm f}}-\binom{N}{M_{\rm f}-1}\ \text{solutions},\\[4pt]
\text{type II BAEs}:  & M_{\rm f}=N-M,   & 2\binom{N}{M_{\rm f}}\ \text{solutions},\\[4pt]
\text{type III BAEs}: & M_{\rm f}=M-2,   & \sum_{k=N-M_{\rm f}-2}^{M_{\rm f}} \binom{N}{k}\ \text{solutions},\\[4pt]
\text{type III BAEs}: & M_{\rm f}=N-1-M, & \sum_{k=0}^{M_{\rm f}} \binom{N}{k}\ \text{solutions},\\[4pt]
G_M^{-}\ \text{BAEs}: & M'_{\rm f}=N-1-M, & \sum_{k=0}^{M'_{\rm f}} \binom{N}{k}\ \text{solutions}.\\[4pt]
\end{array}
\end{align}
}

\begin{figure}[htbp]
\centering
\resizebox{0.95\textwidth}{!}{
\begin{tikzpicture}[node distance=2cm, scale=0.4]
\tikzstyle{startstop} = [rectangle, rounded corners, minimum width=1cm, minimum height=1cm,text centered, draw=black, fill=red!20,font=\scriptsize]
\tikzstyle{io} = [rectangle, rounded corners, minimum width=1cm, minimum height=1cm, text centered, draw=black, fill=blue!20,font=\scriptsize]
\tikzstyle{arrow} = [thick,->,>=stealth]
\tikzstyle{darrow} = [thick,<->,>=stealth]

\node (start1) [align=center] at (0,0) {\footnotesize BAEs \\ for $G_m^+$};

\node (in1) [align=center] at (4,0) {\footnotesize II};
\node (in2) [align=center] at (4,-4){\footnotesize III};
\node (in3) [align=center] at (4,4){\footnotesize  I};

\node (nu1) [io,align=center] at (12,0) {$m\!-\!1$ unfixed finite roots\\without phantom string\\[2pt]$2\binom{N}{m-1}$ solutions};
\node (nu2) [io,align=center] at (12,-4){$m\!-\!2$ unfixed finite roots\\without phantom string\\[2pt]$\sum_{k=0}^{m-2}\binom{N}{k}$ solutions};
\node (nu3) [io,align=center] at (12,4){$m$ unfixed finite roots\\without phantom string\\[2pt]$\binom{N}{m}-\binom{N}{m-1}$ solutions};

\node (in4) [align=center] at (0,-12){\footnotesize BAEs\\ for $G_m^-$};

\node (nu4) [startstop,align=center] at (12,-10) {$m\!-\!2$ unfixed finite roots\\with phantom strings\\[2pt]$\sum_{k=0}^{m-2}\binom{N}{k}$ solutions};

\node (nu5) [startstop,align=center] at (12,-14) {$\bar m$ unfixed finite roots\\without phantom string\\[2pt]$\sum_{k=m-1}^{\bar m-2}\binom{N}{k}$ solutions};

\node (nu6) [io,align=center] at (26,0) {$m\!-\!1$ unfixed finite roots\\with phantom strings\\[2pt]$2\binom{N}{m-1}$ solutions};
\node (nu7) [io,align=center] at (26,-4){$\bar m$ unfixed finite roots\\without phantom string\\[2pt]$\sum_{k=m-1}^{\bar m-2}\binom{N}{k}$ solutions};
\node (nu8) [io,align=center] at (26,4){$m$ unfixed finite roots\\with phantom strings\\[2pt]$\binom{N}{m}-\binom{N}{m-1}$ solutions};
\node (nu9) [io,align=center] at (26,-8){$m\!-\!2$ unfixed finite roots\\with phantom strings\\[2pt]$\sum_{k=0}^{m-2}\binom{N}{k}$ solutions};
\node (nu10) [startstop,align=center] at (26,-14){$m\!-\!2$ unfixed finite roots\\without phantom string\\[2pt]$\sum_{k=0}^{m-2}\binom{N}{k}$ solutions};

\node (in5) [align=center] at (34,0) {\footnotesize II};
\node (in6) [align=center] at (34,-6){\footnotesize III};
\node (in7) [align=center] at (34,4){\footnotesize I};

\node (in8) [align=center] at (39,-14){\footnotesize BAEs for\\ $G_{N+1-m}^-$};

\node (start1) [align=center] at (39,0) {\footnotesize BAEs for\\ $G_{N+1-m}^+$};

\draw[red] (2,0)  -- (3,0);
\draw[red] (2.5,0)  -- (2.5,4)--(3,4);
\draw[red] (2.5,0)  -- (2.5,-4)--(3,-4);

\draw[red] (5,0)  -- (6.5,0);
\draw[red] (5,4)  -- (6.5,4);
\draw[red] (5,-4)  -- (6.5,-4);

\draw[red] (2,-12) --(2.5,-12)--(2.5,-10)--(6,-10);
\draw[red] (2,-12) --(2.5,-12)--(2.5,-14)--(6,-14);

\draw[red] (10,-6) --(10,-8);
\node at (12,-7){\scriptsize degeneracy};
\draw[red] (24,-10) --(24,-12);
\node at (26,-11){\scriptsize degeneracy};

\draw[red] (36,0)--(35,0);
\draw[red] (33,0)--(31.5,0);
\draw[red] (35.5,0)--(35.5,4)--(35,4);
\draw[red] (33,4)--(31.5,4);
\draw[red] (35.5,0)--(35.5,-6)--(35,-6);
\draw[red] (33,-6)-- (32.5,-6)-- (32.5,-4)--(31.5,-4);
\draw[red] (33,-6)-- (32.5,-6)-- (32.5,-8)--(31.5,-8);
\draw[red] (36,-14)-- (32,-14);

\draw[dashed] (17,4)--(21,4);
\draw[dashed] (17,0)--(21,0);
\draw[dashed] (17,-4)--(21,-14);
\draw[dashed] (17,-10)--(21,-8);
\draw[dashed] (17,-14)--(21,-4);

\end{tikzpicture}}
\caption{Schematic diagram summarizing the classification of Bethe-root configurations (the blocks) and the correspondence between the BAEs for the $M=m$ ($m<\frac{N+1}{2}$) and $M=\bar m=N+1-m$ cases (indicated by dashed lines).}
\label{fig2}
\end{figure}

\section{Bethe-root configuration at partially degenerate points}\label{sec:partially:degeneration}
\setcounter{equation}{0}
\setcounter{table}{0}
The fully degenerate points studied in Section \ref{sec:fully:degeneration} correspond to the situation where $F(u)$ and $F(-u)$ share two independent common zeros. A natural question is what happens when only one such common zero survives. We refer to this case as the \textit{partially degenerate points}, where the boundary parameters satisfy
\begin{align}
\begin{aligned}\label{Constraints_bp2}
({\rm i}):&\quad \alpha_2^- +\alpha_2^+ +\eta=\i \pi n_1,\quad \mbox{or} \quad \alpha_1^- +\alpha_1^+ +\eta=\i \pi n_2,\\
({\rm ii}):&\quad \alpha_1^- +\alpha_2^- +\eta=\i \pi m_1, \quad \mbox{or} \quad \alpha_1^+ +\alpha_2^+ +\eta=\i \pi m_2,
\end{aligned}
\end{align}
with $n_1,n_2,m_1,m_2\in \mathbb{Z}$.

The parameters $\{\alpha_k^\pm\}$ play the same role in both the BAEs and the energy expression. Therefore, all the cases in Eq. \eqref{Constraints_bp2} can be unified into one for the BAEs. Without loss of generality, we consider the case $\alpha_2^- +\alpha_2^+ +\eta=\i \pi n_1$. The solutions of the BAEs now fall into two categories.

\paragraph{Type I} All the Bethe roots do not fall on $\alpha_2^-+\frac{\eta}{2}$ and they satisfy the following BAEs
\begin{align}
&\left [ \frac{\sinh(x_j+\frac{\eta}{2})}{\sinh(x_j-\frac{\eta}{2})} \right ]^{2N}\prod_{\sigma=\pm }\, \frac{\sinh(x_j-\alpha_1^\sigma-\frac{\eta}{2})}{\sinh(x_j+\alpha_1^\sigma+\frac{\eta}{2})}\prod_{k\ne j}^{M} S(x_j,x_k)=1,\quad j=1,\ldots,M.\label{iBAE_I_x:2}
\end{align}

\paragraph{Type II} One Bethe root $x_{M}$ is pinned to $\alpha_2^-+\frac{\eta}{2}$ and the remaining roots $\{x_1,\ldots,x_{M-1}\}$ are given by the following BAEs
\begin{align}
&\left[\frac{\sinh(x_j+\frac{\eta}{2})}{\sinh(x_j-\frac{\eta}{2})} \right ]^{2N}\prod_{\sigma=\pm }\, \frac{\sinh(x_j-\alpha_1^\sigma-\frac{\eta}{2})}{\sinh(x_j+\alpha_1^\sigma+\frac{\eta}{2})}\no\\
&\times S(x_j,\alpha_2^-+\tfrac{\eta}{2})\prod_{k\ne j}^{M-1} S(x_j,x_k)=1,\quad j=1,\ldots,M-1.\label{iBAE_I_x:3}
\end{align}

In contrast to the fully degenerate points discussed in Section \ref{sec:fully:degeneration}, the BAEs \eqref{iBAE_I_x:2} and \eqref{iBAE_I_x:3} both contain non-degenerate boundary terms, and the boundary parameters $\alpha_1^\pm$ and $\alpha_2^-$ remain as free parameters. 
{At the partially degenerate points, the phantom strings are absent. This
can be proved easily by contradiction. Suppose that $k$ of the $M$ Bethe roots form a phantom string as in \eqref{phantom:string}. Substituting
this solution into the BAEs \eqref{BAE_G_M^+_x}, for the infinite root $x_j$, the LHS of the BAEs reads
$$
\mathrm{e}^{2(\alpha_1^+ + \alpha_1^- + \alpha_2^+ + \alpha_2^- + 2\eta)},
$$
which contradicts the RHS (right-hand side) of \eqref{BAE_G_M^+_x}, which equals $1$.} The double degeneracy of certain energy levels observed at the fully degenerate points thus does not occur.

Let us denote by $\mathcal{N}_{\mathrm{I}}^{\mathrm{p}}(N,M)$ and $\mathcal{N}_{\mathrm{II}}^{\mathrm{p}}(N,M)$ the numbers of solution sets for the type I BAEs in \eqref{iBAE_I_x:2} and type II BAEs in \eqref{iBAE_I_x:3}, respectively, and by $\mathcal{N}_{-}(N,M)$ the total number of solutions of the BAEs in \eqref{BAE_G_M^-_x}. {Based on numerical results for small $N$, we propose the following conjecture}
\begin{align}
{\mathcal{N}}_{\rm I}^{\rm p}(N,M)&=\binom{N}{M},\label{Number1_p}\\
{\mathcal{N}}_{\rm II}^{\rm p}(N,M)&=\sum_{k=0}^{M-1} \binom{N}{k},\label{Number2_p}\\
{\mathcal{N}}_{-}(N,M)&=\sum_{k=0}^{N-1-M} \binom{N}{k}.\label{Number3_p}
\end{align}
As an example, Table \ref{Counting:p} presents the BAEs \eqref{iBAE_I_x:2}, \eqref{iBAE_I_x:3} and \eqref{BAE_G_M^-_x} for $N=4$ with varying $M$. The total numbers of solutions are consistent with the formulas in \eqref{Number1_p}–\eqref{Number3_p}. More numerical results can be found in Appendix \ref{App:data:2}.


\begin{table}[H]
\centering
\caption{Numerical validation of the counting formulas (\ref{Number1_p}), (\ref{Number2_p}), and (\ref{Number3_p}) for $N=4$ and $N=5$. In the $M=0$ case, the symbol $/$ indicates that there are no Bethe roots, and hence the classification becomes meaningless. Nevertheless, one can still obtain an energy level $E=E_0$.}
\label{Counting:p}
\begin{tabular}{|c|c|c|c|}
\hline
\diagbox[height=2.2em]{$(N,M)$}{}      & ${\mathcal{N}}_{\rm I}^{\rm p}(N,M)$     & ${\mathcal{N}}_{\rm II}^{\rm p}(N,M)$     & ${\mathcal{N}}_{-}(N,M)$\\
\hline
$(4,0)$       & $/$                                   & $/$                                   & $15$\\
\hline
$(4,1)$       & $4$                                   & $1$                                   & $11$\\                 
\hline
$(4,2)$       & $6$                                   & $5$                                   & $5$\\                           
\hline
$(4,3)$       & $4$                                   & $11$                                  & $1$\\                       
\hline
$(5,0)$       & $/$                                   & $/$                                   & $31$\\
\hline
$(5,1)$       & $5$                                   & $1$                                   & $26$\\                 
\hline
$(5,2)$       & $10$                                  & $6$                                   & $16$\\                           
\hline
$(5,3)$       & $10$                                  & $16$                                  & $6$\\                       
\hline
$(5,4)$       & $5$                                   & $26$                                  & $1$\\  
\hline
\end{tabular}
\end{table}

{

\section{Bethe state at degenerate points}\label{sec:BetheState}
\setcounter{equation}{0}

The classification of the Bethe roots also affects the eigenstates. The eigenstates of the XXZ model with the constrained non-diagonal boundary fields \eqref{special_boundaries} can be expanded in terms of a set of factorized chiral vectors \cite{Zhang2021CBA}.
In the generic case (under Eq. \eqref{special_boundaries}), the eigenstates of the Hamiltonian have nonzero projections onto every basis vector, and they can be written in the
form of the conventional coordinate Bethe Ansatz.
At degenerate points, however, some of these projections may vanish or exhibit distinctive features
\cite{Zhang2021CBA,Zhang2021Phantom}.

\subsection{Chiral basis}

Introduce the following local left vector on site $n$
\begin{align}
&\phi_n(x)=\left(1,\,-\i\e^{\theta_-+\alpha_1^-+\alpha_2^-+(2x-n+1)\eta}\right).\label{left;local}
\end{align}
We then construct the following factorized global vectors \cite{Zhang2021CBA}
\begin{align}
&\bra{\underbrace{0,\dots,0}_{m_0},\underbrace{n_1,\dots,n_{k}}_{k},\underbrace{N,\dots,N}_{m_N}}\no\\
&=\e^{\eta (N m_N+\sum_{j=1}^k n_j)}\bigotimes_{l_1=1}^{n_1}\phi_{l_1}(m_0)\bigotimes_{l_2=n_1+1}^{n_2}\phi_{l_2}(m_0+1)\dots
\bigotimes_{l_{k+1}=n_k+1}^{N}\phi_{l_{k+1}}(m_0+k),\label{BraVecs}\\
&0<n_1<n_2<\cdots<n_k<N, \qquad k\geq 0.\no
\end{align}
One can regard $\e^{\gamma}$ in the vector $(1,\e^{\gamma})$ as a phase factor. In the factorized vector defined in Eq. \eqref{BraVecs}, \(m_0\) determines the phase of the local vector  at the first site, and the phases of the local vectors change by a fixed amount \(-\eta\) from site \(l\) to site \(l+1\), except at the kinks \(n_1,\ldots,n_k\).

Under the constraint \eqref{special_boundaries}, the
vectors defined in Eq. \eqref{BraVecs} with the following configuration 
\begin{align}
m_0,m_N=0,1,\ldots M,\quad m_0+k+ m_N = M, \label{extended:basis}
\end{align}
form a symmetric extended basis for the invariant subspace $G_M^+$, and one can use these vectors to expand the Bethe states. In the set \eqref{extended:basis}, the \textit{maximum} number of kinks equals \(M\), and can be treated as a conserved charge.

Analogously, we introduce the local ket vectors
\begin{align}
\tilde\phi_n(x)=\left(\begin{array}{c}
	1\\
	-\i\e^{-\theta_--\alpha_1^--\alpha_2^-+(2x-n+1)\eta}
\end{array}
\right),\label{right:local}
\end{align}
and then construct the factorized global vectors
\begin{align}
&\ket{\underbrace{0,\dots,0}_{m_0},\underbrace{n_1,\dots,n_{k}}_{k} ,\underbrace{N,\dots,N}_{m_N}},\label{KetVecs}
\end{align}
which can be obtained from Eq. \eqref{BraVecs}
by replacing $\phi \rightarrow \tilde{\phi}$.
The ket vectors in Eq. \eqref{KetVecs} with the configuration
\begin{align}
&m_0+k+ m_N = M',\quad m_0,m_N=0,1,\ldots M',\label{M;RightBasis}
\end{align}
form a symmetric extended basis of the invariant subspace $G_M^-$. Notably, it is clarified in Refs. \cite{Zhang2021CBA,Zhang2021Phantom} that any vector in $G_M^+$ is orthogonal to any vector in $G_M^-$.

\subsection{Chiral coordinate Bethe ansatz}

Define the following parameters 
\begin{align}
&a_l^\pm=\frac{\sinh(\alpha_l^\pm+\eta)}{\sinh(\alpha_l^\pm)},\quad l=1,2.
\end{align}
The eigenstate of the Hamiltonian can be constructed as follows
\begin{align}
\bra{\Psi_M(\bm{u})}=\sum_{l_1,\ldots,l_M}\bra{l_1,\ldots,l_M}f_{l_1,\ldots,l_M}(\bm{u}),\label{BetheState}
\end{align}
where the wave function $f_{l_1,\ldots,l_M}(\bm{u})$ is parameterized by $\bm{u}=\{u_1,\ldots,u_M\}$ as \cite{Zhang2021CBA}
\begin{align}
f_{l_1,\ldots,l_M}(\bm{u})=c_{m_0,m_N}\sum_{r_1,\ldots,r_M}\,\sum_{\sigma_1,\ldots,\sigma_{M}=\pm 1}A_{\sigma_{r_1},\ldots,\sigma_{r_M}}^{r_1,\ldots,r_M}\prod_{i=1}^M \left(u_{r_i}\right)^{\sigma_{r_i}l_i}.\label{wavefunction}
\end{align}
In Eq. \eqref{BetheState}, all configurations $\{l_1,\ldots,l_M\}$ allowed by \eqref{extended:basis} are summed over, $(r_1,\ldots,r_M)$ is a permutation of $(1,\ldots,M)$, and the coefficients $c_{m_0,m_N}$ depend only on the numbers of $0$'s and $N$'s in $\{l_1,\ldots,l_M\}$
\begin{align}
c_{m_0,m_N}&=\prod_{j=1}^{m_0}F_-(j)\prod_{j'=1}^{m_N}F_+(j'),\label{c0N}\\
F_{\sigma}(k)&=\begin{cases}
 1, & k=0,\\
 -\frac{\sinh(\alpha_1^\sigma+(k-1)\eta)\sinh(\alpha_2^\sigma+(k-1)\eta)}{\sinh(k\eta)\sinh(\alpha_1^\sigma+\alpha_2^\sigma+k\eta)}, & k\geq 1.
\end{cases}\label{F:plus:minus}
\end{align}

The amplitudes $A_{\sigma_{r_1},\ldots,\sigma_{r_M}}^{r_1,\ldots,r_M}
$ depend on $\{u_1,\ldots,u_M\}$ and are determined by the two-body scattering matrix and the reflection matrices as follows \cite{Zhang2021Phantom,Zhang2021CBA}
\begin{align}
\frac{A^{\ldots,k,j,\ldots}_{\ldots,\sigma_k,\sigma_j,\ldots}}{A^{\ldots,j,k,\ldots}_{\ldots,\sigma_j,\sigma_k,\ldots}}=S_{j,k}(u_j^{\sigma_j},u_k^{\sigma_k}),\quad \frac{A_{-,\ldots}^{j,\ldots}}{A_{+,\ldots}^{j,\ldots}}=S_{\rm L}(u_j),\quad \frac{A_{\ldots,-}^{\ldots,j}}{A_{\ldots,+}^{\ldots,j}}=u_j^{2N}S_{\rm R}(u_j),
\end{align}
where 
\begin{align}
S_{j,k}(u_j,u_k)&=-\frac{1-2\cosh\eta\,u_k+u_j u_k}{1-2\cosh\eta\,u_j+u_j u_k},\\
S_{\rm L}(u_j)&=-\frac{(1-a_1^-u_j)(1-a_2^-u_j)}{(a_1^--u_j)(a_2^--u_j)},\\
S_{\rm R}(u_j)&=-\frac{(a_1^+-u_j)(a_2^+-u_j)}{(1-a_1^+u_j)(1-a_2^+u_j)}.
\end{align}
The parameters $\{u_1,\ldots,u_M\}$ in the wave function \eqref{wavefunction} satisfy the BAEs
\begin{align}
u_j^{2N}S_{\rm R}(u_j)S_{\rm L}(u_j^{-1})\prod_{k\neq j}S_{j,k}(u_j,u_k)S_{k,j}(u_k,u_j^{-1})=1,\quad j=1,\ldots,M,\label{BAE:GM+:momentum}
\end{align}
which are equivalent to \eqref{BAE_G_M^+_x} under the transformation $u_j=\frac{\sinh(x_j+\frac{\eta}{2})}{\sinh(x_j-\frac{\eta}{2})}$. The energy in terms of $\{u_1,\ldots,u_M\}$ reads 
\begin{align}
\mathcal{E}(\bm{u})=2\sum_{j=1}^M(u_j+u_j^{-1}-\e^{\eta}-\e^{-\eta})+E_0.
\end{align}
Analogously, one can construct the ket Bethe state belonging to $G_M^-$. In the generic case (under Eq. \eqref{special_boundaries}), there is no spectral degeneracy, and the bra eigenstate in $G_M^+$ and the ket eigenstate in $G_M^-$ are all independent eigenstates. 

As discussed in Section \ref{sec:fully:degeneration}, at the fully degenerate point \eqref{Constraints_bp} there exist twofold degenerate states, each pair consisting of one state in $G_M^+$, denoted by $\bra{\varphi}$, and one in
$G_M^-$, denoted by $\ket{\bar{\varphi}}$. For a Hermitian system, $\bra{\varphi}$ and $\ket{\bar{\varphi}}$ can still be regarded as linearly independent. In contrast, their independence is not guaranteed in the non-Hermitian case, because the fully degenerate point may be an exceptional point.

\subsection{Bethe states at the degenerate points}

\paragraph{Fully degenerate points}
The fully degenerate point condition in \eqref{Constraints_bp} can be rewritten as
\begin{align}
({\rm i}):\quad & a_1^-a_2^-=a_1^+a_2^+=1,\label{fully:degenerate:case1}\\
({\rm ii}):\quad & a_1^-a_1^+=a_2^-a_2^+=1.\label{fully:degenerate:case2}
\end{align}
The type I, II and III Bethe-root configurations in
Section \ref{sec:classification} correspond to the cases in which none, one and two elements of $\{u_1,\ldots,u_M\}$ lie at specific points $a_j^\sigma$, respectively.

Case (i) in \eqref{fully:degenerate:case1} gives separate constraints on the left and right boundaries, whereas case (ii) in \eqref{fully:degenerate:case2} gives crossed constraints between them. Although the boundary parameters $\{\alpha_1^-,\alpha_2^-,\alpha_1^+,\alpha_2^+\}$ (or $\{a_1^-,a_2^-,a_1^+,a_2^+\}$) play the same role in the spectrum of the Hamiltonian, they affect the eigenstates differently, thus leading to quite different properties at the degenerate points \eqref{fully:degenerate:case1} and \eqref{fully:degenerate:case2}.

In the case $a_1^- a_2^-=a_1^+a_2^+=1$ (i.e., $\sinh(\alpha_1^-+\alpha_2^-+\eta)=\sinh(\alpha_1^++\alpha_2^++\eta)=0$), where the left and right boundary fields are separately constrained, the Bethe states corresponding to the type II and type III BAEs possess the following property
\begin{align}
&\mbox{type II}\quad (u_M=a_1^-):\quad f_{l_1,\ldots,l_M}(\bm u)=0,\quad \mbox{with}\,\, l_1\neq 0,\\
&\mbox{type II}\quad (u_M=a_1^+):\quad f_{l_1,\ldots,l_M}(\bm u)=0,\quad \mbox{with}\,\, l_M\neq N,\\
&\mbox{type III}:\quad f_{l_1,\ldots,l_M}(\bm u)=0,\quad \mbox{with}\,\, l_1\neq 0\,\, \mbox{or}\,\, l_M\neq N.
\end{align}
This property can be understood from Eqs. \eqref{c0N} and \eqref{F:plus:minus} in a simple way. Since $F_-(1)$ and $F_+(1)$ diverge in this case, only the terms containing $F_\pm(1)$ survive after normalization, unless the remaining part of the wave function in
\eqref{wavefunction} goes to zero.  
Obviously, there exist internal subspaces in $G_M^+$. For the Bethe states corresponding to the type II and type III BAEs, the maximum numbers of kinks reduce to $M-1$ and $M-2$, respectively. 

For the type I Bethe roots, the maximum number of kinks still equals $M$, yet the boundary reflection matrices in the corresponding wave functions are greatly simplified
\begin{align}
S_{\mathrm{L}}(u_j)=S_{\mathrm{R}}(u_j)=-1.
\end{align}

For the case where the two boundary fields are cross-constrained (i.e., $a_1^-a_1^+=a_2^-a_2^+=1$), by contrast, no further partitioning of the subspace $G_M^+$ is observed. However, the boundary reflection matrices in the wave function corresponding to different Bethe-root configurations still exhibit the following interesting properties
\begin{align}
&\text{type I}:\quad S_{\mathrm{L}}(u_j)=S_{\mathrm{R}}(u_j),\\
&\text{type II}:\quad S_{\mathrm{L}}(u_M)=S_{\mathrm{R}}(u_M)=0,\quad
u_M=a_1^+ \text{ or } a_2^+,\\
&\text{type III}:\quad S_{\mathrm{L/R}}(u_{M-1})=S_{\mathrm{L/R}}(u_{M})=0,
\quad u_{M-1}=a_1^+,\ u_M=a_2^+.
\end{align}
The existence of boundary-pinned roots simplifies the wave function. Whether the Bethe state at this special point possesses any other characteristic properties remains an open question.

\paragraph{Partially degenerate points}

The Bethe states at the partially degenerate points show similar properties. Equation  \eqref{Constraints_bp2} can be rewritten as
\begin{align}
({\rm i}):\quad & a_1^-a_2^-=1.\quad \mbox{or}\quad  a_1^+a_2^+=1,\label{partially:degenerate:case1}\\
({\rm ii}):\quad & a_1^-a_1^+=1.\quad \mbox{or}\quad  a_2^-a_2^+=1,\label{partially:degenerate:case2}
\end{align}
type I and II Bethe-root configurations in
Section \ref{sec:partially:degeneration} correspond to the cases in which none, and one element of $\{u_1,\ldots,u_M\}$ lies at specific points $a_j^\sigma$, respectively.

At the partially degenerate points $a_1^\sigma a_2^\sigma=1$,  the Bethe state corresponding to the type I and type II BAEs in
Section \ref{sec:partially:degeneration} possesses the following property
\begin{align}
\mbox{type I}\quad (a_1^-a_2^-=1):&\quad  S_{\rm L}(u_j)=-1,\\
\mbox{type I}\quad (a_1^+a_2^+=1):&\quad  S_{\rm R}(u_j)=-1,\\
\mbox{type II}\quad (a_1^-a_2^-=1):&\quad u_M=a_1^-,\quad f_{l_1,\ldots,l_M}(\bm u)=0,\quad \mbox{with}\,\, l_1\neq 0,\\
\mbox{type II}\quad (a_1^+a_2^+=1):&\quad u_M=a_1^+,\quad f_{l_1,\ldots,l_M}(\bm u)=0,\quad \mbox{with}\,\, l_M\neq N.
\end{align}

Analogously to the fully degenerate points, for $a_k^-a_k^+=1$, no further partitioning of the invariant subspace $G_M^+$ is observed, whereas both $S_{\mathrm{L}}(u_M)$ and $S_{\mathrm{R}}(u_M)$ in the wave function vanish when $u_M=a_k^+$.

As an example, we analyze the Bethe state and its properties at degenerate points in the case $M=1$.
Acting the Hamiltonian on the chiral vectors, we get 
\begin{align}
\bra{n}H&=2\bra{n-1}+2\bra{n+1}+\left(-a_1^--a_2^--a_1^+-a_2^+\right)\bra{n},\quad n=1,\ldots,N-1,\\
\bra{0}H&=2(1-a_1^-a_2^-)\bra{1}+\left(a_1^-+a_2^--a_1^+-a_2^+\right)\bra{0},\\
\bra{N}H&=2(1-a_1^+a_2^+)\bra{N-1}+\left(a_1^++a_2^+-a_1^--a_2^-\right)\bra{N}.
\end{align}
The Bethe state is 
\begin{align}
\bra{\Psi_1(u_1)}&=\sum_{n=1}^{N-1}f_n(u_1)\bra{n}+\frac{f_0(u_1)}{1-a_1^-a_2^-}\bra{0}+\frac{f_N(u_1)}{1-a_1^+a_2^+}\bra{N}\no\\
&\overset{\eqref{BAE:GM+:momentum}}{=}\sum_{n=1}^{N-1}f_n(u_1)\bra{n}+(u_1^2-1)\bra{0}+\frac{(u_1-a_1^-)(u_1-a_2^-)}{(a_1^+ u_1-1)(a_2^+ u_1-1)}\left(1-u_1^2\right) u_1^{N}\bra{N},\no\\
&\overset{\eqref{BAE:GM+:momentum}}{=}\sum_{n=1}^{N-1}f_n(u_1)\bra{n}+(u_1^2-1)\bra{0}+\frac{(a_1^- u_1-1)(a_2^- u_1-1)}{(u_1-a_1^+)(u_1-a_2^+)}\left(1-u_1^2\right) u_1^{-N}\bra{N},\\
f_n(u)&=u^n (u-a_1^-)(u-a_2^-)-u^{-n} (a_1^-u-1)(a_2^-u -1).
\end{align}

In the case $a_1^- a_2^-=1$ (or $a_1^+a_2^+=1$), there will be a factorized eigenstate $\bra{0}$ (or $\bra{N}$) in $G_M^+$, which corresponds to the solution $u_1=a_1^-$ (or $u_1=a_1^+$).

By contrast, for the case $a_1^- a_1^+=1$, the Bethe state corresponding to $u_1=a_1^-$ is still a linear combination of $\bra{0},\bra{1},\ldots,\bra{N}$
\begin{align}
\bra{\Psi_1(a_1^-)}\propto\sum_{n=1}^{N-1}(1-a_1^-a_2^-) (a_k^-)^{-n}\bra{n}+\bra{0}-\frac{a_1^- a_2^--1}{a_1^--a_2^+}(a_1^-)^{1-N}\bra{N}.
\end{align}
which shows the feature of a traveling wave, whereas the type I Bethe roots give a standing wave.}

\section{Bethe-root configurations and spectral degeneracies in the diagonal limit}\label{sec:diagonal:boundary}
\setcounter{equation}{0}

The classification scheme developed in Sections \ref{sec:fully:degeneration} and \ref{sec:partially:degeneration} for non-diagonal boundaries has a natural counterpart in the diagonal boundary limit, where the boundary magnetic fields become purely longitudinal and the $U(1)$ symmetry is restored. As we shall see, the phantom-string mechanism and spectral degeneracy also persist in this case, providing a nontrivial consistency check of our analysis.

By letting $\alpha_2^\pm\to \mp\infty$, the boundary fields all point along the $z$-direction with 
\begin{align} 
\vec{h}_1=\frac{\sinh\eta}{\tanh\alpha_1^-}(0,0,1),\quad 
\vec{h}_N=\frac{\sinh\eta}{\tanh\alpha_1^+}(0,0,1).\label{h1N:d}
\end{align}
In this case, due to the restoration of the $U(1)$ symmetry, the exact solution of the model can be given by the following Bethe Ansatz equations 
\begin{align}
\left [\frac{\sinh(\lambda_j+\frac{\eta}{2})}{\sinh(\lambda_j-\frac{\eta}{2})} \right ]^{2N} \prod_{\sigma=\pm }\frac{\sinh(\lambda_j-\alpha_1^\sigma-\frac{\eta}{2})}{\sinh(\lambda_j+\alpha_1^\sigma+\frac{\eta}{2})}\prod_{k\ne j}^{M_0}S(\lambda_j,\lambda_k)=1,\quad j=1,\ldots,M_0,\label{BAEs:diag}
\end{align}
where the integer $M_0$ (different from the parameter $M$ used in the preceding sections) represents the magnon number and ranges from $0$ to $N$.
The energy of the model in terms of the Bethe roots $\bm{\lambda}=\{\lambda_1,\ldots,\lambda_{M_0}\}$ is 
\begin{align}
E_{\rm d}(\bm{\lambda})=\sum_{j=1}^{M_0}\varepsilon(\lambda_j)-\sinh\eta(\coth\alpha_1^-+\coth\alpha_1^+).
\end{align}
The boundary-related terms in the BAEs \eqref{BAEs:diag} are of degree two. The degenerate points are now 
\begin{align}
\alpha_1^+ = -\alpha_1^- - \eta + \i m\pi,\quad m\in\mathbb{Z},\label{degenerate:diag}
\end{align}
under which the boundary fields become 
\begin{align}
\vec{h}_1=\frac{\sinh\eta}{\tanh\alpha_1^-}(0,0,1),\quad 
\vec{h}_N=-\frac{\sinh\eta}{\tanh(\alpha_1^-+\eta)}(0,0,1).\label{h1N:d2}
\end{align}
At the degenerate point \eqref{degenerate:diag}, the BAEs \eqref{BAEs:diag} can be classified 
into the following two categories, as established in Ref.~\cite{Jiang:2024}.
\paragraph{Type I} All the roots satisfy the following equations
\begin{align}
\left [\frac{\sinh(\lambda_j+\frac{\eta}{2})}{\sinh(\lambda_j-\frac{\eta}{2})} \right ]^{2N}\prod_{k\ne j}^{M_0}S(\lambda_j,\lambda_k)=1,\quad j=1,\ldots,M_0,\label{BAE:diag:I}
\end{align}
\paragraph{Type II} One root is pinned with $\lambda_{M_0}=\alpha_1^-+\tfrac{\eta}{2}$ and the remaining roots satisfy
\begin{align}
\left[\frac{\sinh(\lambda_j+\frac{\eta}{2})}{\sinh(\lambda_j-\tfrac{\eta}{2})} \right ]^{2N}S(\lambda_j,\alpha_1^-+\tfrac{\eta}{2})\prod_{k\ne j}^{M_0-1}S(\lambda_j,\lambda_k)=1,\quad j=1,\ldots,M_0-1.\label{BAE:diag:II}
\end{align}

The BAEs \eqref{BAE:diag:I} and \eqref{BAE:diag:II} share the same form as the type I and type II BAEs in \eqref{iBAE_I_x} and \eqref{iBAE_II_x} in the non-diagonal case, and the Bethe roots in both cases satisfy the same selection rules \eqref{selection:rule}. Consequently, the analysis of the Bethe-root configurations can be referred to Section \ref{sec:solution:number}.

When $M_0 <\frac{N+1}{2}$, the BAEs \eqref{BAE:diag:I} and\eqref{BAE:diag:II} 
admit only finite solutions, and the total numbers of their solutions are 
$\binom{N}{M_0}-\binom{N}{M_0-1}$ and $\binom{N}{M_0-1}$, respectively.

In the case $M_0 \geq \frac{N+1}{2}$, the only physical solutions of the type II BAEs \eqref{BAE:diag:II} 
take the same configuration as in \eqref{type2:solution}. Specifically, among $\{\lambda_1,\ldots,\lambda_{M_0-1}\}$, $N-M_0$ roots are finite and satisfy the same equations as the BAEs \eqref{BAE:diag:II} (with $M_0$ replaced 
by $N+1-M_0$), while the remaining $2M_0-1-N$ roots form a phantom string. The BAEs \eqref{BAE:diag:II} thus possess $\binom{N}{M_0}$ sets of solutions. 
Notably, the type II BAEs with $M_0=m,\, m<\frac{N+1}{2}$ and $M_0=N+1-m$ yield identical energy spectra. As a consequence, $\sum_{k=0}^{[\frac{N}{2}]-1} \binom{N}{k}$ of the $2^N$ energy levels are doubly degenerate.

To study the type I BAEs \eqref{BAE:diag:I} with \(M_0 \geq \frac{N+1}{2}\), we first consider the case \(\alpha_1^+ = -\alpha_1^- - \eta + \epsilon\), where \(\epsilon\) is small. Unlike the non-diagonal case in Section \ref{sec:classification}, when \(M_0 \geq \frac{N+1}{2}\), one can only find solutions of the original BAEs \eqref{BAEs:diag} with \(\lambda_{M_0} \approx \alpha_1^- + \frac{\eta}{2}\) while some Bethe roots simultaneously form a quasi-phantom string, as shown in Tables \ref{Tab:diag:1} and \ref{Tab:diag:2}.
This observation suggests that the type I BAEs have no physical solutions for \(M_0 \geq \frac{N+1}{2}\) cases \cite{Jiang:2024}. Nevertheless, from the results in Section \ref{sec:solution:number}, the BAEs \eqref{BAE:diag:I} do possess $\binom{N}{M_0-1}-\binom{N}{M_0}$ ``nontrivial" solutions, which at first glance appear contradictory. The explanation is as follows: in the diagonal case, the Hilbert space decomposes into subspaces labeled by the magnon number \(M_0\). For the case \(M_0 \geq \frac{N+1}{2}\), no physical solution exists in the \(M_0\)-magnon sector. Instead, the solutions of \eqref{BAE:diag:I} yield the correct energy levels, which belong to the \((N+1-M_0)\)-magnon sector.

It is instructive to compare the diagonal case with the non-diagonal fully degenerate case in Section \ref{sec:fully:degeneration}. In both cases, phantom strings induce spectral degeneracies, but the nature of the subspaces involved is fundamentally different.
In the diagonal case, the $U(1)$ symmetry decomposes the Hilbert space into distinct $S^z$ eigensectors labeled by the magnon number $M_0$.  The degeneracy
occurs between two \textit{different} $S^z$ sectors, namely $M_0$ and $M_0'=N+1-M_0$. In the non-diagonal fully degenerate case, by contrast, the $U(1)$ symmetry is broken. The relevant invariant subspaces are $G_M^+$ and $G_M^-$, which correspond to two \textit{distinct sets} of BAEs, namely \eqref{BAE_G_M^+_x} and \eqref{BAE_G_M^-_x}. The degeneracy is therefore a genuine \textit{cross-sector} degeneracy connecting the two globally defined invariant subspaces $G_M^+$ and $G_M^-$.

{Due to the $Z_2$ symmetry, the spectrum of the Hamiltonian is invariant under the substitution $\alpha_1^\pm \to -\alpha_1^\pm$. Consequently, the classification of Bethe roots and the spectral degeneracy can also be studied by applying this substitution to the BAEs \eqref{BAEs:diag} and the degenerate point condition \eqref{degenerate:diag}.

The classification of the Bethe roots at degenerate points also affects the eigenstates. The eigenstate of the Hamiltonian (the Bethe state) in the sector with $M_0$ magnons is a linear combination of the following vectors
\begin{align}
\sigma_{l_1}^+\sigma_{l_2}^+\cdots \sigma_{l_{M_0}}^+\ket{\downarrow\cdots\downarrow},\quad 1\leq l_1<l_2<\cdots <l_{M_0}\leq N,
\end{align}
and the wave functions can be parametrized by the conventional coordinate Bethe ansatz method. It has been shown that the classification of Bethe roots is related to a specific conserved charge and its action on the Bethe state, see Ref. \cite{Jiang:2024}.}

{
\section{Bethe-root configurations at other quasi-degenerate points} \label{sec:quasi:degenerate}
\setcounter{equation}{0}
\setcounter{table}{0}
\setcounter{figure}{0}

In Sections \ref{sec:fully:degeneration},
\ref{sec:partially:degeneration} and \ref{sec:diagonal:boundary}, we
demonstrate that two types of special solutions can exist at degenerate points: (1) boundary-pinned Bethe roots and (2) phantom strings. However, the degenerate points studied above are not the only points in which such special solutions exist. In this section, we investigate the quasi-degenerate point condition,
under which phantom string solutions or boundary-pinned singular solutions of the BAEs exist.

\subsection{Existence of phantom strings}

Let us go back to the BAEs \eqref{BAE_G_M^+_x} and suppose that they admit the following solution
\begin{align}
\{x_1,\ldots,x_{M}\}=\{y_1,\ldots,y_{\bar m}\}\cup\{\mathrm{PS}\},\qquad 0\leq \bar m<M,\label{Generic:string}
\end{align}
where the finite roots are $\{y_1,\ldots,y_{\bar{m}}\}$, and the phantom
string is given in \eqref{phantom:string} with length $M-\bar{m}$. Substituting the above-mentioned solution \eqref{Generic:string} into the BAEs \eqref{BAE_G_M^+_x}, we find that, for a phantom root $x_j$, the left-hand side of \eqref{BAE_G_M^+_x} becomes
\begin{align}
&\e^{-2(\alpha_1^++\alpha_1^-+\alpha_2^++\alpha_2^-)}
\e^{2(N-\bar m-M-1)\eta}
\prod_{k\ne j}^{M-\bar m}
\frac{\sinh\left(\i\pi\frac{j-k}{M-\bar m}-\eta\right)}
     {\sinh\left(\i\pi\frac{j-k}{M-\bar m}+\eta\right)}\nonumber\\
&=\e^{-2(\alpha_1^++\alpha_1^-+\alpha_2^++\alpha_2^-)}
\e^{2(N-\bar m-M-1)\eta}.\label{proof:phantom:2}
\end{align}
To match the right-hand side of \eqref{BAE_G_M^+_x} (i.e., 1), the boundary parameters
should satisfy
\begin{align}
\e^{2(N-\bar m-M-1)\eta}=\e^{2(\alpha_1^++\alpha_1^-+\alpha_2^++\alpha_2^-)},\label{phantom:string:condition}
\end{align}
which is the necessary condition for the existence of phantom strings.

Analogously, the BAEs for $G_M^-$ \eqref{BAE_G_M^-_x} admit phantom
string solutions as in \eqref{phantom:string} once the following equation
\begin{align}
\e^{2(N-\bar{m}'-M'-1)\eta}=
\e^{-2(\alpha_1^++\alpha_1^-+\alpha_2^++\alpha_2^-)},
\qquad 0\leq \bar{m}'<M,\label{phantom:string:condition:2}
\end{align}
is satisfied, where $M'-\bar{m}'$ is the length of the phantom string.

\begin{figure}[htbp]
\centering
\begin{tikzpicture}
\begin{axis}[
  width=0.5\textwidth,
  axis lines=middle,
  xlabel={${\rm Re}$},
  ylabel={${\rm Im}$},
  xlabel style={at={(axis cs:4.8,0)}, anchor=west},
  ylabel style={at={(axis cs:0,2.8)}, anchor=south},
  xmin=0, xmax=4.8,
  ymin=0, ymax=2.8,
  ytick={0, pi/4, pi/2, 3*pi/4},
  yticklabels={$0$, $\frac{\pi}{4}$, $\frac{\pi}{2}$, $\frac{3\pi}{4}$},
  grid=none,
  only marks,
  axis line style={-}
]
\addplot[blue, mark=asterisk, mark size=3pt]  coordinates {
(1.53391,0.00000)
(1.53641,0.75722)
(1.49172,1.57080)
(1.53641,2.38437)
};

\addplot[green, mark=asterisk, mark size=3pt]  coordinates {
(2.10010,0.00000)
(2.09447,0.77717)
(2.08409,1.57080)
(2.09447,2.36442)
};

\addplot[red, mark=asterisk, mark size=3pt]  coordinates {
(2.67130,0.00000)
(2.66898,0.78283)
(2.66618,1.57080)
(2.66898,2.35876)
};

\addplot[black, mark=asterisk, mark size=3pt]  coordinates {
(3.24530,0.00000)
(3.24452,0.78459)
(3.24368,1.57080)
(3.24452,2.35701)
};

\addplot[purple, mark=asterisk, mark size=3pt]  coordinates {
(3.82041,0.00000)
(3.82015,0.78514)
(3.81989,1.57080)
(3.82015,2.35645)
};

\draw[dashed, thick, gray,line width=0.5pt] (axis cs:0,pi/4) -- (axis cs:5,pi/4);

\draw[dashed, thick, gray,line width=0.5pt] (axis cs:0,pi/2) -- (axis cs:5,pi/2);

\draw[dashed, thick, gray,line width=0.5pt] (axis cs:0,3*pi/4) -- (axis cs:5,3*pi/4);

\end{axis}
\end{tikzpicture}
~~~~

\caption{Numerical evidence for the  quasi-phantom string solution of BAEs \eqref{BAE_G_M^+_x} with $N=5$, $M=4$, $\eta=0.66$, $\alpha_1^-=0.81$, $\alpha_1^+=0.33-\alpha_1^- +\epsilon$, $\alpha_2^-=0.32$, $\alpha_2^+=-0.33-\alpha_2^- -1.5\epsilon$, $\theta_+=0.9$ and $\theta_-=(N-1-2M)\eta-(\alpha_1^- +\alpha_2^- +\alpha_1^+ +\alpha_2^+) -\i \pi +\theta_+$. Here, blue, green, red, black and purple dots represent $\epsilon=10^{-2}$, $\epsilon=10^{-4}$, $\epsilon=10^{-6}$, $\epsilon=10^{-8}$ and $\epsilon=10^{-10}$, respectively. The energies corresponding to the Bethe roots (with descending $\epsilon$) are $-0.50405382$, $-0.52037748$, $-0.52052934$, $-0.52053085$ and $-0.52053087$. No spectral degeneracies occur in the limit $\epsilon\to 0$.}\label{Fig:phantom:string:2}
\end{figure}

The phantom string can only lead to spectral degeneracy if the BAEs of the system admit two solutions that share the same finite Bethe roots, one of which contains a phantom string while the other does not. In general, degeneracy is absent in the subspace $G_M^+$, since the total number of Bethe roots is always fixed at $M$ and the phantom string also contributes to this number.
On the other hand, under the condition in Eq. \eqref{phantom:string:condition}, the energy sets of the subspaces $G_M^+$ and $G_M^-$ do not overlap in the generic case. This can be understood from two perspectives: (i) the BAEs for $G_M^+$ and $G_M^-$ are not obviously related; (ii) the background energies associated with the two subspaces are unequal. A specific example is given in Figure \ref{Fig:phantom:string:2}, where we
can observe the quasi-phantom string in the regime sufficiently close to the quasi-degenerate point, where no spectral degeneracy occurs.

\subsection{Existence of boundary-induced strings}

The Bethe-root configuration with boundary-pinned roots is a type of singular solution of the BAEs, since it gives rise to both zeros and poles
in the BAEs. At the degenerate points studied in
Sections~\ref{sec:fully:degeneration}, \ref{sec:partially:degeneration} and \ref{sec:diagonal:boundary}, the zeros and poles all originate from the boundary-reflection terms. There exist, however, other particular singular solutions that coincide with zeros or poles of the BAEs, and the zeros and poles involved may originate in the boundary-reflection terms and the bulk-scattering terms alike.

When two of the four boundary parameters
$\{\alpha_1^-, \alpha_1^+, \alpha_2^-, \alpha_2^+\}$ form a pair 
\begin{align}
\{\zeta,-\zeta-k\eta\pmod{\i\pi}\},\quad k\in\mathbb{N},\label{boundary:string:condition:1}
\end{align}
the BAEs \eqref{BAE_G_M^+_x} admit solutions with $k$ roots forming the following
boundary-induced string \cite{He2025}\footnote{In Ref. \cite{He2025}, the authors introduced the boundary-induced string in the study of the XXX
model with diagonal boundary fields. In the present work, we show that an analogous string configuration can be generalized to the XXZ case.}
\begin{align}
\{x_1,\ldots,x_{k}\}&=\left\{\zeta+\tfrac{\eta}{2},\zeta+\tfrac{3\eta}{2},\ldots,\zeta+\tfrac{(2k-1)\eta}{2}\right\}.\label{boundary:string:0}
\end{align}
The partially degenerate points in Section \ref{sec:partially:degeneration} correspond to the case $k=1$. For $k>1$, the boundary-induced string is evenly distributed in the complex plane with spacing exactly $\eta$, which also gives rise to zeros, poles, or both in the bulk-scattering terms. 

When only two of the boundary parameters form a pair while the remaining two are arbitrary, the entire energy spectrum of the model is non-degenerate.

Moreover, there may be two independent boundary-induced strings once the
boundary parameters form two pairs as follows
\begin{align}
&\{\alpha_1^-, \alpha_1^+, \alpha_2^-, \alpha_2^+\}
=\left\{\zeta_-, -\zeta_--k_1\eta\!\pmod{\i\pi}, \zeta_+, -\zeta_+-k_2\eta\!\pmod{\i\pi}\right\},\no\\
&k_1,k_2\in \mathbb{N}^*,\qquad k_1+k_2\leq M.
\label{boundary:string:condition:2}
\end{align}
Under this condition, the BAEs \eqref{BAE_G_M^+_x} admit the following solutions
\begin{align}
(\text{i}):\quad & \text{none boundary-induced string},\\
(\text{ii-a}):\quad & \{x_1,\ldots,x_{k_1}\}
=
\left\{\zeta_- +\tfrac{\eta}{2},
\zeta_- +\tfrac{3\eta}{2},
\ldots,
\zeta_- +\tfrac{(2k_1-1)\eta}{2}\right\},\label{BIS:type:2:1}\\
(\text{ii-b}):\quad & \{x_{1},\ldots,x_{k_2}\}=\left\{\zeta_+ +\tfrac{\eta}{2},\zeta_+ +\tfrac{3\eta}{2},\ldots,\zeta_+ +\tfrac{(2k_2-1)\eta}{2}\right\},\label{BIS:type:2:2}\\
(\text{iii}):\quad & \{x_1,\ldots,x_{k_1}\}=\left\{\zeta_- +\tfrac{\eta}{2},\zeta_- +\tfrac{3\eta}{2},\ldots,\zeta_- +\tfrac{(2k_1-1)\eta}{2}\right\},\no\\
&\{x_{k_1+1},\ldots,x_{k_1+k_2}\}=\left\{\zeta_+ +\tfrac{\eta}{2},\zeta_+ +\tfrac{3\eta}{2},\ldots,\zeta_+ +\tfrac{(2k_2-1)\eta}{2}\right\}.\label{BIS:type:3}
\end{align}
The Bethe-root configurations in cases (i), (ii-a)/(ii-b) and (iii) are analogous to those of type I, II and III at the fully degenerate points, which correspond to $k_1=k_2=1$.

Under the condition \eqref{boundary:string:condition:1} or
\eqref{boundary:string:condition:2}, further partitioning of the invariant subspace
$G_M^+$ can occur. For instance, if
$\alpha_1^-+\alpha_2^- + k\eta=0\pmod{\i\pi}$, then $F_-(k)$ diverges,
and consequently the wave function corresponding to the vector
$\bra{l_1,\ldots,l_M}$ with fewer than $k$ zeros in
$\{l_1,\ldots,l_M\}$ vanishes in certain Bethe states.

\begin{figure}[H]
\centering
\begin{tikzpicture}
\begin{axis}[
  width=0.5\textwidth,
  axis lines=middle,
  xlabel={${\rm Re}$},
  ylabel={${\rm Im}$},
  xlabel style={at={(axis cs:3.2,0)}, anchor=west},
  ylabel style={at={(axis cs:0,2.8)}, anchor=south},
  xmin=0, xmax=3.2,
  ymin=0, ymax=2.8,
  ytick={0,  pi/4,  pi/2, 3*pi/4},
  yticklabels={$0$,  $\frac{\pi}{4}$, $\frac{\pi}{2}$, $\frac{3\pi}{4}$},
  xtick={0, 0.5, 0.8, 1, 2, 3},
  xticklabels={$0$, $\alpha_1^+$, $\alpha_1^-$, 1,2,3},
  grid=none,
  only marks,
  axis line style={-}
]
\addplot[blue, mark=asterisk, mark size=3pt]  coordinates {
(0.500000,0.300000)
(0.499954,0.899936)
};

\addplot[green, mark=asterisk, mark size=3pt]  coordinates {
(0.800017,0.300002)
(0.800475,0.901193)
(0.768057,1.501916)
};

\addplot[red, mark=asterisk, mark size=3pt]  coordinates {
(1.242965,0.000000)
(1.421671,1.028563)
(1.477194,1.732369)
(1.435724,2.419169)
};

\addplot[blue, mark=square, mark size=3pt]  coordinates {
(0.500000,0.300000)
(0.499999,0.899998)
};

\addplot[green, mark=square, mark size=3pt]  coordinates {
(0.800000,0.300000)
(0.800007,0.900026)
(0.799062,1.500114)
};

\addplot[red, mark=square, mark size=3pt]  coordinates {
(1.786221,0.000000)
(1.829895,0.860673)
(1.876838,1.610570)
(1.858728,2.345478)
};

\addplot[blue, mark=o, mark size=3pt]  coordinates {
(0.500000,0.300000)
(0.500000,0.900000)
};

\addplot[green, mark=o, mark size=3pt]  coordinates {
(0.800000,0.300000)
(0.800000,0.900000)
(0.799991,1.500001)
};

\addplot[red, mark=o, mark size=3pt]  coordinates {
(2.399268,0.000000)
(2.409230,0.806062)
(2.427277,1.581787)
(2.419875,2.348996)
};


\draw[dashed, thick, gray,line width=0.5pt] (axis cs:0,pi/4) -- (axis cs:3,pi/4);



\draw[dashed, thick, gray,line width=0.5pt] (axis cs:0,pi/2) -- (axis cs:3,pi/2);

\draw[dashed, thick, gray,line width=0.5pt] (axis cs:0,3*pi/4) -- (axis cs:3,3*pi/4);

\draw[dashed, thick, gray,line width=0.5pt] (axis cs:0.5,0) -- (axis cs:0.5,pi);

\draw[dashed, thick, gray,line width=0.5pt] (axis cs:0.8,0) -- (axis cs:0.8,pi);

\end{axis}
\end{tikzpicture}

\caption{Numerical evidence for the special solutions of BAEs \eqref{BAE_G_M^+_x} with $N=10$, $M=9$, $\eta=0.6\i$, $\alpha_1^-=0.8$, $\alpha_1^+=0.5$, $\alpha_2^-=-2\eta-\alpha_1^+ +\epsilon$, $\alpha_2^+=-3\eta-\alpha_1^- +1.22\epsilon$, $\theta_+=0.9\i$ and $\theta_-=(N-1-2M)\eta-(\alpha_1^- +\alpha_2^- +\alpha_1^+ +\alpha_2^+) -\i \pi +\theta_+$. Here, the blue, green, and red marks represent the first quasi boundary-induced string, the second quasi boundary-induced string,
and the quasi-phantom strings, respectively. The asterisk, square and circle marks represent $\epsilon=10^{-1.5}$, $\epsilon=10^{-3}$
and $\epsilon=10^{-5}$, respectively. The energies corresponding to the Bethe roots (with decreasing $\epsilon$) are $-0.09147+1.41357\i$, $-0.09521+1.39569\i$ and $-0.09532+1.39512\i$. There is one twofold degenerate level in the limit $\epsilon\to 0$.}\label{Fig:phantom:string:3}
\end{figure}

\subsection{Spectral degeneracy}

For the Bethe-root configuration given in Eq. \eqref{BIS:type:3}, the remaining Bethe roots $$\{{\bar x}_1,\ldots,{\bar x}_{M-k_1-k_2}\}=\{x_{k_1+k_2+1},\ldots,x_{M}\},$$ satisfy the following equations
\begin{align}
&\left[\frac{\sinh(\bar x_j+\frac{\eta}{2})}{\sinh(\bar x_j-\frac{\eta}{2})} \right ]^{2N} \prod_{\sigma=\pm }\, \prod_{l=1,2 }\frac{\sinh(\bar x_j+\alpha_l^\sigma-\frac{\eta}{2})}{\sinh(\bar x_j-\alpha_l^\sigma+\frac{\eta}{2})}\no\\
&\times \prod_{k\ne j}^{M-k_1-k_2}\frac{\sinh(\bar x_j-\bar x_k-\eta)}{\sinh(\bar x_j-\bar x_k+\eta)} \frac{\sinh(
\bar x_j+\bar x_k-\eta)}{\sinh(\bar x_j+\bar x_k+\eta)}=1,\label{BAE:G+:string}\\
&j=1,\ldots,M-k_1-k_2.\no
\end{align}
Notably, the two boundary-induced strings presented in
Eq. \eqref{BIS:type:3} contribute $-2E_0$ to the energy, thereby eliminating the background energy gap between $G_M^-$ and $G_M^+$ and yielding
\begin{align}
E^+(\bm{x})=\sum_{j=1}^{M-k_1-k_2} \varepsilon(\bar{x}_j)-E_0.\label{Energy:BID}
\end{align}
The BAEs \eqref{BAE:G+:string} and the energy expression \eqref{Energy:BID} have the same form as those in \eqref{BAE_G_M^-_x} and \eqref{E_G-_x} for $G_M^-$. This raises an question: are there any solutions with phantom strings that lead to spectral degeneracy under the condition given in \eqref{boundary:string:condition:2}? We address this question by examining the three regimes of $M$ separately.

\paragraph{$M\geq \frac{N+k_1+k_2}{2}$ case}

Here, the BAEs \eqref{BAE:G+:string} admit solutions of the form
\begin{align}
\{\bar{x}_1,\ldots,\bar{x}_{M-k_1-k_2}\}
&=
\{z_1,\ldots,z_{M'}\}\cup\{\text{PS}\},
\end{align}
where the roots $\{z_1,\ldots,z_{M'}\}$ are finite and satisfy the BAEs
\eqref{BAE_G_M^-_x}. The phantom string has length
$2M+1-k_1-k_2-N$. A specific example is given in Figure \ref{Fig:phantom:string:3}, where we
can observe both the quasi-phantom string and the quasi-boundary-induced string in the original set $\{x_1,\ldots,x_M\}$, in the regime sufficiently close to the quasi-degenerate point \eqref{boundary:string:condition:2}.

\paragraph{$M\leq \frac{N+k_1+k_2-2}{2}$ case}

In this regime, the BAEs \eqref{BAE_G_M^-_x} admit solutions of the form
\begin{align}
\{z_1,\ldots,z_{M'}\}
&=
\{\bar{x}_1,\ldots,\bar{x}_{M-k_1-k_2}\}\cup\{\text{PS}\},
\end{align}
where the roots $\{\bar{x}_1,\ldots,\bar{x}_{M-k_1-k_2}\}$ are finite and
satisfy the BAEs \eqref{BAE:G+:string}. The phantom string now has length $N-1-2M+k_1+k_2$.

\paragraph{$M=\frac{N+k_1+k_2-1}{2}$ case}

In this case, the two sets of BAEs \eqref{BAE_G_M^-_x} and
\eqref{BAE:G+:string} coincide and therefore possess the same solutions.

In summary, there exist twofold degenerate levels under the condition in
\eqref{boundary:string:condition:2}. Numerical results imply that the number of such twofold degenerate levels equals
$$\sum_{k=0}^{\min(M-k_1-k_2,\,M')}\binom{N}{k}.$$ This formula is consistent with the result obtained in
Section \ref{sec:degeneracy}, where $k_1=k_2=1$. It should be noted that the spectral degeneracy also exists when $k_1$, $k_2$, or both in Eq. \eqref{boundary:string:condition:2} are 0. For
example, when $k_1=k_2=0$ the boundary-induced string disappears, but the specific choice of boundary parameters makes the BAEs for $G_M^+$ and $G_M^-$ share the same form, which also leads to spectral degeneracy caused by the phantom string.

\subsection{Special Bethe roots and spectral degeneracy in the diagonal case}
In the diagonal case, the existence condition for phantom-string solutions of the BAEs \eqref{BAEs:diag} is given by
\begin{align}
\e^{2(N-m_0-M_0)\eta}=\e^{2(\alpha_1^++\alpha_1^-)},\quad 0\leq m_0<M_0,\label{phantom:condition:diag}
\end{align}
where $M_0$ and $m_0$ denote the total number of Bethe roots and the total number of finite Bethe roots, respectively.
Since $M_0$ is a free integer ranging from 0 to $N$, Eq. \eqref{phantom:condition:diag} consequently yields
\begin{align}
\alpha_1^++\alpha_1^-=(N-1)\eta,(N-2)\eta,\ldots,
(1-N)\eta,
\pmod{\mathrm{i}\pi}.
\end{align}
At the special points given in Eq. \eqref{phantom:condition:diag}, the
BAEs in the $M_0$-magnon sector admit a solution consisting of $m_0$ finite roots and a phantom string of length $M_0-m_0$. Since the phantom string does not affect the energy, this solution has the same energy as a state in the $m_0$-magnon sector, leading to spectral degeneracy. This degeneracy has also been investigated in Ref. \cite{He2025}, and a
specific example is shown in Figure \ref{Fig:degeneracy}.

{As in the non-diagonal case, when
$\alpha_1^+=-\alpha_1^--k_0\eta\pmod{\i\pi}$ with $k_0$ a positive integer, in certain solutions $k_0$ of the $m_0$ regular roots form the boundary-induced string
\begin{align}
\{\lambda_1,\ldots,\lambda_{k_0}\}
=\{\alpha_1^-+\tfrac{\eta}{2},\alpha_1^-+\tfrac{3\eta}{2},\ldots,-\alpha_1^+-\tfrac{\eta}{2}\}.
\end{align}}

\begin{figure}[htbp]
\begin{subfigure}{0.45\textwidth}
\includegraphics[width=\textwidth]{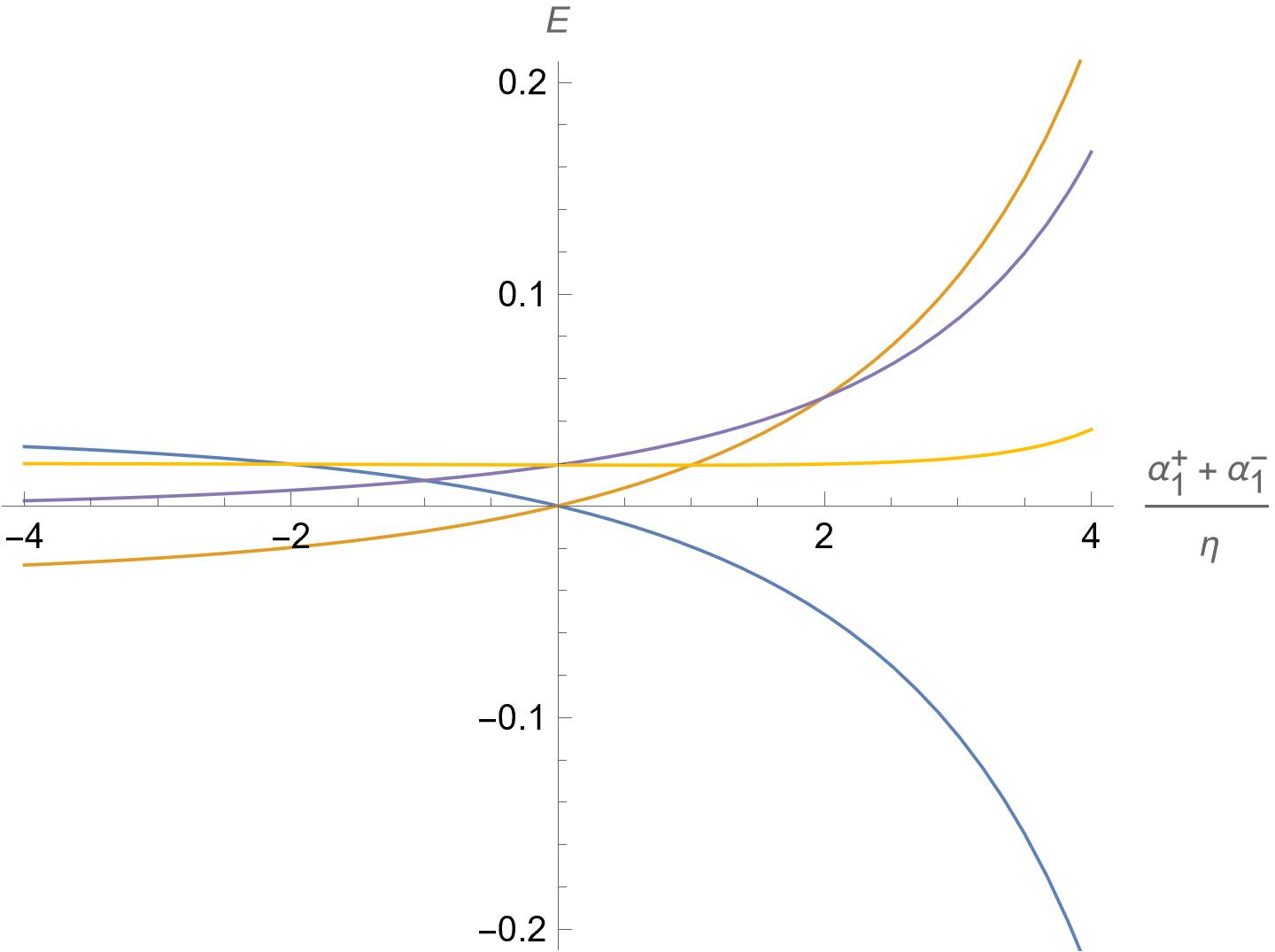}
\caption{$\alpha_1^-=-1.27$, $\eta=0.2$ and $N=3$}
\end{subfigure}
~~~
\begin{subfigure}{0.45\textwidth}
\includegraphics[width=\textwidth]{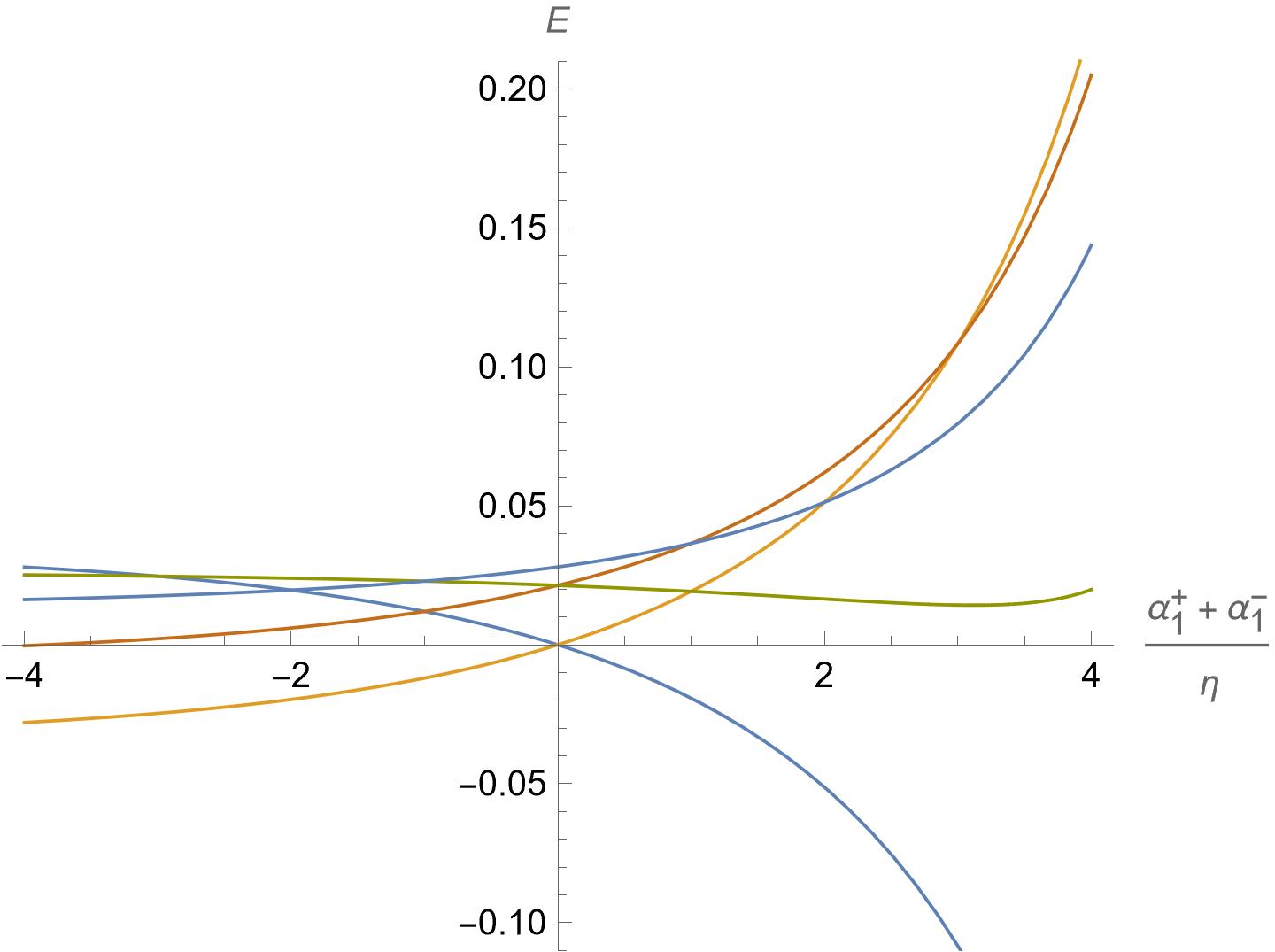}
\caption{$\alpha_1^-=-1.27$, $\eta=0.2$ and $N=4$}
\end{subfigure}

\caption{Parts of the spectral curves of the XXZ chain with diagonal boundary fields given in \eqref{h1N:d}. Spectral degeneracy occurs at $(\alpha_1^-+\alpha_1^+)/\eta=(N-1),(N-2),\ldots,(1-N)$. }\label{Fig:degeneracy}

\end{figure}}

\section{Summary and Outlook}\label{sec:summary:outlook}
\setcounter{equation}{0}

We have investigated the spin-$\frac{1}{2}$ XXZ model with specific non-diagonal boundary magnetic fields, as given in Eq. \eqref{special_boundaries}. The exact solution of this model is parameterized by two sets of conventional Bethe Ansatz equations (Eqs. \eqref{BAE_G_M^+_x} and \eqref{BAE_G_M^-_x}), corresponding to the invariant subspaces $G_M^+$ and $G_M^-$, respectively.

A central aspect of our analysis is the degenerate structure of the BAEs for $G_M^+$ \eqref{BAE_G_M^+_x}, which arises when the numerator $F(u)$ and denominator $F(-u)$ of the boundary term share common zeros. This naturally leads to a classification scheme for the solution of the BAEs according to whether the Bethe roots are located at these common zeros. Both the partially and fully degenerate points are systematically examined. At the partially degenerate points, the BAEs \eqref{BAE_G_M^+_x} decouple into two independent types, all Bethe roots remain finite, and the resulting energy levels are non-degenerate. In contrast, the fully degenerate points exhibit a richer structure: the Bethe roots admit a finer classification, and the BAEs admit solutions with some roots forming phantom strings, which in turn give rise to doubly degenerate energy levels and establish a direct connection between the solutions of the $G_M^+$ and $G_M^-$ sectors. We propose counting formulas for the various types of BAEs that emerge. All these formulas are linear combinations of binomial coefficients that depend on the system length and the number of Bethe roots. {The classification of Bethe roots at degenerate points is also reflected
in the Bethe states, leading to different behaviors of the wave function, as demonstrated in Section \ref{sec:BetheState}.}

{We have also extended the classification of Bethe roots and the associated spectral properties to the diagonal limit and to quasi-degenerate points.
In the diagonal limit, only two free boundary parameters remain. Consequently, there are only the fully degenerate points, and spectral degeneracy occurs. At certain quasi-degenerate points, the boundary-pinned Bethe roots can be generalized to the boundary-induced strings, which are a type of singular solution of the BAEs. Twofold degenerate energy levels also exist at specific quasi-degenerate points, and can be understood in terms of phantom strings.}

{Even for the BAEs in the generic case, the Bethe roots still admit a classification when $N$ is large. Consider, for example, the BAEs in \eqref{BAE_G_M^+_x}. For a large $N$, most roots are distributed along the curve $\left|\frac{\sinh(x_j+\frac{\eta}{2})}{\sinh(x_j-\frac{\eta}{2})}\right|=1$,
whereas in the region
\begin{align}
\left|\frac{\sinh(\alpha_k^\sigma+\eta)}{\sinh(\alpha_k^\sigma)}\right|>1,
\end{align}
one root may approach $\alpha_k^\sigma+\frac{\eta}{2}$, corresponding to
the boundary bound state \cite{Skorik:1995,Kapustin:1996,Grijalva:2019}. These observations allow the Bethe roots to be
classified systematically in terms of how many roots lie off the curve
$\left|\frac{\sinh(x_j+\frac{\eta}{2})}{\sinh(x_j-\frac{\eta}{2})}\right|=1.
$ Moreover, one can also observe boundary strings in the non-degenerate points. When $N$ is large, in certain solutions, the Bethe roots approximately form a string \cite{Skorik:1995,Kapustin:1996}
\begin{align}
x_{l}^{(m)}=\alpha_k^\sigma+\tfrac{\eta}{2}+l\eta,\quad
l=0,\ldots,m,\quad
\text{when}\quad
\left|\frac{\sinh(\alpha_k^\sigma+\eta+m\eta)}
{\sinh(\alpha_k^\sigma+m\eta)}\right|>1.
\end{align}
In open systems, these boundary string solutions play a critical role in the study of phase transitions in the model, the double degeneracy of the spectrum,
and the calculation of physical quantities; see Ref. \cite{Grijalva:2019} for an explicit example.

Although the boundary strings also exist in certain regimes at non-degenerate points, they can only be observed when $N$ is large enough or in
the thermodynamic limit. The farther the parameters are from the degenerate points, the larger the system size required to observe the quasi-boundary strings. By contrast, the boundary-pinned roots (strings) at the
(quasi-)degenerate points studied in this work exist for arbitrary system lengths.}

Beyond the XXZ models considered in this paper, our results admit several generalizations. For open integrable models with conventional BAEs, a classification of Bethe roots emerges naturally at the degenerate points and can be derived analytically from the BAEs themselves. In particular, our findings can be directly extended to the open XYZ model with specific boundary fields, whose Hamiltonian and BAEs are parameterized in terms of elliptic functions \cite{Yang2006,Zhang:2022}. The same classification
scheme is expected to remain applicable to open Heisenberg models with higher spins and ranks.

\section*{Acknowledgment}
We acknowledge financial support from the National Natural Science Foundation of China (Grant Nos. 12434006, 12474287, 12547107, 12575007, T2121001) and the National Key R\& D Program of China (Grant No. 2021YFA1402104). X. Zhang thanks Y. Miao and Y. Jiang for useful discussions.

\appendix

\section{Exact solutions of the XXZ model with generic open boundaries}\label{App:A}

\setcounter{equation}{0}

The $R$-matrix and $K^\pm$-matrices are given by \cite{Vega:1993,Ghoshal:1994,Korepin}
\begin{align} 
&&R_{0,n}(u) =\left(\begin{array}{cccc}
	\sinh(u+\eta) & 0 & 0 & 0 \\[2pt]
	 0 & \sinh u &\sinh\eta & 0 \\[2pt]
	 0 & \sinh\eta & \sinh u & 0\\[2pt]
	 0 & 0 & 0 & \sinh(u+\eta) 
\end{array}\right),\label{R-matrix}
\end{align}
\begin{align} 
&K^-(u) =\left(\begin{array}{cc}
K^-_{11}(u) & K^-_{12}(u) \\[2pt]
K^-_{21}(u) & K^-_{22}(u)
\end{array}\right),\\[4pt]
&K^-_{11}(u)=2\i \left[\sinh(\alpha_1^-)\sinh(\alpha_2^-)\cosh(u)+\cosh(\alpha_1^-)\cosh(\alpha_2^-)\sinh(u)\right],\no\\[4pt]
&K^-_{22}(u)=2\i \left[\sinh(\alpha_1^-)\sinh(\alpha_2^-)\cosh(u)-\cosh(\alpha_1^-)\cosh(\alpha_2^-)\sinh(u)\right],\no\\[4pt]
&K^-_{12}(u)=\e^{\theta_-}\sinh(2u),\qquad K^-_{21}(u)=\e^{-\theta_-}\sinh(2u),\\[2pt]
&K^+(u)=\left.K^-(-u-\eta)\right|_{(\alpha_1^-,\,\alpha_2^-,\,\theta_-)\rightarrow (-\alpha_1^+,\,-\alpha_2^+ -\i\pi,\,\theta_+)}.
\end{align}
The $R$-matrix (\ref{R-matrix}) satisfies the Yang-Baxter equation (YBE) \cite{BaxterBook,Korepin}
\begin{align}
R_{1,2}(u-v)R_{1,3}(u)R_{2,3}(v)=R_{2,3}(v)R_{1,3}(u)R_{1,2}(u-v),\label{R-YBE}
\end{align}
and the $K^\pm$-matrices satisfy the following reflection equation (RE) and the dual reflection equation (dual-RE) \cite{Sklyanin:1988,Cherednik:1984}, respectively
\begin{align}
&R_{1,2}(u-v)K_{1}^{-}(u)R_{2,1}(u+v)K_{2}^{-}(v)\no\\
&=K_{2}^{-}(v)R_{1,2}(u+v)K_{1}^{-}(u)R_{2,1}(u-v),\label{RE}\\
&R_{1,2}(-u+v)K_{1}^{+}(u)R_{2,1}(-u-v-2\eta)K_{2}^{+}(v)\no\\
&=K_{2}^{+}(v)R_{1,2}(-u-v-2\eta)K_{1}^{+}(u)R_{2,1}(-u+v).\label{dual-RE}
\end{align}
The transfer matrix is given by
\begin{align}  
t(u)=&\,{\rm tr}_0\left\{K_0^+(u-\tfrac{\eta}{2})\,R_{0,N}(u-\tfrac{\eta}{2})\cdots R_{0,1}(u-\tfrac{\eta}{2})\right.\no\\
&\,\times\left. K_0^-(u-\tfrac{\eta}{2})R_{1,0}(u-\tfrac{\eta}{2})\cdots R_{N,0}(u-\tfrac{\eta}{2})\right\}.
\label{transfer matrix}
\end{align}
The first order derivative of the logarithm of the transfer matrix (\ref{transfer matrix}) yields the Hamiltonian (\ref{Ham_1})
\begin{align} 
H = \sinh\eta\left.\frac{\partial \ln t(u)}{\partial u}\right|_{u=\frac{\eta}{2}}-\left[(2N-1)\cosh\eta -\tanh\eta\sinh\eta\right]\mathbb{I}.
\end{align}
\paragraph{Inhomogeneous $T$-$Q$ relation}
Define the following functions 
\begin{align} 
&a(u)=4\frac{\sinh(2u+\eta)}{\sinh(2u)}\prod_{k=1}^{2}\prod_{\sigma=\pm}\sinh(u-\alpha_k^\sigma-\tfrac{\eta}{2})\sinh^{2N}(u+\tfrac{\eta}{2}),\label{def;a}\\
&\tilde a(u)=4\frac{\sinh(2u+\eta)}{\sinh(2u)}\prod_{k=1}^{2}\prod_{\sigma=\pm}\sinh(u+\alpha_k^\sigma-\tfrac{\eta}{2})\sinh^{2N}(u+\tfrac{\eta}{2}),\label{def;tilde:a}
\end{align}
The eigenvalues of $t(u)$, denoted as $\Lambda(u)$, can be parameterized by the
following inhomogeneous $T$-$Q$ relation \cite{OffDiagonal}
\begin{align}
\Lambda(u)=&\,a(u)\frac{Q(u-\eta)}{Q(u)}+a(-u)\frac{Q(u+\eta)}{Q(u)}\no\\&
+\,2c\,\frac{\sinh^{2N}(u-\frac{\eta}{2})\sinh^{2N}(u+\frac{\eta}{2})}{Q(u)}\sinh(2u-\eta)\sinh(2u+\eta),\label{TQ}
\end{align}
where 
\begin{align} 
c=-\cosh\left[\alpha_1^- +\alpha_2^- +\alpha_1^+ +\alpha_2^+ +(N+1)\eta\right]-\cosh(\theta_- -\theta_+).
\end{align}
The $Q$ function in (\ref{TQ}) is defined by 
\begin{align} 
Q(u)=\prod_{j=1}^{N}\sinh(u-\nu_j)\sinh(u+
\nu_j).
\end{align}
The Bethe roots $\{\nu_1,\ldots,\nu_N\}$ should satisfy the following Bethe Ansatz equations
\begin{align}
&a(\nu_j)Q(\nu_j-\eta)+a(-\nu_j)Q(\nu_j+\eta)+2c\sinh^{2N}(\nu_j-\tfrac{\eta}{2})\sinh^{2N}(\nu_j+\tfrac{\eta}{2})\no\\
&\times \sinh(2\nu_j-\eta)\sinh(2\nu_j+\eta)=0,\qquad j=1,\ldots,N.
\end{align} 
The energy is given by 
\begin{align}
E(\bm{\nu})=\sum_{j=1}^{N} \varepsilon(\nu_j)+E_0.
\end{align}

\paragraph{Homogeneous $T$-$Q$ relations under the  constraint \eqref{special_boundaries}}
Under the constraint \eqref{special_boundaries}, the $T$-$Q$ relation can also be written in the following homogeneous form \cite{OffDiagonal,Nepomechie:2003_1,Cao2003}
\begin{align}
\Lambda_+(u)=&\,a(u)\prod_{j=1}^M\frac{\sinh(u-x_j-\eta)\sinh(u+x_j-\eta)}{\sinh(u-x_j)\sinh(u+x_j)}\no\\
&\,+a(-u)\prod_{j=1}^M\frac{\sinh(u-x_j+\eta)\sinh(u+x_j+\eta)}{\sinh(u-x_j)\sinh(u+x_j)},\\
\Lambda_-(u)=&\,\tilde a(u)\prod_{j=1}^{M'}\frac{\sinh(u-z_j-\eta)\sinh(u+z_j-\eta)}{\sinh(u-z_j)\sinh(u+z_j)}\no\\
&\,+\tilde a(-u)\prod_{j=1}^{M'}\frac{\sinh(u-z_j+\eta)\sinh(u+z_j+\eta)}{\sinh(u-z_j)\sinh(u+z_j)},
\end{align}
where $\{x_1,\dots,x_M\}$ and  $\{z_1,\dots,z_{M'}\}$ satisfy BAEs \eqref{BAE_G_M^+_x} and \eqref{BAE_G_M^-_x}, respectively.

\section{Hermiticity condition at the fully degenerate points}\label{App:B}
\setcounter{equation}{0}
In the easy-plane regime $\left | \cosh\eta \right | < 1$ (i.e., ${\rm Re}(\eta)=0$), we find the following Hermiticity conditions for the boundary magnetic fields in \eqref{constraint:case1}
\begin{align}
\alpha_1^-=\frac{\i\pi}{2}+\i\pi k_1,\quad \alpha_1^+=\frac{\i\pi}{2}+\i\pi k_2,\quad \i \theta_\pm,\i\eta\in \mathbb{R}.
\end{align}
Consequently, the boundary fields read
\begin{align}
&\vec{h}_1=\tan\gamma\, (\cos\vartheta_-,\, -\sin\vartheta_-,\, 0),\label{h_1_i_Hermiticity}\\[4pt]
&\vec{h}_N=\tan\gamma\,(\cos\vartheta_+,\, -\sin\vartheta_+,\, 0),\label{h_N_i_Hermiticity}
\end{align}
where $\gamma=-\i \eta,\vartheta_\pm=-\i\theta_\pm$ and 
\begin{align}
(N-2M+1)\gamma+\pi=\pm (\vartheta_--\vartheta_+) \mod{2\pi}. \label{constraint:hermitian}
\end{align}
We see that the two boundary magnetic fields in \eqref{h_1_i_Hermiticity} and \eqref{h_N_i_Hermiticity} both lie in the $xy$-plane and have a fixed azimuthal angle difference. From Eq.~\eqref{constraint:hermitian}, the anisotropic parameter $\gamma$ takes the following discrete values
$$\gamma_{M,k}^\pm=\frac{\pm(\vartheta_--\vartheta_+)+(2k-1)\pi}{N+1-2M}.$$
For large $N$, these discrete points are densely distributed on the real axis.

In the easy-axis regime $\left | \cosh\eta \right |> 1$ (i.e., ${\rm Im}(\eta)=0,\pi$), we find the following Hermiticity conditions for the boundary magnetic fields in \eqref{constraint:case2}
\begin{align}
\alpha_1^-,\alpha_2^--\frac{\i\pi}{2},\i\theta_-,\i\theta_+\in \mathbb{R},\quad M=\frac{N+1}{2}\in \mathbb{N}.
\end{align}
This leads to the following boundary fields
\begin{align}
&\vec{h}_1=\frac{\sinh\eta}{\sinh\alpha_1^- \cosh\bar\alpha_2^-}\left(- \cos\vartheta_-,\,\sin\vartheta_-,\,\cosh\alpha_1^- \sin\bar\alpha_2^-\right),\label{h_1_ii_Hermiticity}\\[4pt]
&\vec{h}_N=\frac{\sinh\eta}{\sinh(\alpha_1^- +\eta) \cosh(\bar\alpha_2^- +\eta)}\left(\cos\vartheta_-,\,-\sin\vartheta_-,\, -\cosh(\alpha_1^- +\eta) \sinh(\bar\alpha_2^- +\eta)\right),\label{h_N_ii_Hermiticity}
\end{align}
where $\bar\alpha_2^-=\alpha_2^--\frac{\i\pi}{2}$ and $N$ should be an odd number. The two boundary magnetic fields in \eqref{h_1_ii_Hermiticity} and \eqref{h_N_ii_Hermiticity} also lie in the same plane (with the same polarization). Unlike the easy-plane case, there is no adjustable integer $M$, which can be easily understood since trigonometric functions are singly periodic.

\section{Numerical results}
\setcounter{table}{0}

\subsection{Fully degenerate points}
\label{App:data:1}

The numerical result in this section is for the fully degenerate point or points sufficiently close to the fully degenerate point. Here, the labels I, II, III, and $G_M^-$ in the tables correspond to the BAEs \eqref{iBAE_I_x}, \eqref{iBAE_II_x}, \eqref{iBAE_III_x}, and \eqref{BAE_G_M^-_x}, respectively, while the symbol $/$ denotes the empty set.

\begin{table}[H]
\centering
\caption{Numerical solutions of BAEs \eqref{iBAE_I_x}, \eqref{iBAE_II_x}, \eqref{iBAE_III_x} and \eqref{BAE_G_M^-_x} for $N=4$, $M=2$, $\eta=0.6$, $\alpha_1^-=0.3$, $\alpha_1^+=0.8$, $\alpha_2^-=-\eta-\alpha_1^-$,  $\alpha_2^+=-\eta-\alpha_1^+$, $\theta_+=0.7$ and $\theta_-=(N-1-2M)\eta-(\alpha_1^- +\alpha_2^- +\alpha_1^+ +\alpha_2^+) -\i \pi +\theta_+$.}
\label{Numerical_4_2_x_real}
\begin{tabular}{|c|c|c|c|}
\hline
Class          & $x_1\,\,\mbox{or}\,\,z_1$     & $x_2\,\,\mbox{or}\,\,z_2$     & $E$ \\ 
\hline
$\rm {I}$      & 0.00000$+$0.38057\i           & 0.00000$+$0.13633\i           & $-$12.33982\\ 
\hline
$\rm {I}$      & 0.30357$+$0.39147\i           & 0.30357$-$0.39147\i           & $-$4.95840\\ 
\hline
$\rm {II}$     & 0.00000$+$0.06063\i           & $\alpha_1^- +\frac{\eta}{2}$  & $-$7.35197\\ 
\hline
$\rm {II}$     & 0.00000$+$0.20487\i           & $\alpha_1^- +\frac{\eta}{2}$  & $-$4.98722\\ 
\hline
$\rm {II}$     & 0.00000$+$0.47006\i           & $\alpha_1^- +\frac{\eta}{2}$  & $-$1.66437\\ 
\hline
$\rm {II}$     & 1.21686$+$0.00000\i           & $\alpha_1^- +\frac{\eta}{2}$  & \phantom{+}1.41262\\ 
\hline
$\rm {II}$     & 0.00000$+$0.11498\i           & $\alpha_1^+ +\frac{\eta}{2}$  & $-$8.71225\\ 
\hline
$\rm {II}$     & 0.00000$+$0.26711\i           & $\alpha_1^+ +\frac{\eta}{2}$  & $-$6.04872\\ 
\hline
$\rm {II}$     & 0.00000$+$0.55898\i           & $\alpha_1^+ +\frac{\eta}{2}$  & $-$3.22466\\ 
\hline
$\rm {II}$     & 1.73103$+$0.00000\i           & $\alpha_1^+ +\frac{\eta}{2}$  & $-$0.94723\\ 
\hline
$\rm {III}$    & $\alpha_1^- +\frac{\eta}{2}$  & $\alpha_1^+ +\frac{\eta}{2}$  & \phantom{+}1.53632\\ 
\hline
$G_M^-$        & 0.00000$+$0.05813\i           & $/$                           & $-$6.89855\\ 
\hline
$G_M^-$        & 0.00000$+$0.19410\i           & $/$                           & $-$4.70248\\ 
\hline
$G_M^-$        & 0.00000$+$0.42908\i           & $/$                           & $-$1.51336\\ 
\hline
$G_M^-$        & $\infty$                      & $/$                           & \phantom{+}1.53632\\ 
\hline
$G_M^-$        & 1.14614$+$0.00000\i           & $/$                           & \phantom{+}1.96144\\ 
\hline

\end{tabular}
\end{table}

\begin{table}[H]
\centering
\caption{Numerical solutions of BAEs \eqref{iBAE_I_x}, \eqref{iBAE_II_x}, \eqref{iBAE_III_x} and \eqref{BAE_G_M^-_x} for $N=4$, $M=3$, $\eta=0.4$, $\alpha_1^-=0.9$, $\alpha_2^-=0.7$, $\alpha_1^+=-\eta-\alpha_1^-$, $\alpha_2^+=-\eta-\alpha_2^-$, $\theta_+=0.1$ and $\theta_-=(N-1-2M)\eta-(\alpha_1^- +\alpha_2^- +\alpha_1^+ +\alpha_2^+) -\i \pi +\theta_+$.}
\label{Numerical_4_3_x_real}
\begin{tabular}{|c|c|c|c|c|}
\hline
Class          & $x_1\,\,\mbox{or}\,\,z_1$  & $x_2\,\,\mbox{or}\,\,z_2$     & $x_3\,\,\mbox{or}\,\,z_3$    & $E$ \\ 
\hline
$\rm {I}$      & 0.00000$+$0.26103\i   & 0.00000$+$0.09171\i           & $\infty$                      & $-$10.30991\\ 
\hline
$\rm {I}$      & 0.20359$+$0.27427\i   & 0.20359$-$0.27427\i           & $\infty$                      & $-$3.18955\\ 
\hline
$\rm {II}$     & 0.00000$+$0.08032\i   & $\infty$                      & $\alpha_1^- +\frac{\eta}{2}$  & $-$7.25320\\ 
\hline
$\rm {II}$     & 0.00000$+$0.19023\i   & $\infty$                      & $\alpha_1^- +\frac{\eta}{2}$  & $-$4.49286\\ 
\hline
$\rm {II}$     & 0.00000$+$0.42119\i   & $\infty$                      & $\alpha_1^- +\frac{\eta}{2}$  & $-$1.69445\\ 
\hline
$\rm {II}$     & 1.56624$+$0.00000\i   & $\infty$                      & $\alpha_1^- +\frac{\eta}{2}$  & $-$0.00489\\ 
\hline
$\rm {II}$     & 0.00000$+$0.07933\i   & $\infty$                      & $\alpha_2^- +\frac{\eta}{2}$  & $-$7.13784\\ 
\hline
$\rm {II}$     & 0.00000$+$0.18684\i   & $\infty$                      & $\alpha_2^- +\frac{\eta}{2}$  & $-$4.42685\\ 
\hline
$\rm {II}$     & 0.00000$+$0.40778\i   & $\infty$                      & $\alpha_2^- +\frac{\eta}{2}$  & $-$1.63618\\ 
\hline
$\rm {II}$     & 1.35569$+$0.00000\i   & $\infty$                      & $\alpha_2^- +\frac{\eta}{2}$  & \phantom{+}0.17394\\ 
\hline
$\rm {III}$    & 0.00000$+$0.07815\i   & $\alpha_1^- +\frac{\eta}{2}$  & $\alpha_2^- +\frac{\eta}{2}$  & $-$6.97285\\ 
\hline
$\rm {III}$    & 0.00000$+$0.18268\i   & $\alpha_1^- +\frac{\eta}{2}$  & $\alpha_2^- +\frac{\eta}{2}$  & $-$4.32524\\ 
\hline
$\rm {III}$    & 0.00000$+$0.38860\i   & $\alpha_1^- +\frac{\eta}{2}$  & $\alpha_2^- +\frac{\eta}{2}$  & $-$1.56962\\ 
\hline
$\rm {III}$    & 1.18057$+$0.00000\i   & $\alpha_1^- +\frac{\eta}{2}$  & $\alpha_2^- +\frac{\eta}{2}$  & \phantom{+}0.42143\\ 
\hline
$\rm {III}$    & $\infty$              & $\alpha_1^- +\frac{\eta}{2}$  & $\alpha_2^- +\frac{\eta}{2}$  & \phantom{+}0.26329\\ 
\hline
$G_M^-$        & $/$                   & $/$                           & $/$                           & \phantom{+}0.26329\\ 
\hline
\end{tabular}
\end{table}

\begin{table}[H]
\centering
\caption{Numerical solutions of BAEs \eqref{iBAE_I_x}, \eqref{iBAE_II_x}, \eqref{iBAE_III_x} and \eqref{BAE_G_M^-_x} for $N=4$, $M=2$, $\eta=0.5\rm{i}$, $\alpha_1^-=0.4\rm{i}$, $\alpha_1^+=0.7\rm{i}$, $\alpha_2^-=-\eta-\alpha_1^-$,  $\alpha_2^+=-\eta-\alpha_1^+$, $\theta_+=0.9\rm{i}$ and $\theta_-=(N-1-2M)\eta-(\alpha_1^- +\alpha_2^- +\alpha_1^+ +\alpha_2^+) -\i \pi +\theta_+$.}
\label{Numerical_4_2_x}
\begin{tabular}{|c|c|c|c|}
\hline
Class          & $x_1\,\,\mbox{or}\,\,z_1$     & $x_2\,\,\mbox{or}\,\,z_2$     & $E$ \\ 
\hline
$\rm {I}$      & 0.34762$+$0.00000\i           & 0.11683$+$0.00000\i           & $-$9.73055\\ 
\hline
$\rm {I}$      & 0.38649$+$0.26058\i           & 0.38649$-$0.26058\i           & $-$3.07304\\ 
\hline
$\rm {II}$     & 0.08818$+$0.00000\i           & $\alpha_1^- +\frac{\eta}{2}$  & $-$6.29107\\ 
\hline
$\rm {II}$     & 0.20932$+$0.00000\i           & $\alpha_1^- +\frac{\eta}{2}$  & $-$3.97981\\ 
\hline
$\rm {II}$     & 0.47521$+$0.00000\i           & $\alpha_1^- +\frac{\eta}{2}$  & $-$1.13872\\ 
\hline
$\rm {II}$     & 0.10737$+$1.57080\i           & $\alpha_1^- +\frac{\eta}{2}$  & \phantom{+}0.85440\\ 
\hline
$\rm {II}$     & 0.09906$+$0.00000\i           & $\alpha_1^+ +\frac{\eta}{2}$  & $-$6.84033\\ 
\hline
$\rm {II}$     & 0.23424$+$0.00000\i           & $\alpha_1^+ +\frac{\eta}{2}$  & $-$4.29681\\ 
\hline
$\rm {II}$     & 0.53063$+$0.00000\i           & $\alpha_1^+ +\frac{\eta}{2}$  & $-$1.61239\\ 
\hline
$\rm {II}$     & 0.60303$+$1.57080\i           & $\alpha_1^+ +\frac{\eta}{2}$  & $-$0.02985\\ 
\hline
$\rm {III}$    & $\alpha_1^- +\frac{\eta}{2}$  & $\alpha_1^+ +\frac{\eta}{2}$  & \phantom{+}1.13630\\ 
\hline
$G_M^-$        & 0.08325$+$0.00000\i           & $/$                           & $-$5.60848\\ 
\hline
$G_M^-$        & 0.19233$+$0.00000\i           & $/$                           & $-$3.52326\\ 
\hline
$G_M^-$        & 0.39471$+$0.00000\i           & $/$                           & $-$0.90441\\ 
\hline
$G_M^-$        & 0.00000$-$1.07999\i           & $/$                           & \phantom{+}1.77777\\ 
\hline
$G_M^-$        & $\infty$                      & $/$                           & \phantom{+}1.13630\\ 
\hline

\end{tabular}
\end{table}

\begin{table}[H]
\centering
\caption{Numerical solutions of BAEs \eqref{iBAE_I_x}, \eqref{iBAE_II_x}, \eqref{iBAE_III_x} and \eqref{BAE_G_M^-_x} for $N=4$, $M=3$, $\eta=0.5\rm{i}$, $\alpha_1^-=0.4\rm{i}$, $\alpha_1^+=0.7\rm{i}$, $\alpha_2^-=-\eta-\alpha_1^-$,  $\alpha_2^+=-\eta-\alpha_1^+$, $\theta_+=0.9\rm{i}$ and $\theta_-=(N-1-2M)\eta-(\alpha_1^- +\alpha_2^- +\alpha_1^+ +\alpha_2^+) -\i \pi +\theta_+$.}
\label{Numerical_4_3_x}
\begin{tabular}{|c|c|c|c|c|}
\hline
Class          & $x_1\,\,\mbox{or}\,\,z_1$  & $x_2\,\,\mbox{or}\,\,z_2$     & $x_3\,\,\mbox{or}\,\,z_3$    & $E$ \\ 
\hline
$\rm {I}$      & 0.11683$+$0.00000\rm {i}   & 0.34762$+$0.00000\rm {i}      & $\infty$                     & $-$9.73055\\ 
\hline
$\rm {I}$      & 0.38649$+$0.26058\rm {i}   & 0.38649$-$0.26058\rm {i}      & $\infty$                     & $-$3.07304\\ 
\hline
$\rm {II}$     & 0.08818$+$0.00000\rm {i}   & $\infty$                      & $\alpha_1^- +\frac{\eta}{2}$ & $-$6.29107\\ 
\hline
$\rm {II}$     & 0.20932$+$0.00000\rm {i}   & $\infty$                      & $\alpha_1^- +\frac{\eta}{2}$ & $-$3.97981\\ 
\hline
$\rm {II}$     & 0.47521$+$0.00000\rm {i}   & $\infty$                      & $\alpha_1^- +\frac{\eta}{2}$ & $-$1.13872\\ 
\hline
$\rm {II}$     & 0.10737$+$1.57080\rm {i}   & $\infty$                      & $\alpha_1^- +\frac{\eta}{2}$ & \phantom{+}0.85440\\ 
\hline
$\rm {II}$     & 0.09906$+$0.00000\rm {i}   & $\infty$                      & $\alpha_1^+ +\frac{\eta}{2}$ & $-$6.84033\\ 
\hline
$\rm {II}$     & 0.23424$+$0.00000\rm {i}   & $\infty$                      & $\alpha_1^+ +\frac{\eta}{2}$ & $-$4.29681\\ 
\hline
$\rm {II}$     & 0.53063$+$0.00000\rm {i}   & $\infty$                      & $\alpha_1^+ +\frac{\eta}{2}$ & $-$1.61239\\ 
\hline
$\rm {II}$     & 0.60303$+$1.57080\rm {i}   & $\infty$                      & $\alpha_1^+ +\frac{\eta}{2}$ & $-$0.02985\\ 
\hline
$\rm {III}$    & 0.08325$+$0.00000\rm {i}   & $\alpha_1^- +\frac{\eta}{2}$  & $\alpha_1^+ +\frac{\eta}{2}$ & $-$5.60848\\ 
\hline
$\rm {III}$    & 0.19233$+$0.00000\rm {i}   & $\alpha_1^- +\frac{\eta}{2}$  & $\alpha_1^+ +\frac{\eta}{2}$ & $-$3.52326\\ 
\hline
$\rm {III}$    & 0.39471$+$0.00000\rm {i}   & $\alpha_1^- +\frac{\eta}{2}$  & $\alpha_1^+ +\frac{\eta}{2}$ & $-$0.90441\\ 
\hline
$\rm {III}$    & 0.00000$-$1.07999\rm {i}   & $\alpha_1^- +\frac{\eta}{2}$  & $\alpha_1^+ +\frac{\eta}{2}$ & \phantom{+}1.77777\\ 
\hline
$\rm {III}$    & $\infty$                   & $\alpha_1^- +\frac{\eta}{2}$  & $\alpha_1^+ +\frac{\eta}{2}$ & \phantom{+}1.13630\\ 
\hline
$G_M^-$        & $/$                        & $/$                           & $/$                          & \phantom{+}1.13630\\ 
\hline
\end{tabular}
\end{table}

\begin{table}[H]
\centering
\caption{Numerical solutions of \eqref{BAE_G_M^+_x} and \eqref{BAE_G_M^-_x} for $N=4$, $M=3$, $\eta=0.5\rm{i}$, $\alpha_1^-=0.4\rm{i}$, $\alpha_1^+=0.7\rm{i}$, $\alpha_2^-=-\eta-\alpha_1^- +0.001\rm {i}$,  $\alpha_2^+=-\eta-\alpha_1^+ -0.001\rm {i}$, $\theta_+=0.9\rm{i}$ and $\theta_-=(N-1-2M)\eta-(\alpha_1^- +\alpha_2^- +\alpha_1^+ +\alpha_2^+) -\i \pi +\theta_+$.}
\label{Numerical_4_3_x_quasi}
\begin{tabular}{|c|c|c|c|}
\hline
 $x_1$    & $x_2$    & $x_3$   & $E$ \\ 
\hline
0.11682$-$0.00000\i          & 0.34758$-$0.00000\i         & 13.42289$+$0.41044\i         & $-$9.73078\\ 
\hline
0.38639$+$0.26057\i          & 0.38639$-$0.26057\i         & 11.25174$-$0.74301\i         & $-$3.07336\\ 
\hline
 0.08819$-$0.00000\i         & 13.30744$+$1.54006\i        & 0.00000$+$0.65012\i          & $-$6.29128\\ 
\hline
 0.20931$+$0.00000\i         & 13.22005$-$0.84494\i        & 0.00000$+$0.65005\i          & $-$3.97989\\ 
\hline
0.47517$-$0.00000\i          & 12.50729$+$0.76577\i        & 0.00000$+$0.65002\i          & $-$1.13879\\ 
\hline
0.10553$+$1.57080\i          & 13.34761$-$1.58207\i        & 0.00000$+$0.65000\i          & \phantom{+}0.85483\\ 
\hline
0.09906$+$0.00000\i          & 14.71906$+$1.03505\i        & 0.00000$+$0.94965\i          & $-$6.83983\\ 
\hline
0.23421$+$0.00000\i          & 14.41174$-$0.78581\i        & 0.00000$+$0.94971\i          & $-$4.29665\\ 
\hline
0.53056$-$0.00000\i          & 14.32962$+$0.80682\i        & 0.00000$+$0.94984\i          & $-$1.61227\\ 
\hline
0.60281$+$1.57080\i          & 18.80953$+$2.39651\i        & 0.00000$+$0.94998\i          & $-$0.02952\\ 
\hline
0.08325$+$0.00000\i          & 0.00000$+$0.64997\i         & 0.00000$+$0.95048\i          & $-$5.60871\\ 
\hline
0.19233$-$0.00000\i          & 0.00000$+$0.64999\i         & 0.00000$+$0.95045\i          & $-$3.52346\\ 
\hline
0.39471$+$0.00000\i          & 0.00000$+$0.64999\i         & 0.00000$+$0.95037\i          & $-$0.90460\\ 
\hline
 0.00000$+$1.08021\i         & 0.00000$+$0.65000\i         & 0.00000$+$0.94993\i          & \phantom{+}1.77792\\ 
\hline
 13.28263$-$1.42066\i        & 0.00000$+$0.65000\i         & 0.00000$+$0.95015\i          & \phantom{+}1.13635\\
\hline
\end{tabular}
\end{table}

\begin{table}[H]
\centering
\caption{Numerical solutions of BAEs \eqref{iBAE_I_x}, \eqref{iBAE_II_x}, \eqref{iBAE_III_x} and \eqref{BAE_G_M^-_x} for $N=5$, $M=3$, $\eta=0.5\rm{i}$, $\alpha_1^-=0.8\rm{i}$, $\alpha_1^+=0.7\rm{i}$, $\alpha_2^-=-\eta-\alpha_1^-$,  $\alpha_2^+=-\eta-\alpha_1^+$, $\theta_+=0.9\rm{i}$ and $\theta_-=(N-1-2M)\eta-(\alpha_1^- +\alpha_2^- +\alpha_1^+ +\alpha_2^+) -\i \pi +\theta_+$.}
\label{Numerical_5_3_x}
\begin{tabular}{|c|c|c|c|c|}
\hline
Class          & $x_1\,\,\mbox{or}\,\,z_1$       & $x_2\,\,\mbox{or}\,\,z_2$     & $x_3\,\,\mbox{or}\,\,z_3$       & $E$ \\ 
\hline
$\rm {II}$     & 0.20328$-$0.00000\i             & 0.08685$-$0.00000\i           & $\alpha_1^- +\frac{\eta}{2}$    & $-$11.19345\\ 
\hline
$\rm {II}$     & 0.43289$+$0.00000\i             & 0.08405$+$0.00000\i           & $\alpha_1^- +\frac{\eta}{2}$    & $-$8.54587\\ 
\hline
$\rm {II}$     & 0.07868$+$0.00000\i             & 0.96785$+$1.57080\i           & $\alpha_1^- +\frac{\eta}{2}$    & $-$6.66119\\ 
\hline
$\rm {II}$     & 0.42493$+$0.00000\i             & 0.19155$+$0.00000\i           & $\alpha_1^- +\frac{\eta}{2}$    & $-$6.54197\\ 
\hline
$\rm {II}$     & 0.17415$+$0.00000\i             & 0.95836$+$1.57080\i           & $\alpha_1^- +\frac{\eta}{2}$    & $-$4.84417\\ 
\hline
$\rm {II}$     & 0.20993$+$0.24998\i             & 0.20993$-$0.24998\i           & $\alpha_1^- +\frac{\eta}{2}$    & $-$2.99000\\ 
\hline
$\rm {II}$     & 0.32054$+$0.00000\i             & 0.92962$+$1.57080\i           & $\alpha_1^- +\frac{\eta}{2}$    & $-$2.57362\\ 
\hline
$\rm {II}$     & 0.46408$+$0.25817\i             & 0.46408$-$0.25817\i           & $\alpha_1^- +\frac{\eta}{2}$    & $-$1.72431\\ 
\hline
$\rm {II}$     & 0.64155$+$0.00000\i             & 0.81289$+$1.57080\i           & $\alpha_1^- +\frac{\eta}{2}$    & $-$0.65223\\ 
\hline
$\rm {II}$     & 0.47812$+$1.57080\i             & 1.39102$+$1.57080\i    & $\alpha_1^- +\frac{\eta}{2}$   & \phantom{+}0.43987\\ 
\hline
$\rm {II}$     & 0.20068$+$0.00000\i             & 0.08604$-$0.00000\i           & $\alpha_1^+ +\frac{\eta}{2}$    & $-$11.15377\\ 
\hline
$\rm {II}$     & 0.42406$-$0.00000\i             & 0.08336$-$0.00000\i           & $\alpha_1^+ +\frac{\eta}{2}$    & $-$8.51646\\ 
\hline
$\rm {II}$     & 0.07791$-$0.00000\i             & 0.90543$+$1.57080\i           & $\alpha_1^+ +\frac{\eta}{2}$    & $-$6.55239\\ 
\hline
$\rm {II}$     & 0.41644$+$0.00000\i             & 0.18955$-$0.00000\i           & $\alpha_1^+ +\frac{\eta}{2}$    & $-$6.53826\\ 
\hline
$\rm {II}$     & 0.17205$+$0.00000\i             & 0.89494$+$1.57080\i           & $\alpha_1^+ +\frac{\eta}{2}$    & $-$4.76310\\ 
\hline
$\rm {II}$     & 0.19585$+$0.24999\i             & 0.19585$-$0.24999\i           & $\alpha_1^+ +\frac{\eta}{2}$    & $-$2.95315\\ 
\hline
$\rm {II}$     & 0.31519$-$0.00000\i             & 0.86364$+$1.57080\i           & $\alpha_1^+ +\frac{\eta}{2}$    & $-$2.51140\\ 
\hline
$\rm {II}$     & 0.44718$+$0.25683\i             & 0.44718$-$0.25683\i           & $\alpha_1^+ +\frac{\eta}{2}$    & $-$1.69987\\ 
\hline
$\rm {II}$     & 0.62539$-$0.00000\i             & 0.74171$+$1.57080\i           & $\alpha_1^+ +\frac{\eta}{2}$    & $-$0.57037\\ 
\hline
$\rm {II}$     & 0.40113$+$1.57080\i             & 1.34962$+$1.57080\i   & $\alpha_1^+ +\frac{\eta}{2}$    & \phantom{+}0.57514\\ 
\hline
$\rm {III}$    & 0.07564$+$0.00000\i             & $\alpha_1^- +\frac{\eta}{2}$  & $\alpha_1^+ +\frac{\eta}{2}$    & $-$6.15187\\ 
\hline
$\rm {III}$    & 0.07564$+$0.00000\i             & $\alpha_1^- +\frac{\eta}{2}$  & $\alpha_1^+ +\frac{\eta}{2}$    & $-$4.45305\\ 
\hline
$\rm {III}$    & 0.29828$+$0.00000\i             & $\alpha_1^- +\frac{\eta}{2}$  & $\alpha_1^+ +\frac{\eta}{2}$    & $-$2.29215\\ 
\hline
$\rm {III}$    & 0.57216$-$0.00000\i             & $\alpha_1^- +\frac{\eta}{2}$  & $\alpha_1^+ +\frac{\eta}{2}$    & $-$0.36405\\ 
\hline
$\rm {III}$    & 1.22054$+$1.57080\i      & $\alpha_1^- +\frac{\eta}{2}$ & $\alpha_1^+ +\frac{\eta}{2}$    & \phantom{+}0.85330\\ 
\hline
$\rm {III}$    & 0.00000$-$1.36485\i      & $\alpha_1^- +\frac{\eta}{2}$ & $\alpha_1^+ +\frac{\eta}{2}$    & \phantom{+}1.22783\\ 
\hline
$G_M^-$        & 0.07564$+$0.00000\i             & $/$                           & $/$                             & $-$6.15187\\ 
\hline
$G_M^-$        & 0.16578$+$0.00000\i             & $/$                           & $/$                             & $-$4.45305\\ 
\hline
$G_M^-$        & 0.29828$+$0.00000\i             & $/$                           & $/$                             & $-$2.29215\\ 
\hline
$G_M^-$        & 0.57216$-$0.00000\i             & $/$                           & $/$                             & $-$0.36405\\ 
\hline
$G_M^-$        & 1.22054$+$1.57080\i      & $/$                           & $/$                            & \phantom{+}0.85330\\ 
\hline
$G_M^-$        & 0.00000$-$1.36485\i      & $/$                           & $/$                            & \phantom{+}1.22783\\ 
\hline
\end{tabular}
\end{table}

\begin{table}[H]
\centering
\caption{Numerical solutions of BAEs \eqref{BAE_G_M^+_x} and \eqref{BAE_G_M^-_x} for $N=5$, $M=2$, $\eta=0.5\rm{i}$, $\alpha_1^-=0.4\rm{i}$, $\alpha_1^+=0.7\rm{i}$, $\alpha_2^-=-\eta-\alpha_1^-+0.00017\i$,  $\alpha_2^+=-\eta-\alpha_1^+-0.0001\i$, $\theta_+=0.9\rm{i}$ and $\theta_-=(N-1-2M)\eta-(\alpha_1^- +\alpha_2^- +\alpha_1^+ +\alpha_2^+) -\i \pi +\theta_+$.}\label{quasi:degenerate:1}
\begin{tabular}{|c|c|r|}
\hline 
$x_1$ & $x_2$ & $E$~~~~ \\
\hline 
0.21919$+$0.00000\i & 0.09102$+$0.00000\i& $-$11.92712 \\
\hline
0.08707$+$0.00000\i & 0.52967$+$0.00000\i& $-$9.06275 \\
\hline
0.20163$+$0.00000\i & 0.51716$+$0.00000\i& $-$6.92565 \\
\hline
0.28195$-$0.24948\i & 0.28195$+$0.24948\i & $-$3.74087 \\
\hline
0.63697$-$0.29595\i & 0.63697$+$0.29595\i & $-$2.10785 \\
\hline 
0.07166$+$0.00000\i & 0.00000$+$0.65000\i & $-$6.55739 \\
\hline
0.15950$+$0.00000\i & 0.00000$+$0.65000\i & $-$4.92130 \\
\hline
0.29705$+$0.00000\i & 0.00000$+$0.65000\i & $-$2.65215 \\
\hline
0.64936$+$0.00000\i & 0.00000$+$0.65000\i & $-$0.47182 \\
\hline
0.00000$+$1.27829\i & 0.00000$+$0.65000\i & 0.90803 \\
\hline 
0.07886$+$0.00000\i & 0.00000$+$0.95000\i & $-$7.18695 \\
\hline
0.17495$+$0.00000\i & 0.00000$+$0.95000\i & $-$5.36022 \\
\hline
0.32535$+$0.00000\i & 0.00000$+$0.95000\i & $-$3.06122 \\
\hline
0.70783$+$0.00000\i & 0.00000$+$0.95000\i & $-$1.07600 \\
\hline
0.45849$+$1.57080\i & 0.00000$+$0.95000\i & 0.02426 \\
\hline 
0.00000$+$0.65000\i & 0.00000$+$0.95001\i & 1.13637 \\
\hline 
\end{tabular}
\hfill
\begin{tabular}{|c|c|r|}
\hline 
$z_1$ & $z_2$ & $E$~~~~ \\
\hline 
\hline
0.16857$+$0.00000\i & 0.07403$+$0.00000\i & $-$10.86967 \\
\hline
0.07260$+$0.00000\i & 0.32896$+$0.00000\i & $-$8.42900 \\
\hline
1.02268$+$0.00000\i & 0.06925$+$0.00000\i & $-$6.12879 \\
\hline
0.32506$+$0.00000\i & 0.16292$+$0.00000\i & $-$6.78219 \\
\hline
0.06599$+$0.00000\i & 0.00000$+$1.04258\i & $-$5.20335 \\
\hline
0.15276$+$0.00000\i & 1.01382$+$0.00000\i & $-$4.59729 \\
\hline
0.14359$+$0.00000\i & 0.00000$+$1.04533\i & $-$3.80300 \\
\hline
0.00000$+$0.34413\i & 0.00000$+$0.15587\i & $-$2.51442 \\
\hline
0.27872$+$0.00000\i & 0.98574$+$0.00000\i & $-$2.45622 \\
\hline
0.25246$+$0.00000\i & 0.00000$+$1.05147\i & $-$1.83943 \\
\hline
0.25920$-$0.25010\i & 0.25920$+$0.25010\i & $-$1.56311 \\
\hline
0.63894$-$0.18159\i & 0.63894$+$0.18159\i & $-$0.27625 \\
\hline
0.45084$+$0.00000\i & 0.00000$+$1.06458\i & 0.13958 \\
\hline
3.30193$+$0.00000\i & 3.28965$+$1.57080\i & 1.13627 \\
\hline
1.27129$+$0.00000\i & 0.00000$+$1.09384\i & 1.60102 \\
\hline
0.00000$+$1.15328\i & 0.00000$+$1.75612\i & 2.23791 \\
\hline 
\end{tabular}
\end{table}

\begin{table}[H]
\centering
\caption{Numerical solutions of BAEs \eqref{BAE_G_M^+_x} for $N=5$, $M=4$, $\eta=0.5\rm{i}$, $\alpha_1^-=0.4\rm{i}$, $\alpha_1^+=0.7\rm{i}$, $\alpha_2^-=-\eta-\alpha_1^-+0.00017\i$,  $\alpha_2^+=-\eta-\alpha_1^+-0.0001\i$, $\theta_+=0.9\rm{i}$ and $\theta_-=(N-1-2M)\eta-(\alpha_1^- +\alpha_2^- +\alpha_1^+ +\alpha_2^+) -\i \pi +\theta_+$.}\label{quasi:degenerate:2}
\begin{tabular}{|c|c|c|c|r|}
\hline
\(x_1\) & \(x_2\) & \(x_3\) & \(x_4\) & \(E\)~~~~ \\
\hline 
0.08959$+$0.00000\i & 0.00000$+$1.32779\i & 0.21383$+$0.00000\i & 0.90745$+$0.00000\i & $-$11.92717 \\
\hline
0.00000$+$1.93011\i & 0.08681$+$0.00000\i & 0.53968$-$0.19368\i & 0.53968$+$0.19368\i & $-$9.06277 \\
\hline
0.00000$+$1.19695\i & 0.20274$+$0.00000\i & 0.51036$-$0.20450\i & 0.51036$+$0.20450\i & $-$6.92563 \\
\hline
0.94659$+$0.00000\i & 0.00000$+$1.85569\i & 0.24554$-$0.24986\i & 0.24554$+$0.24986\i & $-$3.74090 \\
\hline
0.00000$+$1.07631\i & 0.36500$+$0.00000\i & 0.36932$-$0.49960\i & 0.36932$+$0.49960\i & $-$2.10786 \\
\hline
0.07166$+$0.00000\i & 2.90885$-$0.77820\i & 2.90885$+$0.77820\i & 0.00000$+$0.65000\i & $-$6.55748 \\
\hline
0.15950$+$0.00000\i & 2.89233$-$0.77815\i & 2.89233$+$0.77815\i & 0.00000$+$0.65000\i & $-$4.92138 \\
\hline
0.29705$+$0.00000\i & 2.83782$-$0.77801\i & 2.83782$+$0.77801\i & 0.00000$+$0.65000\i & $-$2.65221 \\
\hline
0.64935$+$0.00000\i & 2.59513$-$0.78014\i & 2.59513$+$0.78014\i & 0.00000$+$0.65000\i & $-$0.47186 \\
\hline
0.00000$+$1.27831\i & 3.39349$-$0.77946\i & 3.39349$+$0.77946\i & 0.00000$+$0.65000\i & 0.90798 \\
\hline 
0.07886$+$0.00000\i & 3.13257$-$0.77907\i & 3.13257$+$0.77907\i & 0.00000$+$0.94999\i & $-$7.18702 \\
\hline
0.17495$+$0.00000\i & 3.12058$-$0.77907\i & 3.12058$+$0.77907\i & 0.00000$+$0.94999\i & $-$5.36027 \\
\hline
0.32536$+$0.00000\i & 3.08162$-$0.77907\i & 3.08162$+$0.77907\i & 0.00000$+$0.94999\i & $-$3.06125 \\
\hline
0.70784$+$0.00000\i & 2.88047$-$0.78059\i & 2.88047$+$0.78059\i & 0.00000$+$0.94999\i & $-$1.07602 \\
\hline
0.45851$+$1.57080\i & 3.56211$-$0.77966\i & 3.56211$+$0.77966\i & 0.00000$+$0.95000\i & 0.02422 \\
\hline
0.07403$+$0.00000\i & 0.16856$+$0.00000\i & 0.00000$+$0.64998\i & 0.00000$+$0.95007\i & $-$10.86959 \\
\hline
0.32895$+$0.00000\i & 0.07260$+$0.00000\i & 0.00000$+$0.64999\i & 0.00000$+$0.95006\i & $-$8.42896 \\
\hline
0.06925$+$0.00000\i & 1.02257$+$0.00000\i & 0.00000$+$0.65000\i & 0.00000$+$0.95004\i & $-$6.12877 \\
\hline
0.32505$+$0.00000\i & 0.16292$+$0.00000\i & 0.00000$+$0.64999\i & 0.00000$+$0.95006\i & $-$6.78217 \\
\hline
0.06599$+$0.00000\i & 0.00000$+$1.04270\i & 0.00000$+$0.65000\i & 0.00000$+$0.94998\i & $-$5.20327 \\
\hline
1.01370$+$0.00000\i & 0.15276$+$0.00000\i & 0.00000$+$0.65000\i & 0.00000$+$0.95004\i & $-$4.59728 \\
\hline
0.00000$+$1.04545\i & 0.14359$+$0.00000\i & 0.00000$+$0.65000\i & 0.00000$+$0.94998\i & $-$3.80293 \\
\hline
0.00000$+$0.34422\i & 0.00000$+$0.15578\i & 0.00000$+$0.64992\i & 0.00000$+$0.95009\i & $-$2.51428 \\
\hline
0.98563$+$0.00000\i & 0.27871$+$0.00000\i & 0.00000$+$0.65000\i & 0.00000$+$0.95003\i & $-$2.45622 \\
\hline
0.25245$+$0.00000\i & 0.00000$+$1.05158\i & 0.00000$+$0.65000\i & 0.00000$+$0.94998\i & $-$1.83938 \\
\hline
0.25919$-$0.25010\i & 0.25919$+$0.25010\i & 0.00000$+$0.65000\i & 0.00000$+$0.95006\i & $-$1.56306 \\
\hline
0.63887$-$0.18164\i & 0.63887$+$0.18164\i & 0.00000$+$0.65000\i & 0.00000$+$0.95003\i & $-$0.27623 \\
\hline
0.45084$+$0.00000\i & 0.00000$+$1.06469\i & 0.00000$+$0.65000\i & 0.00000$+$0.94999\i & 0.13961 \\
\hline
1.27117$+$0.00000\i & 0.00000$+$1.09395\i & 0.00000$+$0.65000\i & 0.00000$+$0.95000\i & 1.60106 \\
\hline
3.29572$-$0.77926\i & 3.29572$+$0.77926\i & 0.00000$+$0.65000\i & 0.00000$+$0.95001\i & 1.13632 \\
\hline
0.00000$+$1.15345\i & 0.00000$+$1.38548\i & 0.00000$+$0.65000\i & 0.00000$+$0.95000\i & 2.23797 \\
\hline
\end{tabular}
\end{table}

\subsection{Partially degenerate points}
\label{App:data:2}

The numerical result in this section is for the partially degenerate point. 
Here, the labels I, II, and $G_M^-$ in the tables correspond to BAEs \eqref{iBAE_I_x:2}, \eqref{iBAE_I_x:3}, and \eqref{BAE_G_M^-_x}, respectively, while the symbol $/$ denotes the empty set.

\begin{table}[H]
\centering
\caption{Numerical solutions of BAEs \eqref{iBAE_I_x:2}, \eqref{iBAE_I_x:3} and \eqref{BAE_G_M^-_x} for $N=4$, $M=1$, $\eta=0.7$, $\alpha_1^+=0.82$, $\alpha_2^+=0.69$, $\theta_+=0.32$, $\alpha_1^-=0.28$, $\alpha_2^-=-\eta-\alpha_1^-$ and $\theta_-=(N-1-2M)\eta-(\alpha_1^- +\alpha_2^- +\alpha_1^+ +\alpha_2^+) -\i \pi +\theta_+$.}
\label{Numerical_4_1_x_pd_real}
\begin{tabular}{|c|c|c|c|}
\hline
Class & $x_1\,\,\mbox{or}\,\,z_1$              & $x_2\,\,\mbox{or}\,\,z_2$     & $E$ \\ 
\hline
$\rm {I}$      & 0.00000$+$0.15775\i           & $/$                           & $-$11.72313\\ 
\hline
$\rm {I}$      & 0.00000$+$0.40769\i           & $/$                           & $-$8.20590\\ 
\hline
$\rm {I}$      & 1.32804$+$0.00000\i           & $/$                           & $-$3.77450\\
\hline
$\rm {I}$      & 0.99476$+$0.00000\i           & $/$                           & $-$3.23241\\ 
\hline
$\rm {II}$     & $\alpha_1^- +\frac{\eta}{2}$  & $/$                           & $-$0.61995\\ 
\hline
$G_M^-$        & 0.06483$+$0.00000\i           & 0.00000$+$0.19607\i           & $-$12.11629\\ 
\hline
$G_M^-$        & 0.06535$+$0.00000\i           & 0.00000$+$0.45778\i           & $-$8.73307\\ 
\hline
$G_M^-$        & 0.00000$+$0.45161\i           & 0.00000$+$0.18747\i           & $-$6.54436\\ 
\hline
$G_M^-$        & 1.39469$-$0.00000\i           & 0.06718$+$0.00000\i           & $-$4.85436\\ 
\hline
$G_M^-$        & 0.76847$+$0.00000\i           & 0.07003$+$0.00000\i           & $-$3.26253\\ 
\hline
$G_M^-$        & 1.39106$+$0.00000\i           & 0.00000$+$0.15998\i           & $-$3.02367\\ 
\hline
$G_M^-$        & 0.35000$+$0.09396\i           & 0.35000$-$0.09396\i           & $-$0.78041\\ 
\hline
$G_M^-$        & 1.38284$+$0.00000\i           & 0.00000$+$0.33116\i           & $-$0.42606\\ 
\hline
$G_M^-$        & 0.35124$+$0.41940\i           & 0.35124$-$0.41940\i           & \phantom{+}0.29120\\ 
\hline
$G_M^-$        & 1.36673$+$0.00000\i           & 0.00000$+$0.63883\i           & \phantom{+}2.13808\\ 
\hline
$G_M^-$        & 2.13517$+$0.00000\i           & 1.33036$+$0.00000\i           & \phantom{+}4.61925\\ 
\hline
\end{tabular}
\end{table}

\begin{table}[H]
\centering
\caption{Numerical solutions of BAEs \eqref{iBAE_I_x:2}, \eqref{iBAE_I_x:3} and \eqref{BAE_G_M^-_x} for $N=4$, $M=2$, $\eta=0.34$, $\alpha_1^+=0.58$, $\alpha_2^+=0.23$, $\theta_+=0.75$, $\alpha_1^-=0.26$, $\alpha_2^-=-\eta-\alpha_1^-$ and $\theta_-=(N-1-2M)\eta-(\alpha_1^- +\alpha_2^- +\alpha_1^+ +\alpha_2^+) -\i \pi +\theta_+$.}
\label{Numerical_4_2_x_pd_real}
\begin{tabular}{|c|c|c|c|}
\hline
Class & $x_1\,\,\mbox{or}\,\,z_1$         & $x_2\,\,\mbox{or}\,\,z_2$     & $E$ \\ 
\hline
$\rm {I}$      & 0.00000$+$0.38831\i      & 0.00000$+$0.08458\i           & $-$10.92185\\ 
\hline
$\rm {I}$      & 0.30665$+$0.00000\i      & 0.02714$-$0.00000\i           & $-$7.81942\\ 
\hline
$\rm {I}$      & 0.37555$+$0.00000\i      & 0.00000$+$0.14763\i           & $-$5.61643\\
\hline
$\rm {I}$      & 0.52456$+$0.61305\i      & 0.52456$-$0.61305\i           & $-$3.21388\\ 
\hline
$\rm {I}$      & 0.39549$+$0.00000\i      & 0.00000$+$0.61404\i           & $-$1.80707\\ 
\hline
$\rm {I}$      & 0.52298$+$0.00000\i      & 0.40224$+$0.00000\i           & $-$0.32745\\ 
\hline
$\rm {II}$     & 0.00000$-$0.06351\i      & $\alpha_1^- +\frac{\eta}{2}$  & $-$8.71377\\ 
\hline
$\rm {II}$     & 0.00000$-$0.16007\i      & $\alpha_1^- +\frac{\eta}{2}$  & $-$5.88041\\ 
\hline
$\rm {II}$     & 0.00000$+$0.64633\i      & $\alpha_1^- +\frac{\eta}{2}$  & $-$2.09197\\ 
\hline
$\rm {II}$     & 0.53646$+$0.00000\i      & $\alpha_1^- +\frac{\eta}{2}$  & $-$0.64268\\ 
\hline
$\rm {II}$     & 0.40477$+$0.00000\i      & $\alpha_1^- +\frac{\eta}{2}$  & \phantom{+}0.19175\\ 
\hline
$G_M^-$        & 0.00000$+$0.04340\i      & $/$                           & $-$4.82030\\ 
\hline
$G_M^-$        & 0.00000$+$0.10026\i      & $/$                           & $-$3.21502\\ 
\hline
$G_M^-$        & 0.00000$+$0.19148\i      & $/$                           & $-$0.75948\\ 
\hline
$G_M^-$        & 0.00000$+$0.39904\i      & $/$                           & \phantom{+}1.58041\\ 
\hline
$G_M^-$        & 0.77510$+$0.00000\i      & $/$                           & \phantom{+}3.25633\\ 
\hline
\end{tabular}
\end{table}

\begin{table}[H]
\centering
\caption{Numerical solutions of BAEs \eqref{iBAE_I_x:2}, \eqref{iBAE_I_x:3} and \eqref{BAE_G_M^-_x} for $N=4$, $M=1$, $\eta=0.5\rm{i}$, $\alpha_1^-=0.4\rm{i}$, $\alpha_1^+=0.7\rm{i}$, $\theta_+=0.9\i$, $\alpha_2^-=-0.73\i$,  $\alpha_2^+=-\eta-\alpha_2^-$ and $\theta_-=(N-1-2M)\eta-(\alpha_1^- +\alpha_2^- +\alpha_1^+ +\alpha_2^+) -\i \pi +\theta_+$.}
\label{Numerical_4_1_x_pd}
\begin{tabular}{|c|c|c|c|}
\hline
Class & $x_1\,\,\mbox{or}\,\,z_1$              & $x_2\,\,\mbox{or}\,\,z_2$     & $E$ \\ 
\hline
$\rm {I}$      & 0.11502$+$0.00000\i           & $/$                           & $-$9.38561\\ 
\hline
$\rm {I}$      & 0.30428$-$0.00000\i           & $/$                           & $-$6.14886\\ 
\hline
$\rm {I}$      & 0.37794$+$1.57080\i           & $/$                           & $-$2.79270\\
\hline
$\rm {I}$      & 0.00000$-$0.63971\i           & $/$                           & $-$1.65752\\ 
\hline
$\rm {II}$     & $\alpha_2^- +\frac{\eta}{2}$  & $/$                           & $-$0.19129\\ 
\hline
$G_M^-$        & 0.02501$-$0.00000\i           & 0.14611$-$0.00000\i           & $-$9.77733\\ 
\hline
$G_M^-$        & 0.02349$-$0.00000\i           & 0.34894$+$0.00000\i           & $-$6.67352\\ 
\hline
$G_M^-$        & 0.13934$+$0.00000\i           & 0.34403$-$0.00000\i           & $-$4.97198\\ 
\hline
$G_M^-$        & 0.01837$+$0.00000\i           & 0.00000$+$1.33036\i           & $-$3.73299\\ 
\hline
$G_M^-$        & 0.00000$+$1.30032\i           & 0.12237$+$0.00000\i           & $-$2.28313\\ 
\hline
$G_M^-$        & 0.00000$+$0.51922\i           & 0.00000$+$0.02021\i           & $-$1.86107\\ 
\hline
$G_M^-$        & 0.00000$+$0.16297\i           & 0.00000$+$0.33703\i           & $-$0.41463\\ 
\hline
$G_M^-$        & 0.00000$+$1.22809\i           & 0.25132$-$0.00000\i           & \phantom{+}0.11479\\ 
\hline
$G_M^-$        & 0.34291$+$0.25318\i           & 0.34291$-$0.25318\i           & \phantom{+}0.98156\\ 
\hline
$G_M^-$        & 0.00000$+$1.10335\i           & 0.51404$+$0.00000\i           & \phantom{+}2.52472\\ 
\hline
$G_M^-$        & 0.00000$+$0.99718\i           & 0.94065$+$1.57080\i           & \phantom{+}4.14559\\ 
\hline
\end{tabular}
\end{table}

\begin{table}[H]
\centering
\caption{Numerical solutions of BAEs \eqref{iBAE_I_x:2}, \eqref{iBAE_I_x:3} and \eqref{BAE_G_M^-_x} for $N=4$, $M=2$, $\eta=0.5\rm{i}$, $\alpha_1^-=0.4\rm{i}$, $\alpha_1^+=0.7\rm{i}$, $\theta_+=0.9\i$, $\alpha_2^-=-0.73\i$, $\alpha_2^+=-\eta-\alpha_2^-$ and $\theta_-=(N-1-2M)\eta-(\alpha_1^- +\alpha_2^- +\alpha_1^+ +\alpha_2^+) -\i \pi +\theta_+$.}
\label{Numerical_4_2_x_pd}
\begin{tabular}{|c|c|c|c|}
\hline
Class & $x_1\,\,\mbox{or}\,\,z_1$         & $x_2\,\,\mbox{or}\,\,z_2$     & $E$ \\ 
\hline
$\rm {I}$      & 0.12407$+$0.00000\i      & 0.49802$-$0.00000\i           & $-$10.60115\\ 
\hline
$\rm {I}$      & 0.05851$+$0.00000\i      & 0.00000$-$0.50432\i           & $-$7.65875\\ 
\hline
$\rm {I}$      & 0.00000$-$0.58859\i      & 0.22116$+$0.00000\i           & $-$5.49838\\
\hline
$\rm {I}$      & 0.87509$-$0.49477\i      & 0.87509$+$0.49477\i           & $-$3.44477\\ 
\hline
$\rm {I}$      & 0.72883$-$0.00000\i      & 0.00000$+$0.63031\i           & $-$2.27166\\ 
\hline
$\rm {I}$      & 0.00000$-$0.66184\i      & 0.00000$-$0.76295\i           & $-$0.65837\\ 
\hline
$\rm {II}$     & 0.03842$+$0.00000\i      & $\alpha_2^- +\frac{\eta}{2}$  & $-$7.52464\\ 
\hline
$\rm {II}$     & 0.19212$-$0.00000\i      & $\alpha_2^- +\frac{\eta}{2}$  & $-$4.85465\\ 
\hline
$\rm {II}$     & 0.65590$-$0.00000\i      & $\alpha_2^- +\frac{\eta}{2}$  & $-$1.01697\\ 
\hline
$\rm {II}$     & 0.06279$+$0.67908\i      & $\alpha_2^- +\frac{\eta}{2}$  & \phantom{+}1.14570$+$0.24748\rm {i}\\ 
\hline
$\rm {II}$     & 0.06279$-$0.67908\i      & $\alpha_2^- +\frac{\eta}{2}$  & \phantom{+}1.14570$-$0.24748\rm {i}\\ 
\hline
$G_M^-$        & 0.02004$+$0.00000\i      & $/$                           & $-$4.24639\\ 
\hline
$G_M^-$        & 0.12744$-$0.00000\i      & $/$                           & $-$2.71370\\ 
\hline
$G_M^-$        & 0.26954$-$0.00000\i      & $/$                           & $-$0.17412\\ 
\hline
$G_M^-$        & 0.63500$+$0.00000\i      & $/$                           & \phantom{+}2.33373\\ 
\hline
$G_M^-$        & 0.00000$+$1.01143\i      & $/$                           & \phantom{+}3.91447\\ 
\hline
\end{tabular}
\end{table}

\begin{table}[H]
\centering
\caption{Numerical solutions of BAEs \eqref{iBAE_I_x:2}, \eqref{iBAE_I_x:3} and \eqref{BAE_G_M^-_x} for $N=4$, $M=3$, $\eta=0.5\rm{i}$, $\alpha_1^-=0.4\rm{i}$, $\alpha_1^+=0.7\rm{i}$, $\theta_+=0.9\i$, $\alpha_2^-=-0.73\i$, $\alpha_2^+=-\eta-\alpha_2^-$ and $\theta_-=(N-1-2M)\eta-(\alpha_1^- +\alpha_2^- +\alpha_1^+ +\alpha_2^+) -\i \pi +\theta_+$.}
\label{Numerical_4_3_x_pd}
\begin{tabular}{|c|c|c|c|c|}
\hline
Class & $x_1\,\,\mbox{or}\,\,z_1$          & $x_2\,\,\mbox{or}\,\,z_2$ & $x_3\,\,\mbox{or}\,\,z_3$    & $E$ \\ 
\hline
$\rm {I}$      & 0.11412$-$0.00000\i       & 0.18537$+$0.51499\i       & 0.18537$-$0.51499\i          & $-$6.57557\\ 
\hline
$\rm {I}$      & 0.13390$-$0.00000\i       & 0.08772$+$0.51041\i       & 0.08772$-$0.51041\i          & $-$4.55293\\ 
\hline
$\rm {I}$      & 0.43621$+$0.00000\i       & 0.07491$+$0.64075\i       & 0.07491$-$0.64075\i          & $-$2.01206\\
\hline
$\rm {I}$      & 0.00000$-$0.64601\i       & 0.17519$+$0.83385\i       & 0.17519$-$0.83385\i          & $-$0.03155\\ 
\hline
$\rm {II}$     & 0.04548$-$0.00000\i       & 0.24999$+$0.00000\i       & $\alpha_2^- +\frac{\eta}{2}$ & $-$11.13304\\ 
\hline
$\rm {II}$     & 0.03682$-$0.00833\i       & 0.23579$+$0.54543\i       & $\alpha_2^- +\frac{\eta}{2}$ & $-$6.50339$+$1.17262\i\\ 
\hline
$\rm {II}$     & 0.03682$+$0.00833\i       & 0.23579$-$0.54543\i       & $\alpha_2^- +\frac{\eta}{2}$ & $-$6.50339$-$1.17262\i\\ 
\hline
$\rm {II}$     & 0.00000$-$0.43917\i       & 0.00000$+$0.06615\i       & $\alpha_2^- +\frac{\eta}{2}$ & $-$4.43467\\ 
\hline
$\rm {II}$     & 0.18608$+$0.50411\i       & 0.15899$-$0.04741\i       & $\alpha_2^- +\frac{\eta}{2}$ & $-$4.04265$+$0.53869\i\\ 
\hline
$\rm {II}$     & 0.18608$-$0.50411\i       & 0.15899$+$0.04741\i       & $\alpha_2^- +\frac{\eta}{2}$ & $-$4.04265$-$0.53869\i\\ 
\hline
$\rm {II}$     & 0.11258$+$0.60233\i       & 0.37824$-$0.03412\i       & $\alpha_2^- +\frac{\eta}{2}$ & $-$0.80789$+$0.34733\i\\ 
\hline
$\rm {II}$     & 0.11258$-$0.60233\i       & 0.37824$+$0.03412\i       & $\alpha_2^- +\frac{\eta}{2}$ & $-$0.80789$-$0.34733\i\\ 
\hline
$\rm {II}$     & 0.19048$+$0.74311\i       & 0.19048$-$0.74311\i       & $\alpha_2^- +\frac{\eta}{2}$ & \phantom{+}1.68687\\ 
\hline
$\rm {II}$     & 0.01868$+$0.63162\i       & 0.26025$+$0.72703\i       & $\alpha_2^- +\frac{\eta}{2}$ & \phantom{+}2.21093$+$0.68543\i\\ 
\hline
$\rm {II}$     & 0.01868$-$0.63162\i       & 0.26025$-$0.72703\i       & $\alpha_2^- +\frac{\eta}{2}$ & \phantom{+}2.21093$-$0.68543\i\\ 
\hline
$G_M^-$        & $/$                       & $/$                       & $/$                          & \phantom{+}3.21499\\ 
\hline
\end{tabular}
\end{table}

\begin{table}[H]
\centering
\caption{Numerical solutions of BAEs \eqref{iBAE_I_x:2}, \eqref{iBAE_I_x:3} and \eqref{BAE_G_M^-_x} for $N=5$, $M=3$, $\eta=0.4\rm{i}$, $\alpha_1^+=0.3\rm{i}$, $\alpha_2^+=0.75\i$, $\theta_+=0.8\rm{i}$, $\alpha_1^-=0.66\rm{i}$, $\alpha_2^-=-\eta-\alpha_1^-$ and $\theta_-=(N-1-2M)\eta-(\alpha_1^- +\alpha_2^- +\alpha_1^+ +\alpha_2^+) -\i \pi +\theta_+$.}
\label{Numerical_5_3_x_pd}
\begin{tabular}{|c|c|c|c|c|}
\hline
Class          & $x_1\,\,\mbox{or}\,\,z_1$       & $x_2\,\,\mbox{or}\,\,z_2$     & $x_3\,\,\mbox{or}\,\,z_3$       & $E$ \\ 
\hline
$\rm {I}$      & 0.08375$+$0.00000\i             & 0.23526$+$0.20488\i           & 0.23526$-$0.20488\i    & $-$11.01420\\ 
\hline
$\rm {I}$      & 0.35665$-$0.00000\i             & 0.04401$-$0.00000\i           & 0.00000$+$0.41367\i    & $-$8.56296\\ 
\hline
$\rm {I}$      & 0.36470$-$0.00000\i             & 0.13775$-$0.00000\i           & 0.00000$+$0.46116\i    & $-$6.92546\\ 
\hline
$\rm {I}$      & 0.05572$-$0.00000\i             & 0.07170$+$0.51720\i           & 0.07170$-$0.51720\i    & $-$6.34229\\ 
\hline
$\rm {I}$      & 0.11958$-$0.00000\i             & 0.04686$+$0.52378\i           & 0.04686$-$0.52378\i    & $-$4.80486\\  
\hline
$\rm {I}$      & 0.22593$-$0.00000\i             & 0.00000$+$0.58462\i           & 0.00000$+$0.51203\i    & $-$2.62303\\  
\hline
$\rm {I}$      & 0.56495$-$0.00000\i             & 0.44753$+$0.48695\i           & 0.44753$-$0.48695\i    & $-$2.53193\\  
\hline
$\rm {I}$      & 0.00000$+$0.49479\i             & 0.45821$+$0.24613\i           & 0.45821$-$0.24613\i    & $-$1.48848\\  
\hline
$\rm {I}$      & 0.55086$-$0.00000\i             & 0.00000$+$0.67295\i           & 0.00000$+$0.50135\i    & $-$0.31683\\  
\hline
$\rm {I}$      & 0.00000$+$0.49987\i             & 0.12923$+$0.75882\i           & 0.12923$-$0.75882\i    & \phantom{+}0.91763\\  
\hline
$\rm {II}$     & 0.18025$+$0.00000\i             & 0.07392$-$0.00000\i           & $\alpha_1^- +\frac{\eta}{2}$    & $-$12.33596\\  
\hline
$\rm {II}$     & 0.47908$-$0.00000\i             & 0.07019$-$0.00000\i           & $\alpha_1^- +\frac{\eta}{2}$    & $-$9.28015\\  
\hline
$\rm {II}$     & 0.06403$-$0.00000\i             & 0.00000$+$0.76057\i           & $\alpha_1^- +\frac{\eta}{2}$    & $-$7.65745\\  
\hline
$\rm {II}$     & 0.46590$-$0.00000\i             & 0.16327$+$0.00000\i           & $\alpha_1^- +\frac{\eta}{2}$    & $-$7.07566\\  
\hline
$\rm {II}$     & 0.05685$-$0.00000\i             & 0.00000$+$0.50284\i           & $\alpha_1^- +\frac{\eta}{2}$    & $-$6.92206\\  
\hline
$\rm {II}$     & 0.14307$+$0.00000\i             & 0.00000$+$0.76963\i           & $\alpha_1^- +\frac{\eta}{2}$    & $-$5.75983\\  
\hline
$\rm {II}$     & 0.12769$-$0.00000\i             & 0.00000$+$0.50140\i           & $\alpha_1^- +\frac{\eta}{2}$    & $-$5.23932\\  
\hline
$\rm {II}$     & 0.25164$+$0.19876\i             & 0.25164$-$0.19876\i           & $\alpha_1^- +\frac{\eta}{2}$    & $-$4.01041\\  
\hline
$\rm {II}$     & 0.27217$+$0.00000\i             & 0.00000$+$0.78939\i           & $\alpha_1^- +\frac{\eta}{2}$    & $-$3.36881\\  
\hline
$\rm {II}$     & 0.24246$-$0.00000\i             & 0.00000$+$0.50055\i           & $\alpha_1^- +\frac{\eta}{2}$    & $-$2.85480\\  
\hline
$\rm {II}$     & 0.60978$+$0.28224\i             & 0.60978$-$0.28224\i           & $\alpha_1^- +\frac{\eta}{2}$    & $-$2.10206\\  
\hline
$\rm {II}$     & 0.67961$-$0.00000\i             & 0.00000$+$0.83610\i           & $\alpha_1^- +\frac{\eta}{2}$    & $-$1.32567\\ 
\hline
$\rm {II}$     & 0.59240$-$0.00000\i             & 0.00000$+$0.50014\i           & $\alpha_1^- +\frac{\eta}{2}$    & $-$0.50100\\  
\hline
$\rm {II}$     & 0.10157$+$0.93782\i             & 0.10157$-$0.93782\i           & $\alpha_1^- +\frac{\eta}{2}$    & $-$0.42934\\  
\hline
$\rm {II}$     & 0.15231$+$0.79913\i             & 0.00002$+$0.49999\i           & $\alpha_1^- +\frac{\eta}{2}$    & \phantom{+}0.77720$+$0.18813\i\\  
\hline
$\rm {II}$     & 0.15231$-$0.79913\i             & 0.00002$-$0.49999\i           & $\alpha_1^- +\frac{\eta}{2}$    & \phantom{+}0.77720$-$0.18813\i\\  
\hline
$G_M^-$        & 0.05170$-$0.00000\i             & $/$                           & $/$    & $-$5.23612\\  
\hline
$G_M^-$        & 0.11310$+$0.00000\i             & $/$                           & $/$    & $-$3.83692\\  
\hline
$G_M^-$        & 0.19919$+$0.00000\i             & $/$                           & $/$    & $-$1.84627\\  
\hline
$G_M^-$        & 0.35072$+$0.00000\i             & $/$                           & $/$    & \phantom{+}0.15081\\  
\hline
$G_M^-$        & 0.79908$+$0.00000\i             & $/$                           & $/$    & \phantom{+}1.59324\\  
\hline
$G_M^-$        & 0.00000$-$1.46384\i             & $/$                           & $/$    & \phantom{+}2.27997\\  
\hline
\end{tabular}
\end{table}

\subsection{Diagonal case}
\label{App:data:3}

\begin{table}[htbp]
\caption{Numerical solutions of the BAEs \eqref{BAEs:diag} with $N=5$, $M=4$, $\eta=0.95$, and $\alpha_-=-0.3$. Upper panel: $\alpha_+=-\alpha_--\eta+0.001$. Lower panel: $\alpha_+=-\alpha_--\eta+0.0001$.}\label{Tab:diag:1}
\centering
\begin{tabular}{|c|c|c|c|c|}
\hline 
 $\lambda_1$& $\lambda_2$& $\lambda_3$& $\lambda_4$& $E$ \\
 \hline
 0.77601$+$0.00000\i & 3.25581+0.79748\i & 3.25581$-$0.79748\i & 0.17399$+$0.00000\i & $-$0.74354 \\
 \hline 
 1.12520$+$0.00000\i & 3.62913+0.79837\i & 3.62913$-$0.79837\i & 0.17500$+$0.00000\i & $-$4.23577 \\
 \hline 
 0.00000$+$0.74106\i & 2.01110+0.74275\i & 2.01110$-$0.74275\i & 0.17500$+$0.00000\i & $-$9.14377 \\
 \hline
 0.00000$+$0.37585\i & 2.31632+0.79009\i & 2.31632$-$0.79009\i & 0.17500$+$0.00000\i & $-$12.10113 \\
 \hline 
 0.00000$+$0.16458\i & 2.59433$-$0.79340\i & 2.59433+0.79340\i & 0.17500$+$0.00000\i & $-$14.65478 \\
 \hline
\end{tabular}

\vspace{0.5cm}

\begin{tabular}{|c|c|c|c|c|}
\hline 
 $\lambda_1$& $\lambda_2$& $\lambda_3$& $\lambda_4$& $E$ \\
 \hline
 0.77510$+$0.00000\i & 3.83142+0.78922\i & 3.83142$-$0.78922\i & 0.17490$+$0.00000\i & $-$0.74150 \\
 \hline 
 1.12520$+$0.00000\i & 4.20558$-$0.78950\i & 4.20558+0.78950\i & 0.17500$+$0.00000\i & $-$4.23746 \\
 \hline 
 0.00000$+$0.74390\i & 2.58613$-$0.77192\i & 2.58613+0.77192\i & 0.17500$+$0.00000\i & $-$9.14189 \\
 \hline 
 0.00000$+$0.37563\i & 2.89406+0.78689\i & 2.89406$-$0.78689\i & 0.17500$+$0.00000\i & $-$12.09923 \\
 \hline 
 0.00000$+$0.16466\i & 3.17076+0.78793\i & 3.17076$-$0.78793\i & 0.17500$+$0.00000\i & $-$14.65279 \\
 \hline
\end{tabular}
\end{table}

\begin{table}[htbp]
\centering
\caption{Numerical solutions of BAEs \eqref{BAEs:diag} with $N=5,M=5$. From top to bottom: (1) $\eta =0.95$, $\alpha _-=-0.3$ and $\alpha _+=-\alpha _--\eta+0.001$; (2) $\eta =0.95$, $\alpha _-=-0.3$ and $\alpha _+=-\alpha _--\eta +0.0001$; (3) $\eta =0.63\i$, $\alpha _-=-0.3\i$ and $\alpha _+=-\alpha _--\eta+0.001\i$; (4) $\eta =0.63\i$, $\alpha _-=-0.3\i$ and $\alpha _+=-\alpha _--\eta+0.00001\i$.}\label{Tab:diag:2}
\begin{tabular}{|c|c|c|c|}
\hline 
$\lambda_{1,2}$& $\lambda_{3,4}$  &  $\lambda_5$& $E$ \\
\hline
1.64276$\pm$0.36732\i & 1.57428$\pm$1.13619\i & 0.17500$+$0.00000\i & $-$5.69980 \\
\hline
\end{tabular}

\vspace{0.3cm}

\begin{tabular}{|c|c|c|c|}
\hline 
$\lambda_{1,2}$& $\lambda_{3,4}$  & $\lambda_5$ & $E$ \\
\hline
1.91677$\pm$0.37636\i & 1.87866$\pm$1.15655\i & 0.17500$+$0.00000\i & $-$5.69775 \\
\hline
\end{tabular}

\vspace{0.3cm}

\begin{tabular}{|c|c|c|c|}
\hline 
$\lambda_{1,2}$& $\lambda_{3,4}$  & $\lambda_5$ & $E$ \\
\hline
0.63004$\pm$0.32752\i & 0.47025$\pm$0.94893\i & 0.00000+0.01566\i & $-$3.63018 \\
\hline
\end{tabular}

\vspace{0.3cm}

\begin{tabular}{|c|c|c|c|}
\hline 
$\lambda_{1,2}$& $\lambda_{3,4}$  & $\lambda_5$ & $E$ \\
\hline
1.13697$\pm$0.36042\i &  1.05809$\pm$1.12597\i & 0.00000+0.01501\i & $-$3.62461\\
\hline
\end{tabular}
\end{table}


\end{document}